\pdfoutput=1
\PassOptionsToPackage{table}{xcolor}
\documentclass[]{alaya}
\usepackage{makecell}
\usepackage{wrapfig}
\usepackage{tabularx}
\usepackage{textcomp}
\usepackage{stfloats}
\usepackage{url}
\usepackage{verbatim}
\usepackage{titlesec}
\usepackage{adjustbox}
\usepackage{multirow}
\usepackage{pifont}
\usepackage[sc]{mathpazo}
\usepackage{tikz}
\usetikzlibrary{arrows.meta,positioning,fit,backgrounds,calc,shapes.geometric,shadows.blur,decorations.pathreplacing}
\usepackage{comment}
\usepackage{amsmath,amssymb}
\usepackage{colortbl}
\usepackage[numbers]{natbib}
\usepackage{color}
\usepackage{booktabs} 
\usepackage{hyperref}
\usepackage{graphicx}
\RequirePackage{xspace}
\makeatletter
\DeclareRobustCommand\onedot{\futurelet\@let@token\@onedot}
\def\@onedot{\ifx\@let@token.\else.\null\fi\xspace}
\usepackage[most]{tcolorbox}
\usepackage{xcolor}
\usepackage{array}
\usepackage{siunitx} 
\definecolor{headerpurple}{HTML}{d8d2fc}
\definecolor{rowgray}{gray}{0.95}

\definecolor{tablered}{rgb}{1, 0.7, 0.7}
\definecolor{tableorange}{rgb}{1, 0.85, 0.7}
\definecolor{tableyellow}{rgb}{1, 1, 0.7}
\newcommand{\best}{\cellcolor{tablered}}
\newcommand{\sbest}{\cellcolor{tableorange}}
\newcommand{\tbest}{\cellcolor{tableyellow}}

\makeatother

\definecolor{adptorange}{RGB}{248, 205, 172}
\definecolor{cmpblue}{RGB}{189, 215, 238}
\definecolor{cmpblue}{RGB}{189, 215, 238}

\definecolor{our_red}{RGB}{232,157,160}
\definecolor{our_blue}{RGB}{136,206,230}
\definecolor{our_orange}{RGB}{246,200,168}
\definecolor{our_green}{RGB}{178,211,164}

\definecolor{attn_code0}{RGB}{247,215,200}
\definecolor{attn_code1}{RGB}{238,169,139}
\definecolor{mlp_code0}{RGB}{204,201,221}
\definecolor{mlp_code1}{RGB}{102,95,153}
\definecolor{mygray}{HTML}{f0f0f0}

\definecolor{token_blue}{RGB}{84, 120, 140}

\usepackage{bbding}
\usepackage{fontawesome}
\usepackage{xspace}
\newcommand{\xmark}{\ding{55}}
\newcommand{\cmark}{\ding{51}}
\usepackage{float}

\newlength\savewidth
\newcommand{\tablestyle}[2]{\setlength{\tabcolsep}{#1}\renewcommand{\arraystretch}{#2}\centering\footnotesize}
\newcommand{\bnum}[1]{{\normalfont\bfseries #1}}

\newcolumntype{x}[1]{>{\centering\arraybackslash}p{#1pt}}
\newcolumntype{y}[1]{>{\raggedright\arraybackslash}p{#1pt}}
\newcolumntype{z}[1]{>{\raggedleft\arraybackslash}p{#1pt}}

\renewcommand{\paragraph}[1]{\vspace{1mm}\noindent\textbf{#1}}

\renewcommand{\paragraph}[1]{\vspace{1.25mm}\noindent\textbf{#1}}

\usepackage{algorithm}
\usepackage{listings}

\definecolor{codeblue}{rgb}{0.25, 0.5, 0.5}
\definecolor{codekw}{rgb}{0.35, 0.35, 0.75}
\lstdefinestyle{Pytorch}{
    language = Python,
    backgroundcolor = \color{white},
    basicstyle = \fontsize{9pt}{8pt}\selectfont\ttfamily\bfseries,
    columns = fullflexible,
    aboveskip=1pt,
    belowskip=1pt,
    breaklines = true,
    captionpos = b,
    commentstyle = \color{codeblue},
    keywordstyle = \color{codekw},
}

\definecolor{green}{HTML}{009000}
\definecolor{red}{HTML}{ea4335}

\newsavebox{\KaiTitleWord}
\DeclareRobustCommand{\KaiNinjaGradient}{%
  \begingroup
  \sbox{\KaiTitleWord}{KaiNinja}%
  \raisebox{-\dp\KaiTitleWord}{%
    \includegraphics[width=\wd\KaiTitleWord,
      height=\dimexpr\ht\KaiTitleWord+\dp\KaiTitleWord\relax]{assets/kaininja-title-gradient.pdf}%
  }%
  \endgroup
}

\newif\ifvtwo
\vtwotrue
\ifvtwo
  
  \title{\texorpdfstring{\KaiNinjaGradient}{KaiNinja}: Extending Native 3D Generators to the Part Level}
\else
  
  \title{\texorpdfstring{\KaiNinjaGradient}{KaiNinja}: Mask- and Segmentation-Free Part-Aware 3D Generation via Dual Sparse Voxel Flow
  }
\fi
\author[1,2]{Ruihan Yu}
\author[1,2]{Lian Fu}
\author[1,2]{Muyao Niu}
\author[1]{Zheng-hui Huang}
\author[1,3]{Yu-Ju Tsai}
\author[1,2]{Sho Kuno}
\author[1]{Fengbo Lan}
\author[1]{Yonghao Yu}
\author[1,4]{Erwin Wu}
\author[3]{Ming-Hsuan Yang}
\author[1, \dagger]{Kaipeng Zhang}
\author[1,\dagger]{Zhixiang Wang}

\affiliation[1]{Alaya Lab\\}
\affiliation[2]{The University of Tokyo}
\affiliation[3]{University of California, Merced}
\affiliation[4]{Institute of Science Tokyo}

\abstract{
Native 3D generators turn one image into a single mesh. TRELLIS.2 and its
peers deliver high-fidelity non-watertight geometry with materials, but the
output is one fused object, while downstream work such as editing, rigging and
simulation operates on part-level assets. A naive idea is to run a 3D
segmentation network on the fused mesh that TRELLIS.2 generates, but such
pipelines are slow and bounded by the accuracy of the segmentation. We want a
simple way to extend an existing native 3D generator to the part level. But we face a critical problem: the O-Voxel grid stores one sheet of surface per voxel, so a single volume
cannot represent the interface where two parts touch, at any resolution. We introduce a dual-volume representation to solve this problem and put
forward KaiNinja, a part-level extension of TRELLIS.2 built on a dual-volume
form of its O-Voxel representation.
KaiNinja keeps the generation speed and quality of TRELLIS.2 while
extending it to the part level, with no mask or segmenter in the pipeline. Its
training data come from sources of many kinds, including CAD models and assets
authored by an LLM-driven agent; to our knowledge it is the first 3D generative
model trained on agent-authored part data. Surprisingly, we also find that whole-object
fidelity improves over the same backbone fine-tuned on the same dataset.
Against part generation pipelines of different paradigms, it lowers
whole-object Chamfer distance by $40\%$ and raises strict part F-score by
$16\%$.
}

\project{\url{https://alaya-lab.github.io/KaiNinja}}
\code{\url{https://github.com/AlayaLab/KaiNinja}}
\correspondence{Zhixiang Wang (Project Lead), Kaipeng Zhang}
\date{\today} 

\newlength{\QualGridCellWidth}
\newcommand{\QualGridSetup}[2]{%
  \setlength{\tabcolsep}{1.2pt}%
  \renewcommand{\arraystretch}{0.5}%
  \setlength{\QualGridCellWidth}{\dimexpr #2/#1-2\tabcolsep\relax}%
}
\newcommand{\QualGridImage}[1]{\includegraphics[width=\QualGridCellWidth]{#1}}
\newcommand{\QualGridHeader}[1]{%
  \parbox[b]{\QualGridCellWidth}{\centering\sffamily\footnotesize\bfseries #1}%
}

\begin{document}
\maketitle

\begin{figure}[!ht]
    \centering
  \includegraphics[width=\linewidth]{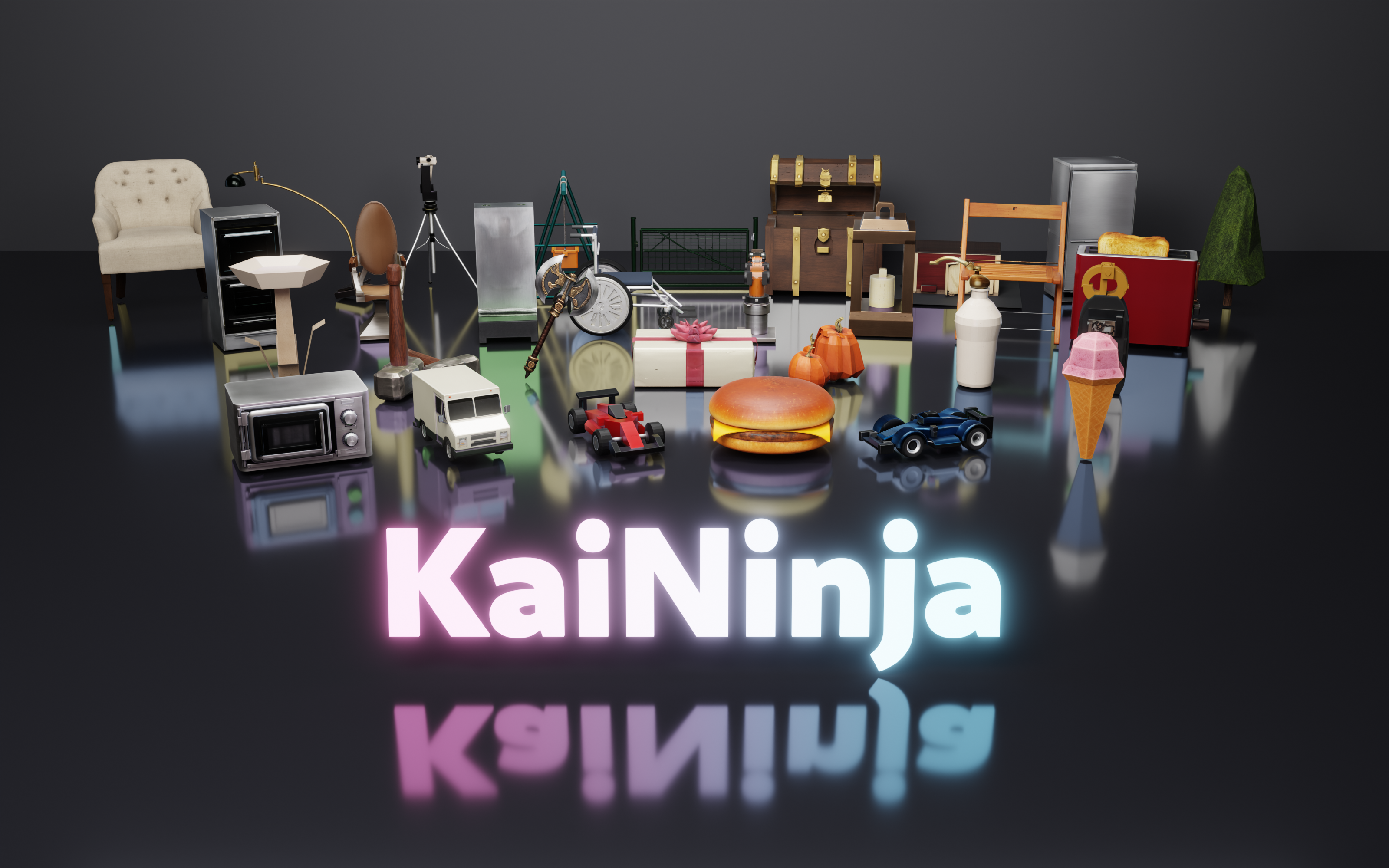}\vspace{5pt}
  \captionof{figure}{\textbf{Part-aware image-to-3D generation with KaiNinja.}
  Each mesh is generated by our model from a single RGB image, with no 2D mask,
  no part segmenter, and no per-object optimization in the geometry pipeline.
  Materials are added with TRELLIS.2 for visualization; mirror reflections use
  distinct colors to show individual parts. Each part is a separate,
  self-contained sub-mesh that can be moved, retextured, or rigged on its own.}
  \label{fig:teaser}
\end{figure}

\section{Introduction}
\label{sec:intro}

Single-image 3D generation has become a practical tool recently. Native generators such
as TRELLIS~\cite{trellis}, TRELLIS.2~\cite{trellis2} and
Hunyuan3D~\cite{hunyuan3d} denoise a 3D latent directly, and the best of them
produce high-fidelity geometry with UVs, materials and PBR appearance.
However, all of them return the object as one entire fused mesh, while most
downstream tasks operate on parts. Editing or retexturing a component, rigging
it for animation, simulating it, and reusing it in another scene all require
each part to be a separate, self-contained mesh. In this paper we study \emph{part-level image-to-3D generation}: from one image,
produce an asset in which every part is its own clean mesh.

Existing methods usually obtain parts in two ways. The first way segments
after generation. A 3D generator produces the whole mesh, a segmentation
network such as P3-SAM~\cite{p3sam} splits it into parts, and a part
generator such as X-Part~\cite{xpart} regenerates every part from the mesh
under the guidance of that segmentation. The second way builds
part structure into the generation process itself, so that parts are
generated together with the mesh from the input image. OmniPart~\cite{omnipart}
is a representative of this way. Methods in this family usually rely on part masks or predicted bounding
boxes, and they generate the parts either in parallel (OmniPart) or one after
another (AutoPartGen~\cite{autopartgen}).
The first family and many methods in the second family depend heavily on the accuracy of a 3D mask or segmentation.
When the mask is wrong, the generated parts overlap or fuse, and the
generator cannot correct it. Besides, the first family also faces a serious
speed problem: measured end to end, the strongest generate-then-segment
cascade is an order of magnitude slower than the methods of the second family
(Section~\ref{sec:exp-main}).

We want a fast and simple way, and therefore a more scalable one, to form
parts during the generation process. Suppose we remove the part bounding
boxes, which predict and constrain the number of parts, three problems then
appear. The first problem is that the number of parts $N$ is not fixed, so
the model has to decide how to split the object and how many parts to
generate, and this still calls for some form of segmentation or clustering.
The second problem is that $N$ can be large. AutoPartGen~\cite{autopartgen}
generates parts one by one in an autoregressive manner, and
PartGen~\cite{partgen} completes and reconstructs each part separately, so
both need $N$ passes through the network. PAct~\cite{pact} instead allocates
one latent volume per part and denoises them jointly, which
multiplies the computation by $N$ and caps $N$ at a fixed maximum. We want to reduce the computation and at the same time leave the
maximum $N$ unlimited. The third problem comes from the representation of
TRELLIS.2 itself. Its O-Voxel grid stores one dual vertex per voxel, solved
by a quadratic error function from the surface crossings of that voxel, and
it extracts the mesh by dual contouring, which connects the dual vertices of
neighboring voxels across each crossed edge. A single voxel therefore holds a
single sheet of surface. When two nearly parallel surfaces fall into the same
voxel, the quadratic error function fits one vertex between them, and the two
sheets collapse into one. This is exactly the situation at a part contact,
where the faces of two parts lie parallel and close.

Faced with these three problems, we notice that the dual-volume representation of
PartPacker~\cite{partpacker} solves all three at once. It packs all parts
into two volumes such that touching parts never share a volume. First, parts can be read off as the connected components of each
volume, so the model never predicts $N$ and never runs a segmenter. Second, any number of parts fits into a fixed pair of streams, so the
backbone runs once and $N$ has no upper bound. Third, the contact faces of touching parts never share a voxel, so nothing collapses.

The remaining question is how to inherit the pretrained prior. The
pretrained flow understands images and generates geometry from them, and we
do not want to retrain these abilities. However, three gaps separate its
prior from our task. First, its prior was learned on whole objects voxelized
as one volume, where the contact faces between parts collapse under the
one-sheet-per-voxel rule, so it has never seen a part interface. Second,
each packed volume is roughly half an object with its contact faces exposed,
which is a distribution it has never seen. Third, a decomposition is a joint
decision, so the two volumes must interact, and two streams that each learn
their own distribution cannot agree on where one part ends and the next
begins.

KaiNinja closes these gaps in order. Each stream starts from the
pretrained weights and is first adapted to its own volume, so the
distribution gap is crossed before any joint structure is asked of it. The
two streams are then connected by new cross-volume attention that is
zero-initialized, so the model begins as exactly the adapted pair and learns
to coordinate from there. The two stages of the cascade take this recipe
differently, because they group their latents differently: the layout flow,
which decides the part structure, receives per-stream weights, while the
refinement flow keeps the pretrained backbone untouched and separates the
streams only through the scope of attention.

The extension keeps the speed and quality of the backbone, and it improves
whole-object fidelity. At $512$ generation resolution, it produces a
part-separated asset in about $24$ seconds per object on one H100.
Surprisingly, it also generates
better whole objects than the same backbone fine-tuned on the same corpus,
including a $38\%$ reduction in Chamfer distance. This suggests that packing is not only a
container for parts but also a better representation of the whole objects.

We train on one corpus assembled from four 3D datasets, which include CAD
models, artist-made assets and agent-authored assets. One of the four is
Articraft-10K~\cite{articraft}, a large collection of assets authored by an LLM-driven 3D
creation agent. Each asset is built by a program that assembles parts from
primitives, so its part labels are exact by construction. On a held-out test
set spanning all four sources, KaiNinja achieves state-of-the-art performance.

In summary, our contributions are:

\begin{itemize}
  \item The first native part-level extension of the state-of-the-art 3D
  object generation foundation model TRELLIS.2. The recipe starts exactly at
  the pretrained model and treats the two stages differently, with fit, merge
  and warm up for the layout flow and an untouched backbone with
  volume-restricted attention for the refinement flow.
  KaiNinja demonstrates state-of-the-art
  performance on part-level image-to-3D generation.
  \item A representation-level argument for why the extension must change the
  representation, namely that one volume cannot hold a part interface, and its
  minimal remedy: dual-volume packing on the generator's own sparse voxel
  grid, which keeps open surfaces, UVs, materials and PBR attributes intact.
  \item The first use of agent-authored 3D assets for training a 3D
  generative model. We train on the released Articraft-10K~\cite{articraft}, whose assets are
  built by programs that assemble each part from primitives, so part labels
  come from authoring rather than from annotation and carry no labeling
  noise.
\end{itemize}

\section{Related Work}
\label{sec:related}

\paragraph{Native image-to-3D generation.}
Most current image-to-3D methods encode shapes into a compact latent and
denoise it with a diffusion or rectified flow model.
3DShape2VecSet~\cite{shape2vecset} and Michelangelo~\cite{michelangelo}
introduced vector-set latents over neural fields, and Dora~\cite{dora}
improved the VAE behind them with sharp-edge sampling. CLAY~\cite{clay},
CraftsMan3D~\cite{craftsman}, Direct3D~\cite{direct3d},
TripoSG~\cite{triposg}, Hunyuan3D~\cite{hunyuan3d} and Hi3DGen~\cite{hi3dgen}
scaled this recipe to high-fidelity generation from a single image.
TRELLIS~\cite{trellis} moved the latent onto a sparse structured grid with
decoders for several output formats, and TRELLIS.2~\cite{trellis2} replaced
the underlying field with the field-free O-Voxel, which supports open,
non-manifold and enclosed surfaces together with PBR appearance. These models
make whole objects easy to obtain, but every one of them outputs one fused
geometry with no parts. KaiNinja extends this line rather than replacing
it. It takes the most recent member, TRELLIS.2, and generates parts on the same
grid the backbone already denoises.

\paragraph{Multi-view and optimization-based 3D generation.}
Before native latents, image-to-3D methods either distilled 2D diffusion
priors through score distillation (DreamFusion~\cite{dreamfusion}) or
generated several views (Zero123++~\cite{zero123pp}, Wonder3D~\cite{wonder3d})
and reconstructed a mesh from them (One-2-3-45~\cite{one2345},
InstantMesh~\cite{instantmesh}). This route is slow and often inconsistent
across views, which is why later work denoises 3D latents directly. Like the
native line, it returns whole objects with no part structure.

\paragraph{Part-aware and compositional 3D generation.}
Structure-aware generation predates the current wave.
StructureNet~\cite{structurenet} generates part hierarchies with graph
networks, SDM-NET~\cite{sdmnet} generates deformable part meshes,
SPAGHETTI~\cite{spaghetti} supports part-aware implicit edits, and
SALAD~\cite{salad} runs a cascaded diffusion from part layout to part
geometry. These works show that parts are worth modeling explicitly, but they
operate on small single-category collections rather than on open-domain
images. Among current methods, the main difference is \emph{where part
structure enters the pipeline}. \emph{(i) Segment, then regenerate.} These
methods take a whole 3D object as input and split it before completing or
regenerating each part. PartGen~\cite{partgen} segments multi-view renders
and completes each part separately, HoloPart~\cite{holopart} completes the
fragments that an external segmenter provides, and X-Part~\cite{xpart} runs
its companion segmenter P3-SAM~\cite{p3sam} on the object and regenerates
all parts at once, guided by the segmentation boxes and features. The
boundaries belong to the segmenter, and part reasoning cannot begin until a
whole mesh exists. \emph{(ii) Masks or boxes before generation.}
OmniPart~\cite{omnipart} segments the input image in 2D and lifts the masks
into a bounding-box layout that steers generation, so the layout is fixed
before any 3D reasoning happens. Part123~\cite{part123} reconstructs a
part-aware shape from a single image with 2D masks, and
ComboVerse~\cite{comboverse} and PhyCAGE~\cite{phycage} segment the image
into components, generate each component separately and then compose them.
\emph{(iii) Autoregressive, one part at a time.}
AutoPartGen~\cite{autopartgen} generates parts in sequence, each conditioned
on the parts already generated, and decides on its own when to stop. This
handles a variable number of parts but costs one pass per part. \emph{(iv)
End-to-end, with no mask and no box.} These methods generate all parts from
the image in one pass and differ in how they encode part identity.
PartPacker~\cite{partpacker} packs all parts into two complementary volumes
that one flow denoises jointly, and PartCrafter~\cite{partcrafter} groups the
latent tokens by part, with the number of parts given as input. Both are
built on vecset backbones that decode a field, which requires watertight
parts and models geometry alone. KaiNinja belongs to this group. It keeps
the dual-volume packing of PartPacker, but moves it onto the sparse voxel
grid that TRELLIS.2 already denoises, through both levels of its structured
latent, and inherits the pretrained prior rather than retraining it.

A separate line reaches parts from the opposite direction. Native mesh
generators model vertex connectivity and face existence explicitly.
MeshGPT~\cite{meshgpt} and MeshAnything~\cite{meshanything} tokenize faces
autoregressively, and Nexus~\cite{nexus} replaces the sequence with diffusion
over vertices and over topology, but all three generate one whole mesh. Two
recent models obtain parts from the connectivity itself. LATO.2~\cite{lato2}
factorizes generation into a flow over vertices and a flow over topology and
adds a part-wise mode, in which a structure planner partitions the scaffold
with part bounding boxes, vertices are generated per part, and connectivity
is predicted per part or jointly and then stitched. Meshy T2~\cite{meshyt2}
generates vertices and edges jointly with flow matching, and its vertex-set
VAE keeps coincident vertices as distinct tokens instead of welding them, so
touching parts are not fused and a multi-part asset comes out as connected
components with no part-wise generation at all. Two things keep this line
from being a comparable baseline today. First, the part-capable models are
not yet available: LATO.2 publishes weights for whole meshes but not for its
part-wise variant, and Meshy T2 has announced a code and weight release. Second, explicit
connectivity remains hardest exactly where part structure is decided, at
dense contacts and at interior surfaces the input view never shows. We treat
this line as complementary.

\paragraph{3D part segmentation.}
Splitting an existing shape is the basic operation that group \emph{(i)}
relies on. PartSLIP~\cite{partslip} uses image-language models.
SAMPart3D~\cite{sampart3d} and Segment Any Mesh~\cite{samesh} lift 2D masks
from SAM~\cite{sam} and SAM~2~\cite{sam2} into 3D. P3-SAM~\cite{p3sam} trains
a promptable native 3D segmenter on part supervision. PartField~\cite{partfield}
learns a feature field for grouping. When these methods run after generation,
they are limited by segmentation quality and cannot recover structure the
generator has already fused.

\paragraph{Program-driven and agentic asset generation.}
A separate route produces part-structured assets by writing programs rather
than by denoising geometry. Infinigen~\cite{infinigen} and Infinite
Mobility~\cite{infinitemobility} generate scenes and articulated objects from
hand-written procedural rules, CAGE~\cite{cage} generates part boxes and joint
parameters from a connectivity graph and retrieves part geometry,
Articulate-Anything~\cite{articulateanything} lets a vision-language model
write the assembly code and retrieve parts with iterative self-correction,
Articraft~\cite{articraft} lets an LLM coding agent build each part from
primitives and join the parts with physical joints, and
img2threejs~\cite{img2threejs} lets a coding agent rebuild the object in a
single reference image as procedural Three.js code, with validation scripts
gating each stage. Part structure, and articulation where it is modeled, are
correct by construction, and the articulated assets can be simulated, but
geometry is bounded by the primitives and the retrieval library. We treat
this route as complementary to generative models and as a data source: the
released Articraft-10K is one of our four training corpora, and its part
labels come from the authoring program rather than from annotation.

\paragraph{Positioning.}
KaiNinja belongs to group \emph{(iv)}, together with PartPacker and
PartCrafter, and it is the first method in this group built on TRELLIS.2 and
its O-Voxel representation. Compared with \emph{(i)} and \emph{(ii)}, no
segmenter, no mask and no box sit on the critical path, so part boundaries
are decided by the generator itself. Compared with \emph{(iii)}, every part
is generated in one pass. Compared with the other members of \emph{(iv)},
the extension keeps the two-level structured latent of the backbone and its
support for open surfaces and PBR materials, which a vecset/SDF
representation does not offer. We compare
against both X-Part cascades, OmniPart, PartPacker and AutoPartGen in
Section~\ref{sec:experiments}.

\section{Method}
\label{sec:method}

\paragraph{Overview.}
KaiNinja extends TRELLIS.2 to the part level, and the design follows the
problems raised in Section~\ref{sec:intro}. Section~\ref{sec:method-repr}
changes the representation: an object with any number of parts is packed into
two interleaved O-Voxel volumes, streams $A$ and $B$, so that parts in the
same volume do not touch. As a result, the generated geometry within each volume can be decomposed into individual parts directly by connected-component analysis, without requiring a separate segmentation network. Section~\ref{sec:method-arch} extends the backbone:
the two-stage flow of TRELLIS.2 becomes a two-stream, two-stage flow that runs
once for any number of parts, and each stage is adapted in the way its latent
grouping asks for. Section~\ref{sec:method-train} organizes training around
the three gaps between the pretrained prior and our task, so that every phase
starts from the pretrained weights and moves them as little as the task
requires. Section~\ref{sec:method-data} describes the corpus, and
Section~\ref{sec:method-infer} describes inference and the two post-processing
steps that turn the two volumes into parts.

\input{figures/fig_pipeline}

\subsection{The Dual-Volume O-Voxel Representation}
\label{sec:method-repr}

\paragraph{O-Voxel.}
TRELLIS.2 stores geometry in O-Voxel, a sparse voxel structure built on dual
contouring~\cite{dualcontouring}. Each active voxel holds one dual vertex and
one crossing flag per axis edge, so it stores a patch of surface rather than a
sample of a field. Because no global signed distance or occupancy field is
fitted, O-Voxel can represent open, non-manifold and enclosed surfaces
directly, together with UVs, materials and PBR attributes. We keep all of
this. An object is voxelized directly from its original textured mesh after a
rigid normalization (centering, scaling into the unit cube and aligning the up
axis), with no remeshing and no watertight conversion. As a result, what the
backbone can represent, the extension can represent too.

\paragraph{One sheet per voxel.}
One property of O-Voxel shapes the design: a voxel holds at most one sheet of
surface. Where two parts touch, their outer surfaces lie against each other as
two nearly coincident sheets. At any finite resolution both sheets fall into
the same voxels, and voxelization keeps one and discards the other. This is a
limit of capacity rather than of resolution. A single volume that spans the
whole object cannot represent a part interface, and a part interface is the
structure a part-level generator has to preserve. Packing parts that touch
into \emph{different} volumes restores one sheet per voxel in each volume, so
both contact faces are kept where parts meet. This is why the extension
begins at the representation. We do not expect any architecture on top of a
single volume to recover surfaces that the input never contained.

\paragraph{Packing parts into two volumes.}
We pack by two-coloring a contact graph $G$. Each part is a node, and two
nodes are joined when the parts touch. PartPacker~\cite{partpacker} detects
contact by dilating SDF grids and measuring interpenetration. We instead read
contact off the O-Voxel grid itself and make it \emph{strict}: two parts are
adjacent if and only if some voxel carries facets of both, and we weight
each edge by the number of such shared voxels. When $G$ is bipartite, its two
color classes are the two volumes. When it is not, we contract edges greedily
until it is: while an odd cycle remains, we take one, pick the edge on it
with the largest weight, merge its two parts into one node, and rebuild the
graph. The result is then two-colored by breadth-first search. This is the
greedy odd cycle contraction of PartPacker (their Algorithm~1), a heuristic
for the bipartite contraction problem~\cite{bipartitecontraction}; it makes
no optimality claim and merges the parts that are in closest contact first. In
the end every part belongs to stream $A$ or $B$, and parts within a stream
do not touch by construction (Figure~\ref{fig:packing}).

\input{figures/fig_packing}

Two-coloring is a heuristic rather than a guarantee, since it works only when
the contact graph is bipartite, and objects whose parts interlock densely are
merged more than their annotation intends. PartPacker suggests, as a possible
remedy, converting the contact graph into a planar graph and applying the
four color theorem. We believe this suggestion does not hold in general: a 3D
contact graph need not be planar, and making it planar deletes contact edges,
which puts touching parts back into one volume. Appendix~\ref{app:coloring}
gives the argument and explains why no fixed number of volumes removes this
limit.

\subsection{Two Streams, Two Stages}
\label{sec:method-arch}

\paragraph{One recipe per stage.}
TRELLIS.2 generates in a cascade of two rectified flows, and the two stages
group their latents differently. The first stage denoises a coarse structure
latent that decides \emph{where} surface exists. The second stage denoises a
structured latent that decides \emph{what} the surface is at each occupied
voxel. Part structure lives mostly in the first decision, because which parts
exist, where they sit and which volume each belongs to is a joint allocation
over both streams. The second stage, in contrast, is local refinement that
the pretrained prior already performs well. For this reason we treat the two
stages differently. The layout flow gets per-stream weights connected by new
cross-volume blocks, while the refinement flow keeps the pretrained backbone
unchanged and separates the streams only through the scope of attention. Both
choices are inheritance decisions, and we explain each below.

\paragraph{Conditioning.}
A DINOv3 ViT-L/16 encoder~\cite{dinov3} reads the image at $512{\times}512$
and produces conditioning tokens. These tokens enter both stages through
cross-attention, as in the backbone. We use classifier-free
guidance~\cite{cfg}: one denoising pass with the image and one with a null
condition, extrapolating past the conditional prediction. Nothing in either
stage is specific to images, so other conditioning signals could be used
instead.

\paragraph{Stage 1: the layout flow.}
The first stage denoises the coarse occupancy of both volumes together. It
works on the sparse structure latent, a $16^3$ grid with $8$ channels per
stream, which the pretrained decoder expands to a $64^3$ occupancy grid. Each
stream is a full copy of the pretrained transformer ($30$ blocks, width
$1536$, $12$ heads, rotary position encoding), with its own input and output
projections and its own timestep modulation. Stream identity is therefore
carried by the weights and needs no extra embedding. After the per-stream
blocks at depths $\{6,12,18,24,29\}$, one shared \emph{cross-volume attention}
block attends over the tokens of both streams at once, with a modulation
branch of its own. These blocks are where the model reasons jointly, for
example when it decides where two touching parts should split. They are
zero-initialized, so at the start of training the network is two independent
pretrained denoisers, each seeing one volume. The training schedule of
Section~\ref{sec:method-train} relies on this property.

\paragraph{Stage 2: the refinement flow.}
Given the predicted layout and the same image tokens, the second flow denoises
the structured latent of both streams, which carries fine geometry and
appearance. Occupancy is fixed by now, so the task is close to the one the
backbone was pretrained for, and we leave the backbone unchanged. Its single
weight set is loaded as released, and the two streams coexist by
\emph{scoping attention}. In $20$ of the $30$ blocks each self-attention layer
attends only over the tokens of one volume, so each stream refines on its own.
In the remaining $10$ blocks ($2,5,8,\dots,29$) it attends over both volumes
at once. Self-attention handles a variable token count natively, so this
regrouping changes no layer width and adds no backbone parameters. In the
merged blocks, stream identity comes from a learned volume embedding, which
is initialized at zero and added to the timestep modulation.

\subsection{Training: Inheriting a Whole-Object Prior}
\label{sec:method-train}

The prior we inherit was trained on whole objects. It understands images and
generates geometry from them, and these are the abilities we want to keep.
However, three gaps separate it from our task (Section~\ref{sec:intro}): its
training data were voxelized as whole objects, so it has never seen a part
interface; a packed volume is roughly half an object with its contact faces
exposed, which is a distribution it has never seen; and the two volumes have
to interact. We therefore organize training so that each phase moves the
weights the least distance the task requires. Every phase uses a flow
matching objective~\cite{flowmatching,rectifiedflow}.

\paragraph{The layout flow: fit, merge, warm up, fine-tune.}
Stage~1 is trained in three phases. \emph{(i) Fit each volume.} Starting from
the pretrained single-stream denoiser, we fine-tune two copies independently,
one on the volume-$A$ latents of our corpus and one on the volume-$B$ latents.
This step carries the prior across the distribution gap. Each copy adapts to
the marginal distribution of its own stream, that is, half objects with
exposed contacts, before any joint structure is asked of it. \emph{(ii) Merge
without loss.} The two fitted copies are loaded as the two streams. Every
component they touch (blocks, projections and timestep modulation) belongs to
one stream, so the fitted weights transfer without reinterpretation. Together
with the zero-initialized cross-volume blocks, the merged model starts out
behaving like the two fitted denoisers. \emph{(iii) Warm up, then fine-tune
jointly.} A frozen warmup holds both streams fixed and trains only the
cross-volume attention and its modulation branch, so the streams learn to
coordinate before their fitted priors are disturbed. Joint fine-tuning then
unfreezes everything. This order is meant to protect the inheritance.
Skipping the warmup would let untrained cross-volume blocks push gradients
through fitted streams, and skipping the per-volume fit would ask one set of
pretrained weights to serve two marginals it has never seen.

Throughout joint fine-tuning we add a \emph{disjointness penalty} to the flow
matching loss. It targets the layer where part structure is decided. Packing
puts every part in exactly one volume, so the two streams should not claim
the same voxel, but nothing in a per-stream flow loss says so. At each step we
recover the predicted clean latent $\hat{x}_0$ from the velocity, decode both
streams with the frozen sparse structure decoder into soft occupancies
$o_A,o_B\in[0,1]$ on the $64^3$ grid, and penalize the overlap,
$\mathcal{L}_{\text{ov}}=\mathbb{E}\,[\,o_A\odot o_B\,]$, added to the flow
loss with weight $\lambda_{\text{ov}}=5$. It costs one decoder pass per step,
and we examine its effect in Section~\ref{sec:exp-abl}.

\paragraph{The refinement flow: warm up, fine-tune.}
Stage~2 needs no per-volume fitting. Its backbone is unchanged and both
streams reuse the pretrained weights as released, so there is no distribution
gap for the weights to cross before joint training. It trains on the same
two-phase schedule: a frozen warmup that updates only the volume embedding,
followed by full fine-tuning. In our experience this is enough, because
scoping attention disturbs the pretrained computation far less than changing
the stream topology would. The two stages therefore receive different
treatment, which is what inheriting one whole-object prior into a cascade
seems to require when the stages group their latents differently.

\subsection{Data Curation}
\label{sec:method-data}
We assemble one corpus from four 3D datasets: a large collection authored by a
3D creation agent rather than by human artists
(\textsc{articraft}~\cite{articraft}), a CAD corpus
(\textsc{fusion360}~\cite{fusion360}), and two collections made by artists
(\textsc{partnext}~\cite{partnext} and \textsc{trellis}~\cite{trellis}).
Articraft turns the creation of an articulated asset into code generation
driven by a language model. The agent writes a program that builds each
semantic part out of primitives and joins the parts with physical joints. Its
parts are therefore not annotations but the structure the agent wrote, so the
part supervision it provides is exact. We use the released Articraft-10K~\cite{articraft}
collection and pass it through the same filters as the other sources. To our
knowledge this is the first time assets authored by an agent have been used
to train a 3D generative model.
For each object we read part annotations off the mesh scene graph, falling
back to connected components when one geometry node fuses everything. We then
voxelize the raw textured mesh into O-Voxel and run the merge-and-two-color
procedure of Section~\ref{sec:method-repr} to obtain the packing.

\paragraph{Quality filtering.}
The corpus passes two filters. Before voxelization, on the normalized meshes,
we keep objects with $2$ to $32$ parts, drop objects whose parts interpenetrate
badly, measured both by the depth of pairwise collisions and by the overlap
between parts under a generalized winding number~\cite{gwn}, and remove
assets dominated by display planes baked in as ground or backdrop quads. After
packing, we require a valid latent and a rendered condition for every object.
We also apply a balance filter, a minimum voxel count and a minimum occupancy
ratio between $A$ and $B$, which discards degenerate packings that leave one
volume empty or nearly so. PartPacker applies no such check. Because O-Voxel
is field-free, we skip the watertight repair and shell dilation that packing
over an SDF~\cite{partpacker} requires, and we keep the original open surfaces
and materials.

\subsection{Inference and Post-processing}
\label{sec:method-infer}
At inference the layout flow samples the structure latents of both volumes
from the image, and the refinement flow fills them in. Two light
post-processing steps then turn the two volumes into parts, and neither
involves a segmenter or a mask.

\paragraph{Merging duplicated occupancy.}
Nothing forces the two streams to stay disjoint at inference, and they
occasionally place the same surface in both volumes. Between the stages we
therefore compare the decoded occupancies. For every pair of connected
components, one from each volume, we measure the fraction of shared voxels and
the depth of embedding. When a pair overlaps by more than half and is
embedded at least three voxels deep, the voxels of the smaller component move
into the volume of the larger. Nothing is deleted, so the union of the two
volumes is unchanged. The step adds negligible cost. Thin shells sometimes
escape it (Section~\ref{sec:conclusion}).

\paragraph{Parts from volumes.}
The frozen TRELLIS.2 VAE decodes the predicted latent of each volume into an
O-Voxel volume, and a textured mesh is extracted from each. Within a volume,
parts do not touch by construction, so we take the connected components of
each mesh as parts. Because the decoder sometimes leaves small gaps, we treat
components whose surfaces lie within two fine voxels of each other as one
part, and attach fragments below $0.5\%$ of the volume's surface area to the
nearest part. The two meshes are then assembled into the finished object,
followed by a light geometric cleanup. Neither step predicts the number of
parts; it follows from the layout the model generated. Section~\ref{sec:exp-abl}
measures what each step contributes.

\section{Experiments}
\label{sec:experiments}

\newcommand{\tbd}{\textcolor{gray}{--}}

\subsection{Setup}
\label{sec:exp-setup}

\paragraph{Data.}
We train on one corpus of $19{,}132$ objects drawn from four 3D datasets
(\textsc{articraft}~\cite{articraft} $8{,}968$,
\textsc{partnext}~\cite{partnext} $6{,}358$, \textsc{trellis}~\cite{trellis}
$1{,}931$, \textsc{fusion360}~\cite{fusion360} $1{,}875$). Each object is
packed into two O-Voxel volumes as in Section~\ref{sec:method-data}. We treat
the four sources as one corpus rather than as separate domains: every object
goes through the same pipeline, and the flows are trained jointly on the
union. We hold $1000$ objects out of training as a test set that spans all
four sources (\textsc{articraft} $513$, \textsc{partnext} $297$,
\textsc{fusion360} $102$, \textsc{trellis} $88$), and we report a breakdown
by source. One baseline, OmniPart, produces no output on $14$ of the $1000$
objects because its sparse convolution backend runs out of resources on very
dense voxel grids. To keep every method on identical inputs, we report all
methods on the $986$ objects that every method completes.

\paragraph{Baselines.}
We compare against a representative of each family in
Section~\ref{sec:related}: X-Part~\cite{xpart}, which segments a generated
mesh and regenerates its parts; OmniPart~\cite{omnipart}, which plans a
layout from masks; AutoPartGen~\cite{autopartgen}, which generates one part
after another; and PartPacker~\cite{partpacker}, which packs dual volumes
over a vecset latent and is the method closest to ours. X-Part takes an
existing mesh rather than an image, so it needs a generator in front of it.
We use TRELLIS.2~\cite{trellis2}, the backbone our own flows extend, at its
released default \texttt{1024\_cascade} setting, and write this cascade as
\mbox{TRELLIS.2@1024 + X-Part}. The X-Part paper feeds it Hunyuan3D-2.5,
whose weights are not released, so we also report it on
Hunyuan3D-2.1~\cite{hunyuan3d21}, the released model closest to its own
setup. The generator in front matters: moving X-Part from TRELLIS.2 to
Hunyuan3D-2.1 raises it by $0.038$ F1$_P^{0.05}$, and we report both
pairings throughout. Neither choice handicaps X-Part, since our own Stage~2 starts from
the $512$ shape flow of TRELLIS.2 and never runs the $1024$ cascade. Every
method therefore starts from the same single image and is scored by the same
protocol. Each baseline runs its own released pipeline, with mesh
simplification and decimation switched off so that geometry is scored as the
model decodes it.

\paragraph{Metrics.}
Following OmniPart~\cite{omnipart}, we normalize every shape into
$[-0.5,0.5]^3$ and report Chamfer distance (CD) and F-score at thresholds
$0.1$ and $0.05$, for the whole object ($W$) and for its parts ($P$). Methods
use different canonical frames, so for each object we align the prediction to
the ground truth by picking, out of $48$ signed axis permutations, the one
with the smallest CD on the whole object, and we reuse that transform when
scoring parts. Each method contributes the set of parts its own pipeline
produces, and we impose no further splitting or filtering on any method. For
KaiNinja a part is a connected component of a decoded volume
(Section~\ref{sec:method-infer}). Methods that emit parts directly contribute
those parts as released. Both routes end at the same kind of object, a set of
parts whose size is not tied to the ground truth, which is what the metric
consumes.

Ground-truth and predicted parts are matched \emph{one-to-one} by Hungarian
assignment, at cost $1-$IoU on surface labels transferred by nearest
neighbor. We match on IoU rather than CD because IoU is bounded, so one badly
placed part cannot dominate the assignment. Predictions left unmatched and
ground-truth parts left uncovered both count against the method, so over- and
under-segmentation are both penalized. On the matched pairs we report the mean
IoU (mIoU$_P$) and CD / F-score. OmniPart's paper does not specify how
predicted parts are matched to ground truth, and no compared method releases
part-metric code, so the protocol above is our own choice.

\paragraph{Implementation.}
DINOv3 ViT-L/16~\cite{dinov3} encodes the input image at $512{\times}512$.
Both flows use the two-stream configuration of
Section~\ref{sec:method-arch}, are trained on the frozen-warmup then
joint-fine-tuning schedule, and run with classifier-free guidance at
inference. The structured latent autoencoder is the pretrained TRELLIS.2 one
and stays frozen throughout. We train only the two flows, so Stage~2 operates
in exactly the released latent space and no autoencoder is retrained.
Appendix~\ref{app:hparams} lists the hyperparameters.

\subsection{Comparison with State of the Art}
\label{sec:exp-main}

\paragraph{Part-level quality.}
Table~\ref{tab:main} gives the overall comparison and
Table~\ref{tab:perdataset} the breakdown by source, both on the $986$ objects
every method completes. KaiNinja is best on all seven metrics of
Table~\ref{tab:main} and on every entry of the breakdown. Against the
strongest baseline, Hunyuan3D-2.1 + X-Part, it gains $0.10$ F1$_W^{0.05}$ and $0.09$ F1$_P^{0.05}$ at
the strict threshold, and its whole-object Chamfer distance is $40\%$ lower.
The gap is small at the loose threshold and widest at the strict one, which
suggests that the difference lies in fine geometry and boundary placement
rather than in coarse layout.

\providecolor{ptAd}{HTML}{8E3D47}
\providecolor{ptAl}{HTML}{D48A92}
\providecolor{pxSlaT}{HTML}{2F5D7E}
\providecolor{pxSlaM}{HTML}{C2DAEB}
\providecolor{pxTanB}{HTML}{A08468}
\providecolor{pxTanL}{HTML}{F7F3EC}
\providecolor{ptCd}{HTML}{A06B28}
\providecolor{ptCl}{HTML}{EFC98E}
\providecolor{ptTd}{HTML}{2F7A6B}
\providecolor{ptTl}{HTML}{A8D5C8}
\providecolor{pxPlmD}{HTML}{6B4E7D}
\providecolor{pxPlmL}{HTML}{D9CBE4}
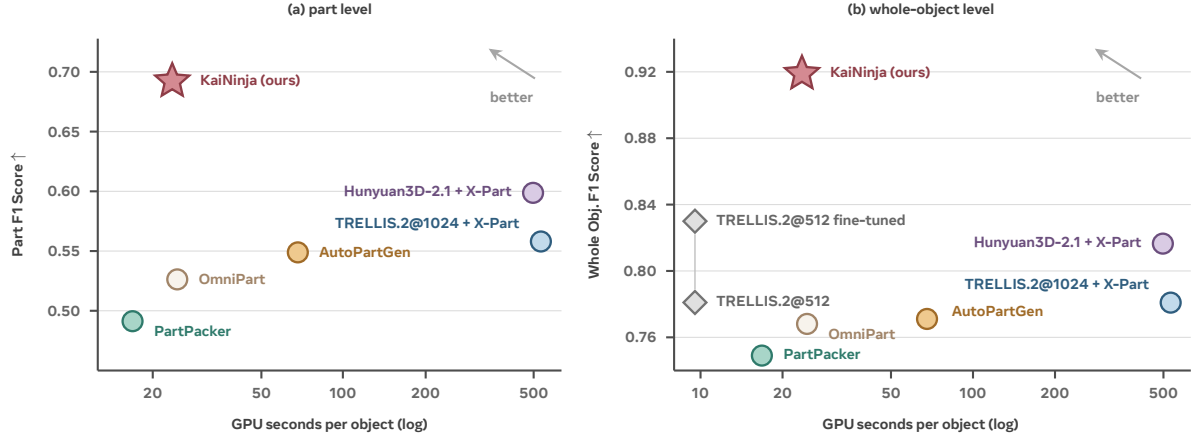
\begin{figure*}[t]
\centering
\resizebox{\textwidth}{!}{%
\begin{tikzpicture}[
  font=\sffamily\small,
  >={Stealth[length=2mm, width=1.8mm]},
  ax/.style={line width=0.9pt, black!70},
  grid/.style={line width=0.4pt, black!15},
  tk/.style={font=\sffamily\scriptsize, black!60},
  lb/.style={font=\sffamily\scriptsize, black!75},
  ours/.style={draw=ptAd, fill=ptAl, line width=1pt},
  xp/.style={draw=pxSlaT, fill=pxSlaM, line width=1pt},
  op/.style={draw=pxTanB, fill=pxTanL, line width=1pt},
  pp/.style={draw=ptTd, fill=ptTl, line width=1pt},
  ag/.style={draw=ptCd, fill=ptCl, line width=1pt},
  hy/.style={draw=pxPlmD, fill=pxPlmL, line width=1pt},
  ft/.style={draw=black!55, fill=black!12, line width=1pt},
]
\def\W{70mm}
\def\H{45mm}

\begin{scope}
\foreach \y in {9,18,27,36,45}{\draw[grid] (0,\y mm) -- (\W,\y mm);}
\draw[ax] (0,0) -- (\W,0);
\draw[ax] (0,0) -- (0,50mm);
\foreach \x/\t in {8.3/20, 24.7/50, 37.0/100, 49.5/200, 65.8/500}{
  \draw[ax] (\x mm,0) -- (\x mm,-1.2mm); \node[tk, below] at (\x mm,-1.4mm) {\t};}
\foreach \y/\t in {9/0.50, 18/0.55, 27/0.60, 36/0.65, 45/0.70}{
  \draw[ax] (0,\y mm) -- (-1.2mm,\y mm); \node[tk, left] at (-1.4mm,\y mm) {\t};}
\node[lb, below] at ($(0.5*\W,-6mm)$) {GPU\,seconds per object (log)};
\node[lb, rotate=90] at (-12mm,25mm) {Part F1 Score $\uparrow$};
\node[pp, circle, minimum size=3mm, inner sep=0pt] (aPP) at (5.25mm,7.43mm) {};
\node[op, circle, minimum size=3mm, inner sep=0pt] (aOP) at (12.0mm,13.73mm) {};
\node[ag, circle, minimum size=3mm, inner sep=0pt] (aAG) at (30.2mm,17.80mm) {};
\node[xp, circle, minimum size=3mm, inner sep=0pt] (aXP) at (66.89mm,19.44mm) {};
\node[hy, circle, minimum size=3mm, inner sep=0pt] (aHX) at (65.71mm,26.77mm) {};
\node[ours, star, star points=5, star point ratio=2.2, minimum size=5.4mm, inner sep=0pt]
      (aUS) at (11.24mm,43.60mm) {};
\node[lb, ptTd, anchor=west]   at ($(aPP)+(2mm,-1.4mm)$) {PartPacker};
\node[lb, pxTanB, anchor=west] at ($(aOP)+(2mm,0)$) {OmniPart};
\node[lb, ptCd, anchor=west]   at ($(aAG)+(2mm,0)$) {AutoPartGen};
\node[lb, pxSlaT, anchor=east] at ($(aXP)+(-2mm,2.6mm)$) {TRELLIS.2@1024 + X-Part};
\node[lb, pxPlmD, anchor=east] at ($(aHX)+(-2mm,0)$) {Hunyuan3D-2.1 + X-Part};
\node[lb, ptAd, anchor=west]   at ($(aUS)+(3mm,0)$) {\textbf{KaiNinja (ours)}};
\draw[->, line width=0.9pt, black!35] (66mm,44mm) -- (59mm,48.5mm);
\node[lb, black!45, anchor=north] at (62.5mm,43.2mm) {better};
\node[lb, anchor=south] at ($(0.5*\W,52mm)$) {\textbf{(a) part level}};
\end{scope}

\begin{scope}[xshift=87mm]
\foreach \y in {5,15,25,35,45}{\draw[grid] (0,\y mm) -- (78mm,\y mm);}
\draw[ax] (0,0) -- (78mm,0);
\draw[ax] (0,0) -- (0,50mm);
\foreach \x/\t in {3.99/10, 16.37/20, 32.74/50, 45.12/100, 57.50/200, 73.87/500}{
  \draw[ax] (\x mm,0) -- (\x mm,-1.2mm); \node[tk, below] at (\x mm,-1.4mm) {\t};}
\foreach \y/\t in {5/0.76, 15/0.80, 25/0.84, 35/0.88, 45/0.92}{
  \draw[ax] (0,\y mm) -- (-1.2mm,\y mm); \node[tk, left] at (-1.4mm,\y mm) {\t};}
\node[lb, below] at (37.25mm,-6mm) {GPU\,seconds per object (log)};
\node[lb, rotate=90] at (-12mm,25mm) {Whole Obj.\ F1 Score $\uparrow$};
\node[ft, diamond, minimum size=3.6mm, inner sep=0pt] (bT2a) at (3.15mm,10.25mm) {};
\node[ft, diamond, minimum size=3.6mm, inner sep=0pt] (bFT)  at (3.15mm,22.5mm) {};
\draw[line width=0.6pt, black!25] (bT2a) -- (bFT);
\node[pp, circle, minimum size=3mm, inner sep=0pt] (bPP) at (13.25mm,2.25mm) {};
\node[op, circle, minimum size=3mm, inner sep=0pt] (bOP) at (20.07mm,7.02mm) {};
\node[xp, circle, minimum size=3mm, inner sep=0pt] (bXP) at (74.96mm,10.23mm) {};
\node[hy, circle, minimum size=3mm, inner sep=0pt] (bHX) at (73.78mm,19.12mm) {};
\node[ag, circle, minimum size=3mm, inner sep=0pt] (bAG) at (38.18mm,7.77mm) {};
\node[ours, star, star points=5, star point ratio=2.2, minimum size=5.4mm, inner sep=0pt]
      (bUS) at (19.30mm,44.73mm) {};
\node[lb, ptTd, anchor=west]   at ($(bPP)+(2mm,0.2mm)$) {PartPacker};
\node[lb, pxTanB, anchor=west] at ($(bOP)+(2mm,-1.6mm)$) {OmniPart};
\node[lb, ptCd, anchor=west]   at ($(bAG)+(2.5mm,1.2mm)$) {AutoPartGen};
\node[lb, pxSlaT, anchor=east] at ($(bXP)+(-2mm,2.6mm)$) {TRELLIS.2@1024 + X-Part};
\node[lb, pxPlmD, anchor=east] at ($(bHX)+(-2mm,0)$) {Hunyuan3D-2.1 + X-Part};
\node[lb, black!60, anchor=west] at ($(bT2a)+(2mm,0)$) {TRELLIS.2@512};
\node[lb, black!60, anchor=west] at ($(bFT)+(2mm,0)$) {TRELLIS.2@512 fine-tuned};
\node[lb, ptAd, anchor=west]   at ($(bUS)+(3mm,0)$) {\textbf{KaiNinja (ours)}};
\draw[->, line width=0.9pt, black!35] (70.5mm,44mm) -- (63.5mm,48.5mm);
\node[lb, black!45, anchor=north] at (67mm,43.2mm) {better};
\node[lb, anchor=south] at (37.25mm,52mm) {\textbf{(b) whole-object level}};
\end{scope}

\path (175mm,0);
\end{tikzpicture}}
\par\vspace{1.6mm}
\caption{\textbf{Quality against generation cost}, at (a) the part level and
(b) the whole-object level. Both vertical axes are F-scores at the strict
$0.05$ threshold (F1$_P^{0.05}$ and F1$_W^{0.05}$ of Table~\ref{tab:main}).
The horizontal axis is GPU\,seconds per object on a logarithmic scale, so up
and to the left is better. Gray diamonds in (b) are two whole-object
references of Table~\ref{tab:whole}: TRELLIS.2@512 zero-shot, and the same
model with both flow denoisers fine-tuned on our corpus. Fine-tuning leaves
the architecture unchanged, so the pair shares one measured cost. They emit
one undivided mesh, so they have no point in (a), and the x~axis of (a)
therefore starts later. KaiNinja sits in the upper left of both panels:
close to PartPacker in cost, and above every method in quality.}
\label{fig:cost}
\end{figure*}

\paragraph{Cost.}
Figure~\ref{fig:cost} reports the measured cost of every method. Cost is
wall-clock time on H100s over the same $986$ objects, converted to
GPU\,seconds. We run one process per GPU for every method except X-Part,
whose segmenter occupies a whole four-GPU node, so its $117$\,s wall clock is
charged as $469$ GPU\,seconds. The measurement covers the whole pipeline,
generation and mesh export together. For the $1024$ cascade the export
dominates: $51.2$ of its $62.4$\,s go to writing and decimating a mesh of
$692$K vertices, against $11.1$\,s of generation. Both X-Part cascades pay
the segmenter charge and differ only in the generator in front, so they sit
close together on the axis. PartPacker is the cheapest method and the weakest
on parts; our model costs $40\%$ more and gains $0.20$ F1$_P^{0.05}$.
AutoPartGen spends $3\times$ our budget, because it generates one part after
another, and the segment-then-regenerate cascade costs an order of magnitude
more than any other method, because it cannot begin part reasoning until a
whole object exists. The panels also show that whole-object quality
separates these methods very little: four of the five baselines sit within
$0.032$ of one another, and the fifth is $0.035$ above the best of them. The
methods separate once parts are scored.

\paragraph{Whole-object quality.}
Table~\ref{tab:whole} shows what the extension costs and what it improves.
Four of the five part-producing baselines score at or below every zero-shot
whole-object reference at the strict threshold, $0.749$ to $0.781$ against
$0.781$ to $0.807$. In other words, reaching parts from outside the generator
usually costs fidelity. Only the Hunyuan3D-2.1 cascade is higher, at $0.816$,
and only $0.009$ above its own upstream. The fine-tuned row then separates
what the corpus contributes from what the extension does. Giving
TRELLIS.2@512 the same fine-tuning data raises it to $0.830$ and cuts its
failure rate to $1.8\%$. KaiNinja starts from the same pretrained weights
and sees the same corpus, yet it is another $0.089$ above that at about
$2.5\times$ the generation cost (Figure~\ref{fig:cost}b), and it delivers
parts. At $0.919$ it is above every reference in the table, and it misses the
shape outright on $0.7\%$ of objects, against $3.9$ to $6.8\%$ for the
zero-shot references. We did not expect this gain, and we read it as
evidence that packing is a better representation of the same objects rather
than only a container for parts. Two details qualify this reading. Running
TRELLIS.2 as a $1024$ cascade does not help on this corpus ($0.0364$ against
$0.0363$). Also, TRELLIS and Hunyuan3D-2.1 share a mean but not a
distribution: Hunyuan3D-2.1 has the better median CD$_W$, $0.0209$ against
$0.0235$ (not shown in the table), and nearly twice the failure rate.

\paragraph{X-Part and its generator.}
X-Part's score tracks the generator in front of it. Paired with Hunyuan3D-2.1
it reaches $0.816$ F1$_W^{0.05}$, within $0.009$ of what Hunyuan3D-2.1
reaches alone (Table~\ref{tab:whole}), and the same holds on TRELLIS.2.
Segmenting after the fact changes the whole-object score very little, and it
improves the part score about as little, so the cascade scores roughly what
its generator hands it. To see \emph{which} property is inherited, we reran
the cascade from the $512$ shape flow of TRELLIS.2 instead of the $1024$
cascade, paired over the $985$ objects X-Part completes from both.\footnote{One
\textsc{fusion360} object crashes X-Part on the $512$ mesh but not on the
$1024$ one, and is dropped from this paired comparison only.} The part scores
came out indistinguishable: the per-object differences scatter far more widely
than the gap between the two means. Swapping the \emph{model} in front, by
contrast, moves X-Part by $0.035$ F1$_W^{0.05}$. The cascade therefore
appears to inherit the shape of the upstream generator rather than its
compute budget, and it cannot exceed the mesh it is handed.

\paragraph{The role of the representation.}
PartPacker shares our packing principle and is likewise free of masks and
segmenters, so the main difference between the two is where the volumes
live: in a vecset latent decoded as an SDF field there, and in the backbone's
own sparse voxel grid here. On the whole object the gap is moderate (CD$_W$
$0.0390$ against $0.0186$), but on parts it is the widest in the table:
F1$_P^{0.05}$ $0.492$ against $0.692$, with CD$_P$ more than $60\%$ larger.
The SDF representation also requires watertight preprocessing and drops the
materials. We take this as support for building the packing on the backbone
it extends, with the caveat that the two systems also differ in training
corpus, so this comparison does not isolate the representation on its own.

\begin{table}[t]
\centering
\caption{\textbf{Part-aware image-to-3D.} Whole-object ($W$) and Hungarian-matched
part ($P$) metrics of Section~\ref{sec:exp-setup}, on the $986$ held-out objects
all six methods complete.}
\label{tab:main}
\tablestyle{5pt}{1.15}
\begin{tabular}{l ccccccc}
\toprule
Method & CD$_W\!\downarrow$ & F1$_W^{0.1}$ & F1$_W^{0.05}$ & mIoU$_P$ & CD$_P\!\downarrow$ & F1$_P^{0.1}$ & F1$_P^{0.05}$ \\
\midrule
TRELLIS.2@1024~\cite{trellis2} + X-Part~\cite{xpart} & 0.0350 & 0.900 & 0.780 & 0.449 & 0.0804 & 0.703 & 0.559  \\
Hunyuan3D-2.1~\cite{hunyuan3d21} + X-Part~\cite{xpart} & 0.0314 & 0.917 & 0.814 & 0.489 & 0.0764 & 0.728 & 0.597  \\
OmniPart~\cite{omnipart} & 0.0366 & 0.901 & 0.769 & 0.456 & 0.0913 & 0.664 & 0.518  \\
PartPacker~\cite{partpacker} & 0.0389 & 0.892 & 0.749 & 0.449 & 0.0986 & 0.640 & 0.492  \\
AutoPartGen~\cite{autopartgen} & 0.0373 & 0.901 & 0.771 & 0.466 & 0.0919 & 0.678 & 0.535  \\
\rowcolor{rowgray}
KaiNinja (ours)          & \bnum{0.0186} & \bnum{0.975} & \bnum{0.919} & \bnum{0.521} & \bnum{0.0613} & \bnum{0.785} & \bnum{0.692}  \\
\bottomrule
\end{tabular}
\end{table}

\begin{table}[t]
\centering
\caption{\textbf{Whole-object generators, for reference.} One undivided mesh
each, so only the whole-object columns are defined. \emph{Fail} is the fraction of
objects with CD$_W>0.1$.}
\label{tab:whole}
\tablestyle{5pt}{1.15}
\begin{tabular}{l cccc}
\toprule
Method & CD$_W\!\downarrow$ & F1$_W^{0.1}$ & F1$_W^{0.05}$ & Fail\,\% $\downarrow$ \\
\midrule
TRELLIS~\cite{trellis}                    & 0.0340 & 0.916 & 0.807 & 3.9  \\
TRELLIS.2@512~\cite{trellis2}             & 0.0363 & 0.897 & 0.781 & 5.3  \\
TRELLIS.2@1024~\cite{trellis2}            & 0.0364 & 0.898 & 0.781 & 5.4  \\
TRELLIS.2@512, fine-tuned on our corpus   & 0.0300 & 0.940 & 0.830 & 1.8  \\
Hunyuan3D-2.1~\cite{hunyuan3d21}          & 0.0340 & 0.912 & 0.807 & 6.8  \\
\midrule
\rowcolor{rowgray}
KaiNinja (ours, part-aware)               & \bnum{0.0186} & \bnum{0.975} & \bnum{0.919} & \bnum{0.7}  \\
\bottomrule
\end{tabular}
\end{table}

\begin{table}[t]
\centering
\caption{\textbf{Breakdown by source} (whole-object F1$_W^{0.1}$ /
Hungarian-matched part F1$_P^{0.05}$)
on the $986$ objects every method completes (\textsc{articraft} $499$,
\textsc{partnext} $297$, \textsc{fusion360} $102$, \textsc{trellis} $88$). Best
in \bnum{bold}.}
\label{tab:perdataset}
\tablestyle{4pt}{1.15}
\begin{tabular}{l cc cc cc cc}
\toprule
& \multicolumn{2}{c}{\textsc{articraft}} & \multicolumn{2}{c}{\textsc{partnext}} & \multicolumn{2}{c}{\textsc{fusion360}} & \multicolumn{2}{c}{\textsc{trellis}} \\
\cmidrule(lr){2-3}\cmidrule(lr){4-5}\cmidrule(lr){6-7}\cmidrule(lr){8-9}
Method & F1$_W$ & F1$_P$ & F1$_W$ & F1$_P$ & F1$_W$ & F1$_P$ & F1$_W$ & F1$_P$ \\
\midrule
TRELLIS.2@1024~\cite{trellis2} + X-Part~\cite{xpart} & 0.911 & 0.559 & 0.898 & 0.584 & 0.851 & 0.512 & 0.899 & 0.529  \\
Hunyuan3D-2.1~\cite{hunyuan3d21} + X-Part~\cite{xpart} & 0.928 & 0.614 & 0.922 & 0.612 & 0.866 & 0.548 & 0.901 & 0.509  \\
OmniPart~\cite{omnipart} & 0.913 & 0.519 & 0.896 & 0.536 & 0.844 & 0.470 & 0.917 & 0.499  \\
PartPacker~\cite{partpacker} & 0.913 & 0.517 & 0.874 & 0.466 & 0.835 & 0.463 & 0.899 & 0.466  \\
AutoPartGen~\cite{autopartgen} & 0.914 & 0.544 & 0.901 & 0.538 & 0.828 & 0.481 & 0.912 & 0.531  \\
\rowcolor{rowgray}
KaiNinja (ours)          & \bnum{0.985} & \bnum{0.738} & \bnum{0.967} & \bnum{0.666} & \bnum{0.956} & \bnum{0.627} & \bnum{0.964} & \bnum{0.595}  \\
\bottomrule
\end{tabular}
\end{table}

\subsection{Generalization Beyond the Fine-Tuning Corpus}
\label{sec:exp-realmat}

To test generalization beyond our curated splits, we also evaluate on objects
from Sketchfab and GitHub, filtered down to a single connected component. We
score the $95$ of them that every method completes, so every number in the
table comes from the same objects. Neither source feeds our fine-tuning corpus
or our test split. That separation is what this set is designed to test, and
it holds for every method compared. We do not call these objects unseen,
because the pretrained TRELLIS.2 weights our flows start from were trained on
data derived from Objaverse~\cite{objaverse,objaversexl}, which draws on both
sources, and the same holds for the generators in front of the baselines.
Every method receives the same rendered view. Table~\ref{tab:realmat} gives
the comparison. KaiNinja keeps a clear lead on the whole object, by $0.075$
F1$_W^{0.05}$ over the next method. On parts it is ahead on Chamfer distance
and on the strict F1$_P^{0.05}$, where the margin over the closest baseline is
$0.036$. The two X-Part cascades stay with us only at the loose threshold,
F1$_P^{0.1}$ $0.862$ against our $0.861$, and both carry a higher mIoU$_P$
($0.644$ and $0.624$ against our $0.619$). A dedicated segmenter seems to
place boundaries a little more conservatively on irregular inputs, which
helps once a coarse tolerance is allowed. The ordering at the strict
threshold is the one that matters for downstream use, and there a model with
no segmenter is ahead.

\begin{table}[t]
\centering
\caption{\textbf{Sources outside the fine-tuning corpus}: the $95$ Sketchfab and
GitHub assets, single connected component, that every method completes. Same
method set and same row order as Table~\ref{tab:main}.}
\label{tab:realmat}
\tablestyle{5pt}{1.15}
\begin{tabular}{l ccccccc}
\toprule
Method & CD$_W\!\downarrow$ & F1$_W^{0.1}$ & F1$_W^{0.05}$ & mIoU$_P$ & CD$_P\!\downarrow$ & F1$_P^{0.1}$ & F1$_P^{0.05}$ \\
\midrule
TRELLIS.2@1024~\cite{trellis2} + X-Part~\cite{xpart} & 0.0277 & 0.939 & 0.826 & 0.624 & 0.0459 & \bnum{0.862} & 0.717  \\
Hunyuan3D-2.1~\cite{hunyuan3d21} + X-Part~\cite{xpart} & 0.0274 & 0.937 & 0.826 & \bnum{0.644} & 0.0507 & 0.835 & 0.706  \\
OmniPart~\cite{omnipart} & 0.0318 & 0.928 & 0.804 & 0.569 & 0.0579 & 0.815 & 0.659  \\
PartPacker~\cite{partpacker} & 0.0317 & 0.926 & 0.791 & 0.607 & 0.0614 & 0.792 & 0.651  \\
AutoPartGen~\cite{autopartgen} & 0.0306 & 0.942 & 0.812 & 0.614 & 0.0543 & 0.833 & 0.678  \\
\rowcolor{rowgray}
KaiNinja (ours)          & \bnum{0.0195} & \bnum{0.969} & \bnum{0.901} & 0.619 & \bnum{0.0444} & 0.861 & \bnum{0.753}  \\
\bottomrule
\end{tabular}
\end{table}

\begin{table}[t]
\centering
\caption{\textbf{HY3D-Bench}, $200$ part-annotated objects, every method reading
the same released view. Shading marks \best{}first, \sbest{}second and
\tbest{}third per column.}
\label{tab:hy3d}
\tablestyle{4pt}{1.15}
\begin{tabular}{l ccc cccc}
\toprule
Method & CD$_W\!\downarrow$ & F1$_W^{0.1}$ & F1$_W^{0.05}$ & mIoU$_P$ & CD$_P\!\downarrow$ & F1$_P^{0.1}$ & F1$_P^{0.05}$ \\
\midrule
TRELLIS.2@1024~\cite{trellis2} + X-Part~\cite{xpart} & 0.0516 & 0.807 & 0.639 & 0.344 & \tbest 0.0852 & \tbest 0.666 & 0.476 \\
Hunyuan3D-2.1~\cite{hunyuan3d21} + X-Part~\cite{xpart} & \sbest 0.0441 & \sbest 0.844 & \sbest 0.699 & \best 0.412 & \best 0.0728 & \best 0.729 & \best 0.555 \\
OmniPart~\cite{omnipart} & \tbest 0.0475 & \tbest 0.834 & \tbest 0.685 & 0.354 & 0.0897 & 0.665 & 0.490 \\
PartPacker~\cite{partpacker} & 0.0566 & 0.788 & 0.632 & \tbest 0.381 & 0.0904 & 0.657 & 0.476 \\
AutoPartGen~\cite{autopartgen} & 0.0512 & 0.821 & 0.666 & \sbest 0.399 & 0.0905 & 0.665 & \tbest 0.498 \\
\rowcolor{rowgray}
KaiNinja (ours) & \best 0.0413 & \best 0.865 & \best 0.724 & 0.356 & \sbest 0.0787 & \sbest 0.688 & \sbest 0.505 \\
\bottomrule
\end{tabular}
\end{table}

\paragraph{An external benchmark.}
The set above still shares its sources with the pretraining data. HY3D-Bench
does not. It is an outside benchmark with its own part annotations, and no
method here was trained on it. Every method reads the same conditioning
image, one of the views HY3D-Bench itself publishes, and we score all $200$
annotated objects. Table~\ref{tab:hy3d} reports the result, and the result
splits.

On the whole object KaiNinja is first on all three metrics, by $2.6$
points of F1$_W^{0.05}$ over the next method. On parts it is not.
Hunyuan3D-2.1 + X-Part leads every part metric, and on mIoU$_P$ we sit
fourth of six, behind AutoPartGen and PartPacker as well. This reverses the
ordering of Table~\ref{tab:main}, and we attribute it to granularity rather
than geometry. HY3D-Bench annotates parts more coarsely than our corpus does.
Our model is trained to separate at contacts, so it cuts a chair into more
pieces than the benchmark labels, and Hungarian matching charges for every
extra piece. A pipeline that segments after generating a whole object
inherits its granularity from the segmenter, which is easier to tune toward
whatever convention a benchmark uses. We report the split as it is: on this
benchmark our geometry is the most faithful, and our part decomposition is
finer than the labels reward.

\subsection{Qualitative Results}
\label{sec:exp-qual}
\begin{figure*}[!tp]
\centering
\QualGridSetup{7}{\linewidth}
\begin{tabular}{ccccccc}
\QualGridHeader{Input} &
\QualGridHeader{Ground truth} &
\QualGridHeader{KaiNinja\\(ours)} &
\QualGridHeader{TRELLIS.2\\+X-Part} &
\QualGridHeader{Hunyuan3D-2.1\\+X-Part} &
\QualGridHeader{OmniPart} &
\QualGridHeader{PartPacker} \\
\QualGridImage{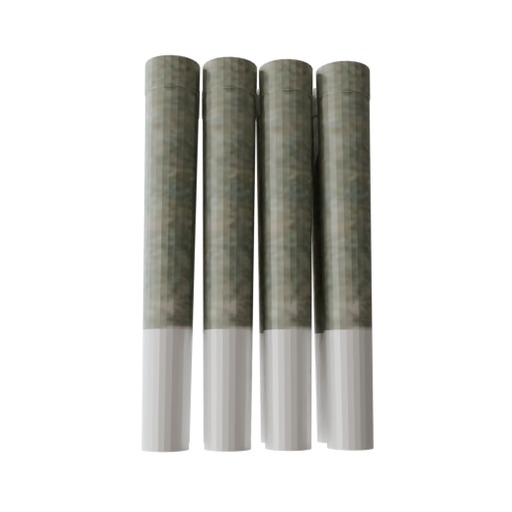} &
\QualGridImage{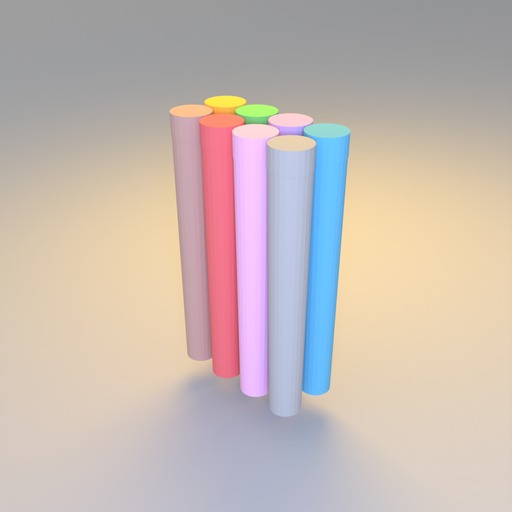} &
\QualGridImage{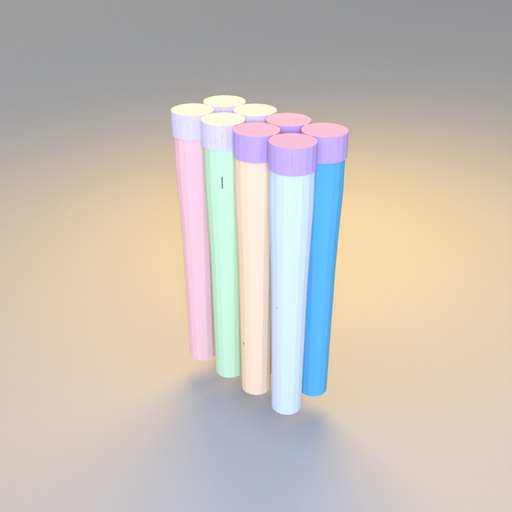} &
\QualGridImage{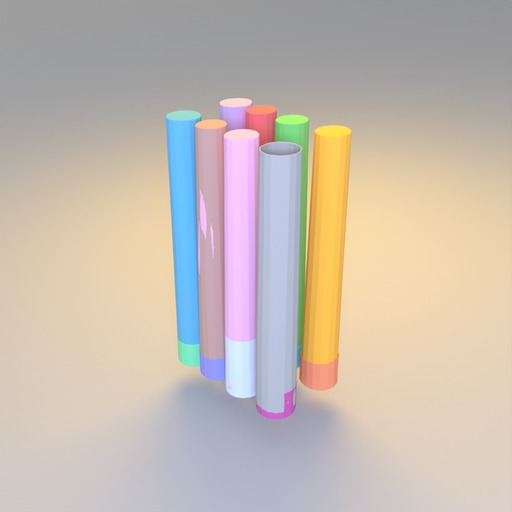} &
\QualGridImage{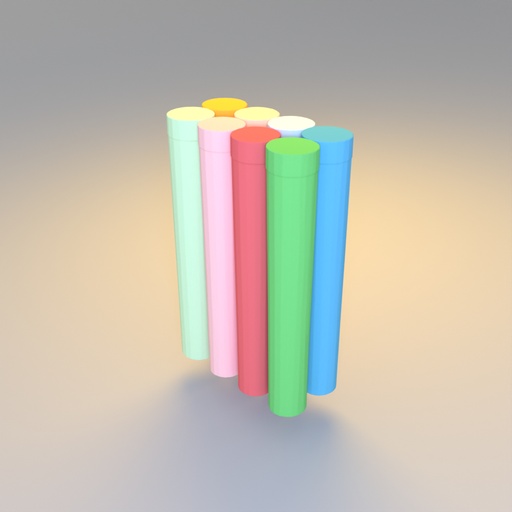} &
\QualGridImage{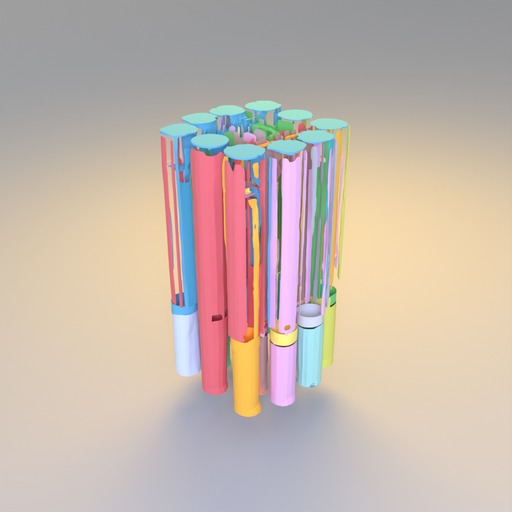} &
\QualGridImage{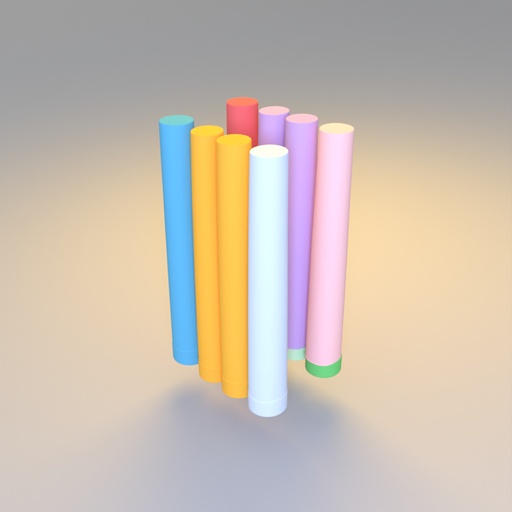} \\
\QualGridImage{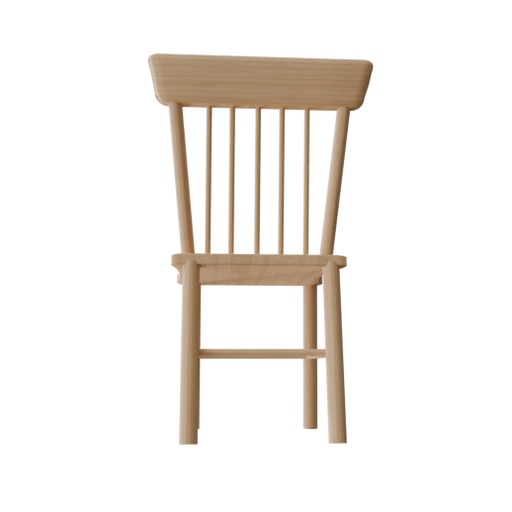} &
\QualGridImage{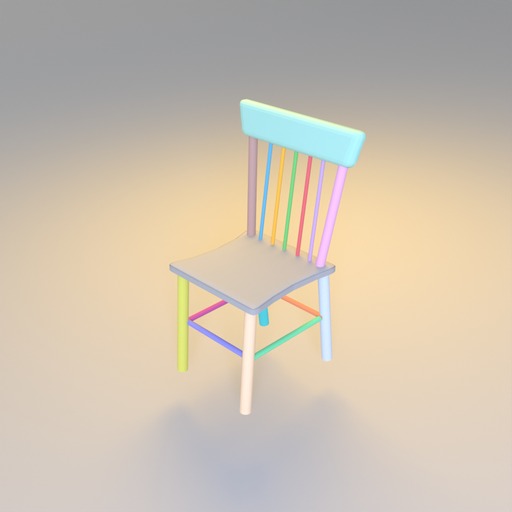} &
\QualGridImage{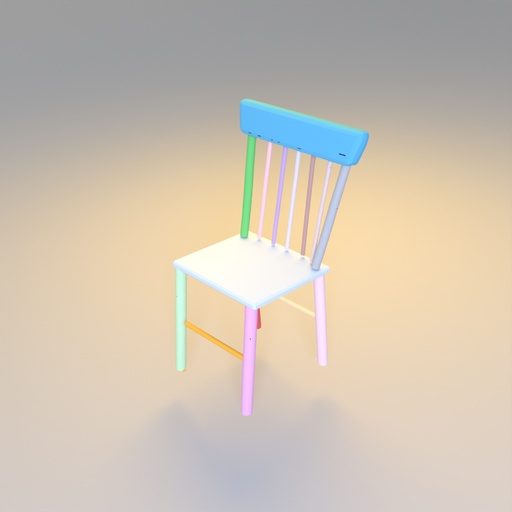} &
\QualGridImage{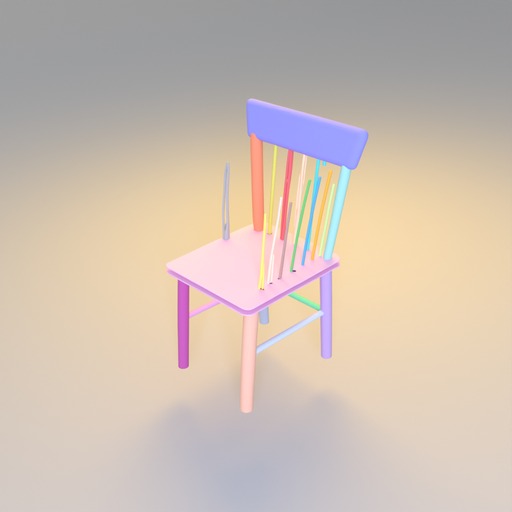} &
\QualGridImage{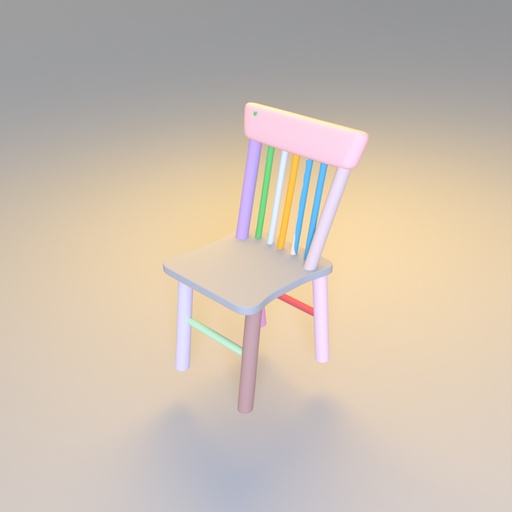} &
\QualGridImage{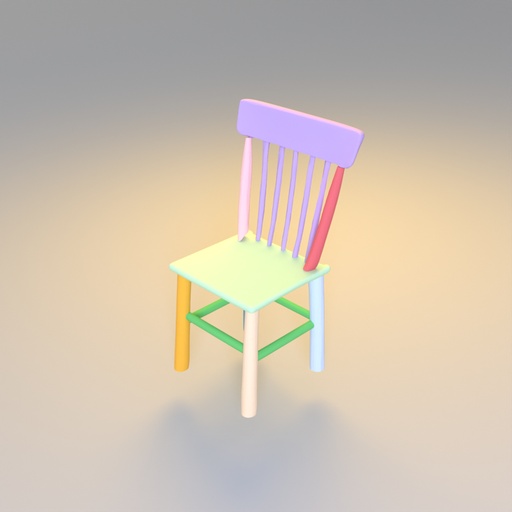} &
\QualGridImage{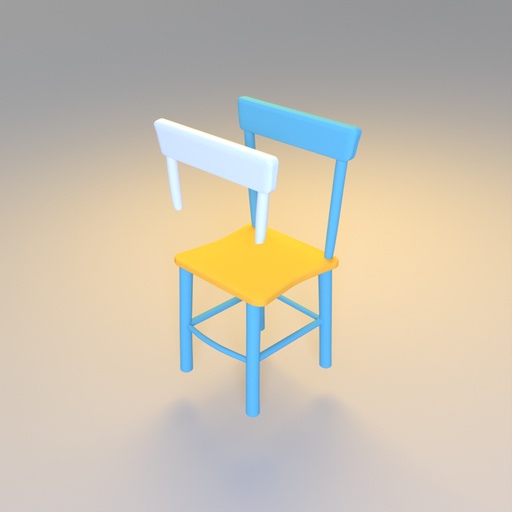} \\
\QualGridImage{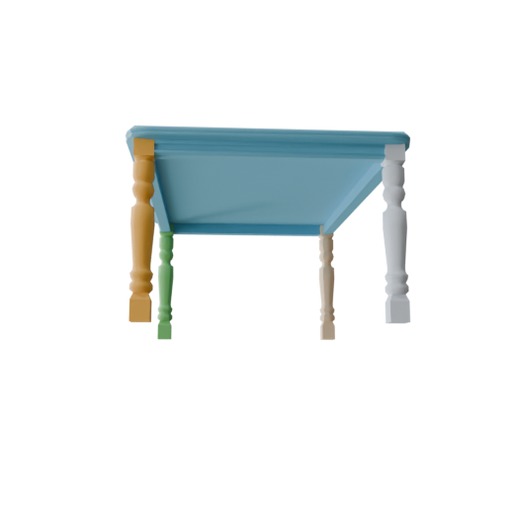} &
\QualGridImage{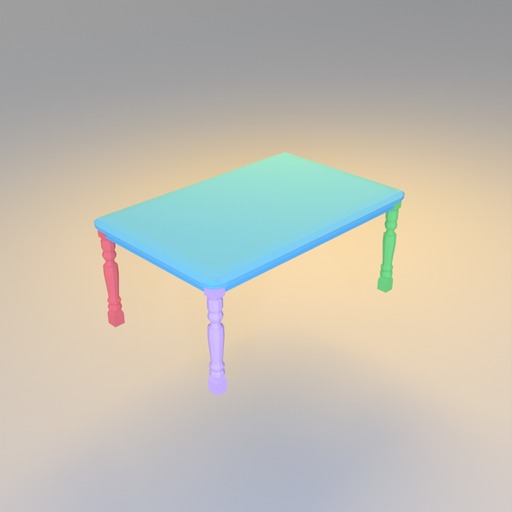} &
\QualGridImage{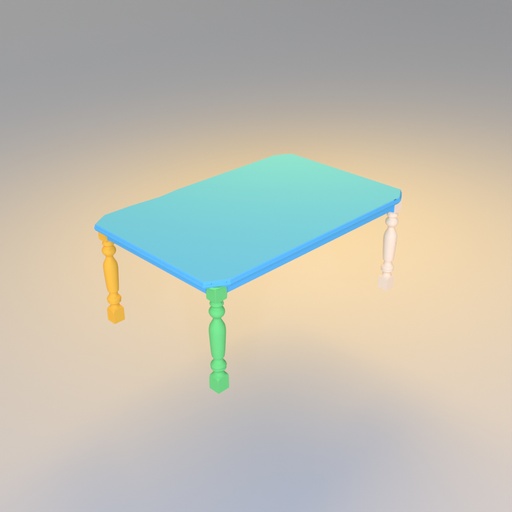} &
\QualGridImage{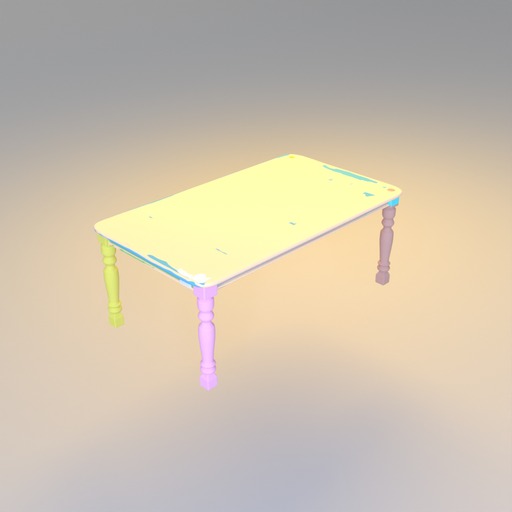} &
\QualGridImage{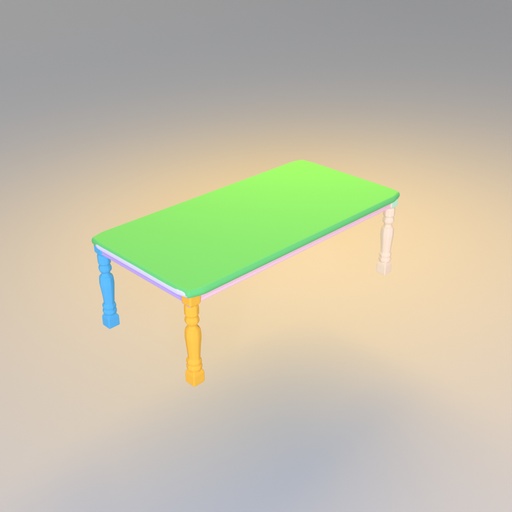} &
\QualGridImage{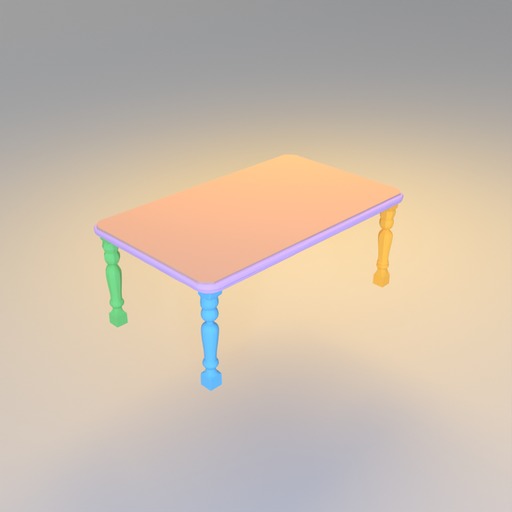} &
\QualGridImage{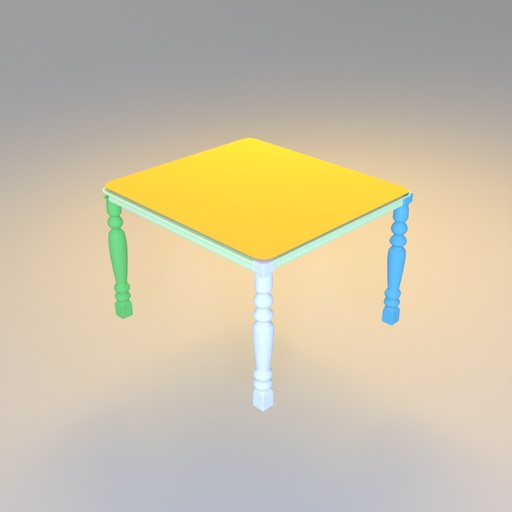} \\
\QualGridImage{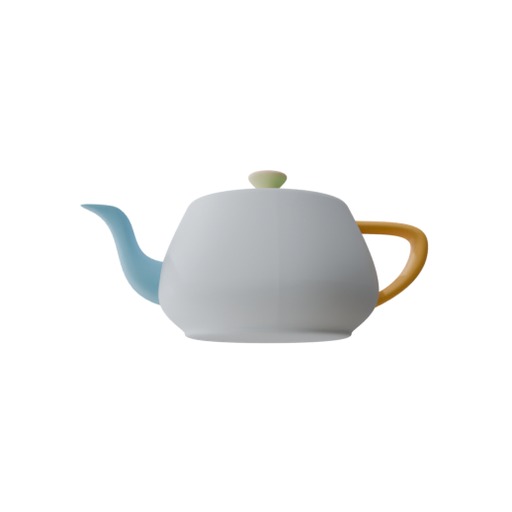} &
\QualGridImage{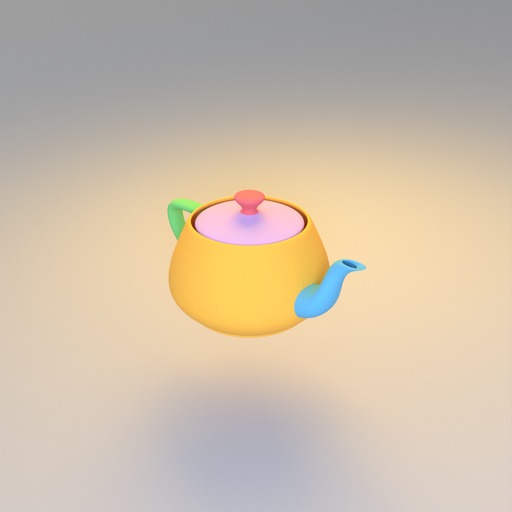} &
\QualGridImage{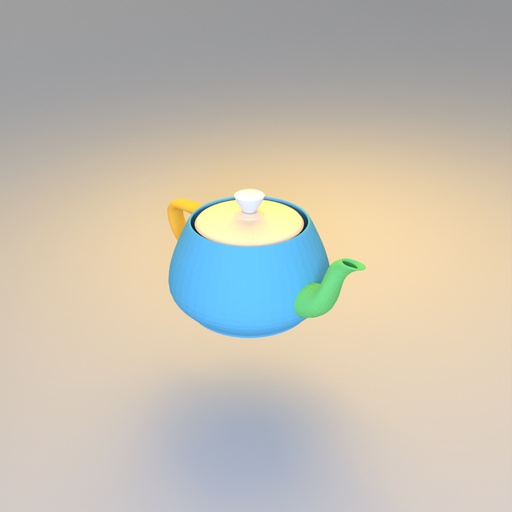} &
\QualGridImage{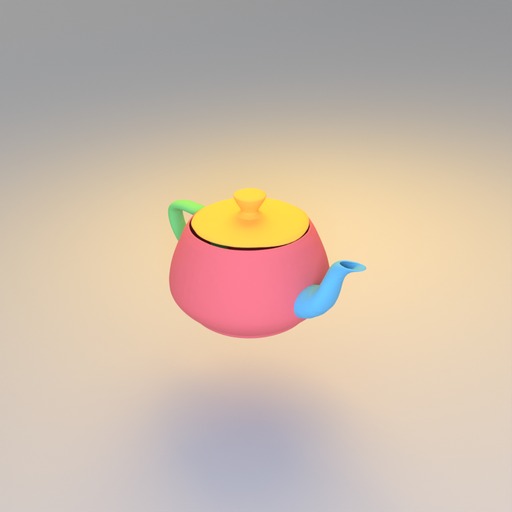} &
\QualGridImage{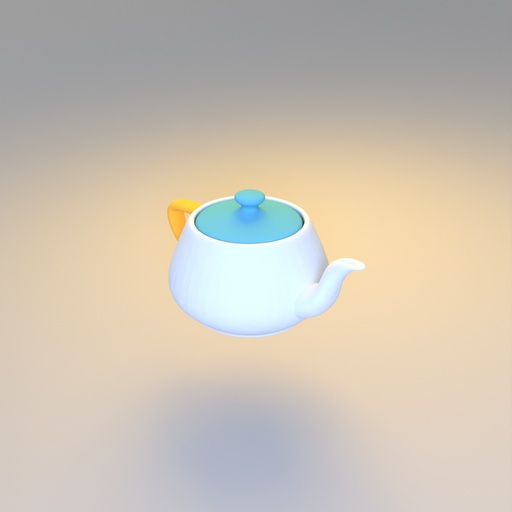} &
\QualGridImage{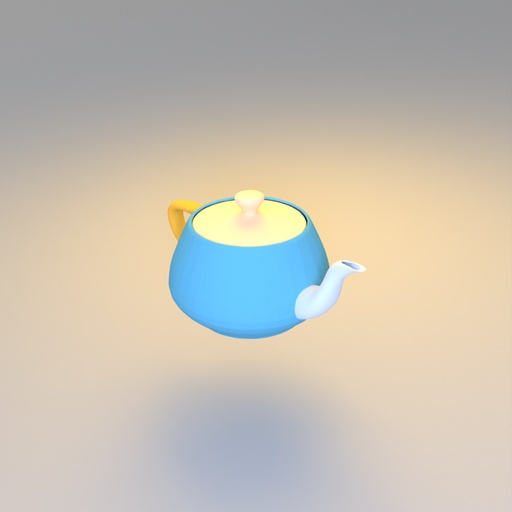} &
\QualGridImage{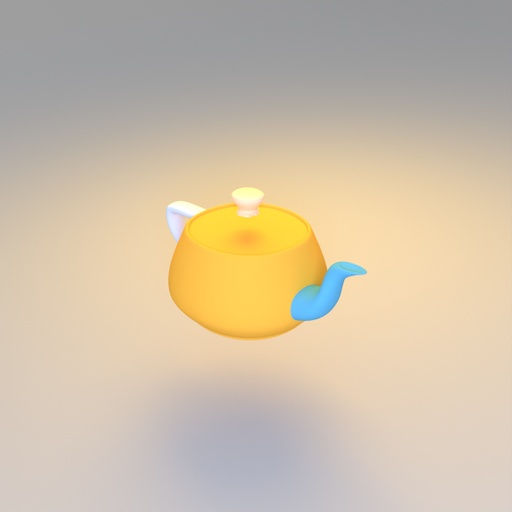} \\
\QualGridImage{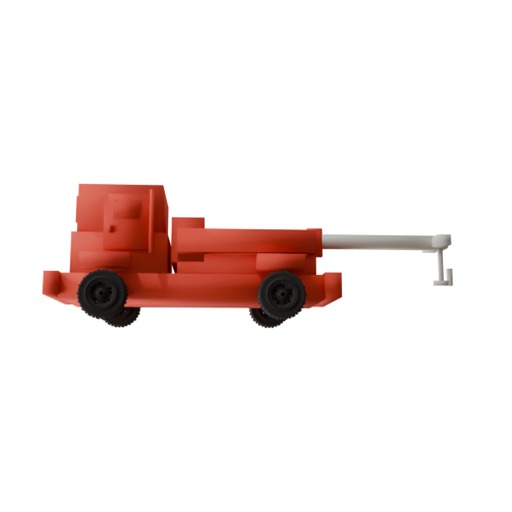} &
\QualGridImage{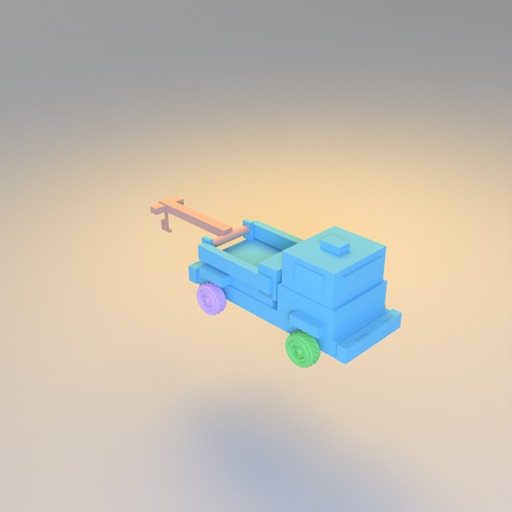} &
\QualGridImage{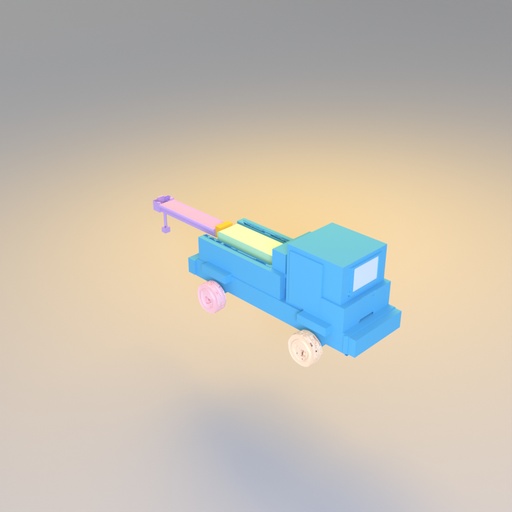} &
\QualGridImage{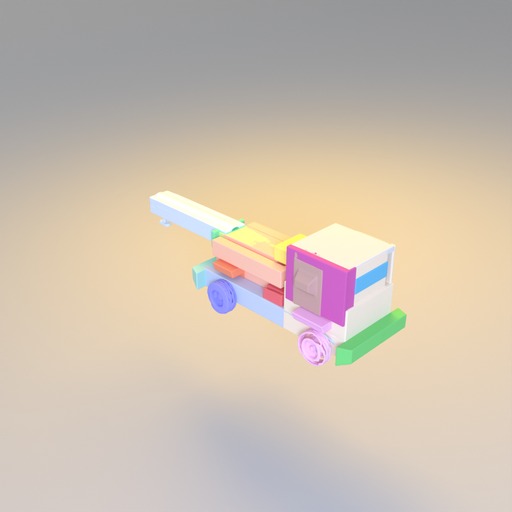} &
\QualGridImage{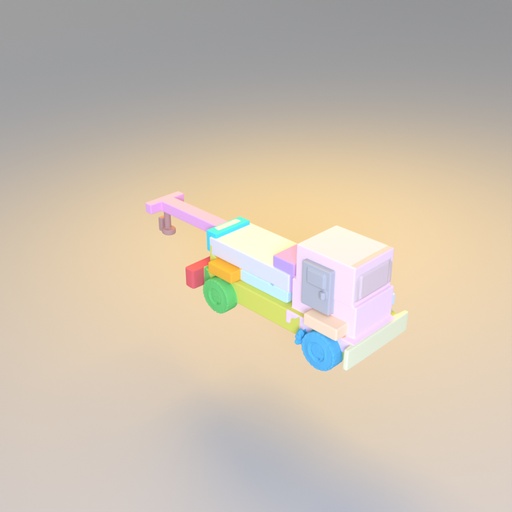} &
\QualGridImage{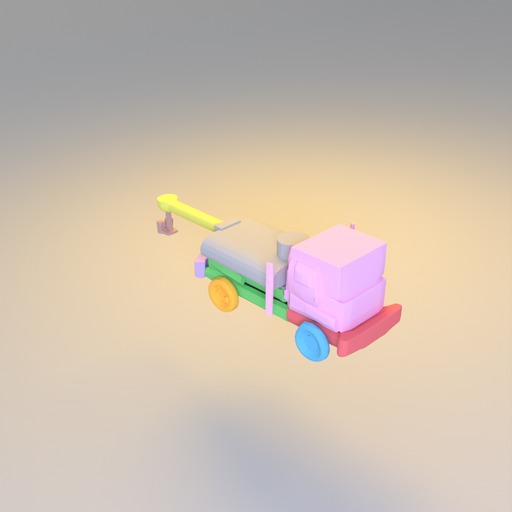} &
\QualGridImage{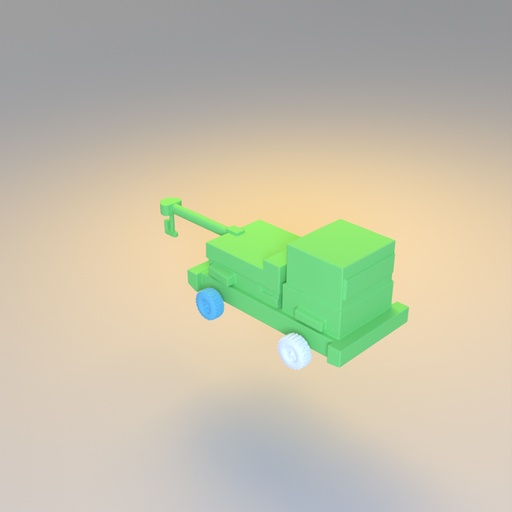} \\
\end{tabular}
\caption{\textbf{Held-out test set.} Columns: the input image, the ground truth
parts, and the five methods. Colors mark parts, not materials. Each method
uses its own canonical frame, so every shape is rotated into the pose of the
ground truth cell before rendering, and all cells then share one camera, one
light and one scale. KaiNinja keeps the part count close to the ground
truth where the baselines fuse or shatter parts. SPACE PROMPT}
\label{fig:qual-test}
\end{figure*}

\begin{figure*}[!tp]
\centering
\QualGridSetup{7}{\linewidth}
\begin{tabular}{ccccccc}
\QualGridHeader{Input} &
\QualGridHeader{Ground truth} &
\QualGridHeader{KaiNinja\\(ours)} &
\QualGridHeader{TRELLIS.2\\+X-Part} &
\QualGridHeader{Hunyuan3D-2.1\\+X-Part} &
\QualGridHeader{OmniPart} &
\QualGridHeader{PartPacker} \\
\QualGridImage{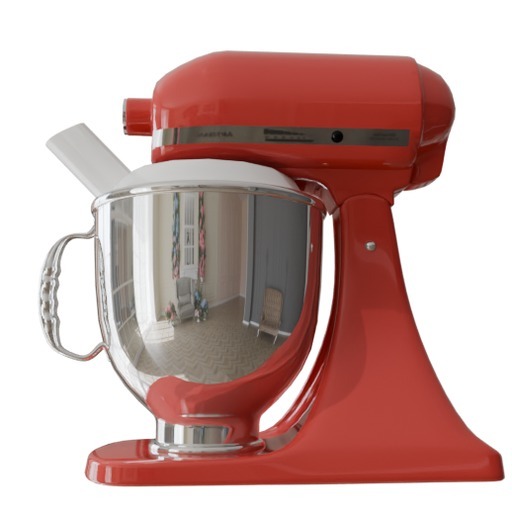} &
\QualGridImage{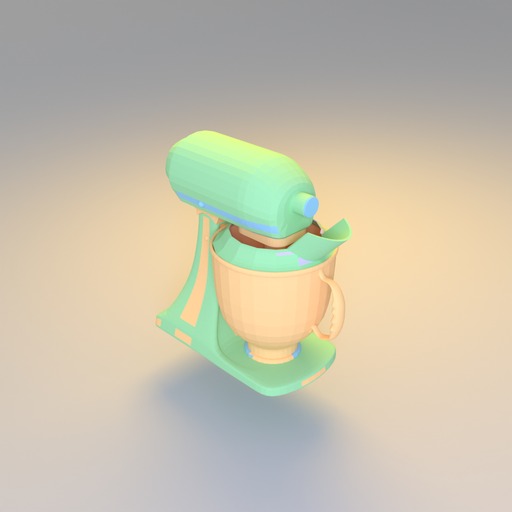} &
\QualGridImage{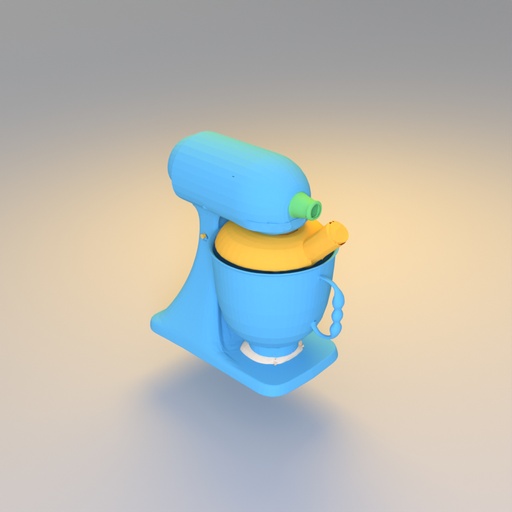} &
\QualGridImage{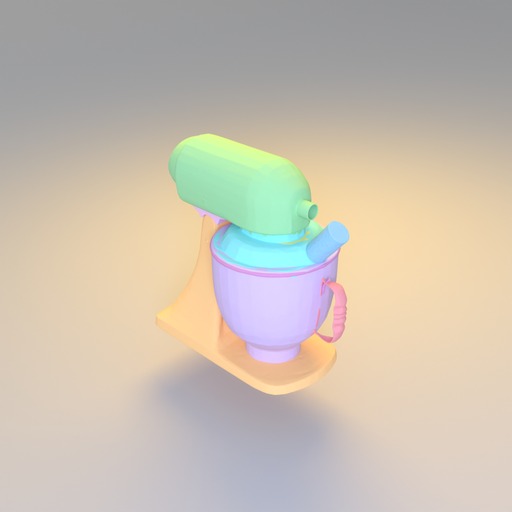} &
\QualGridImage{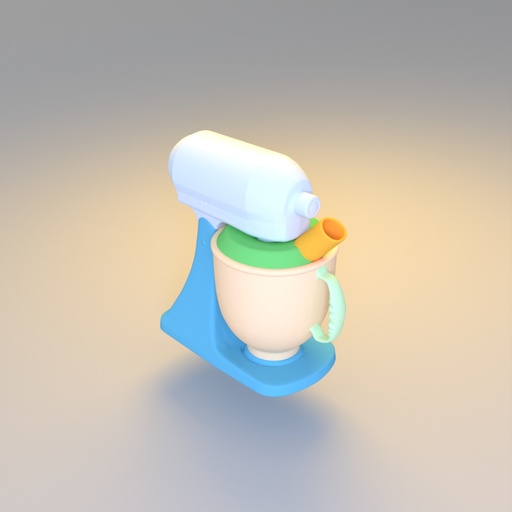} &
\QualGridImage{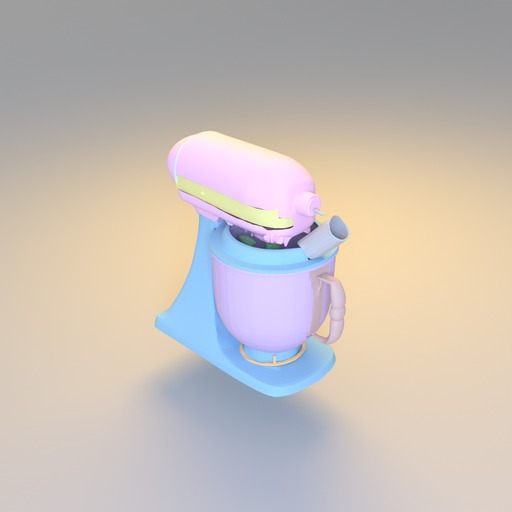} &
\QualGridImage{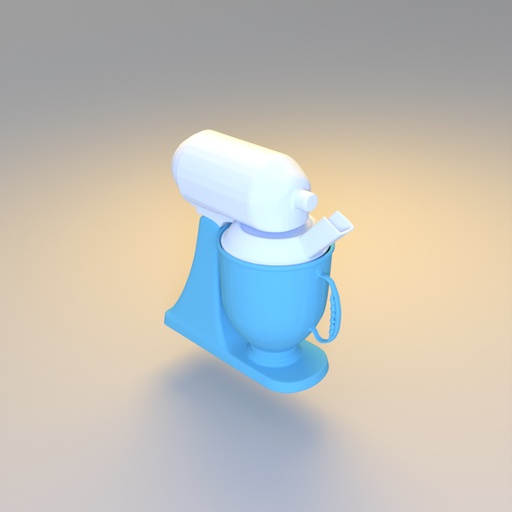} \\
\QualGridImage{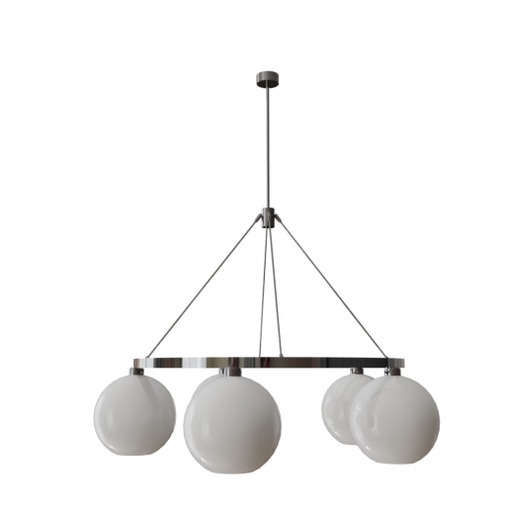} &
\QualGridImage{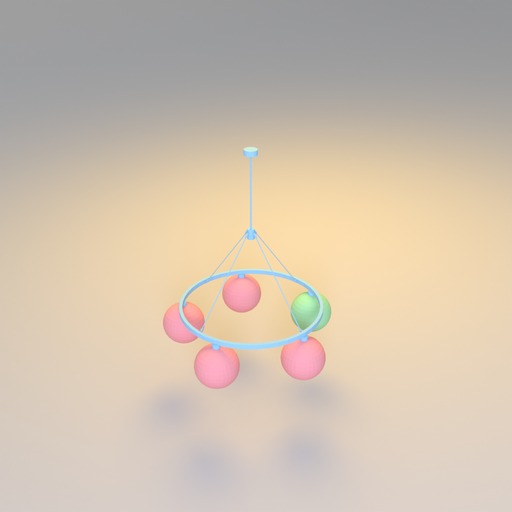} &
\QualGridImage{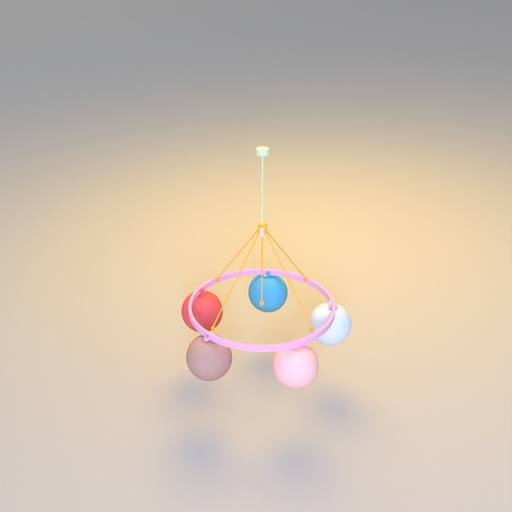} &
\QualGridImage{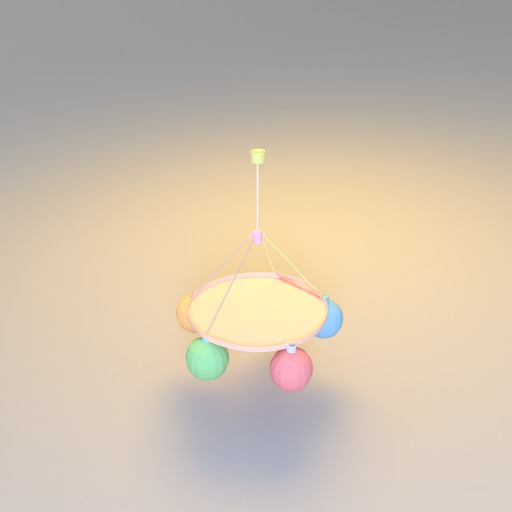} &
\QualGridImage{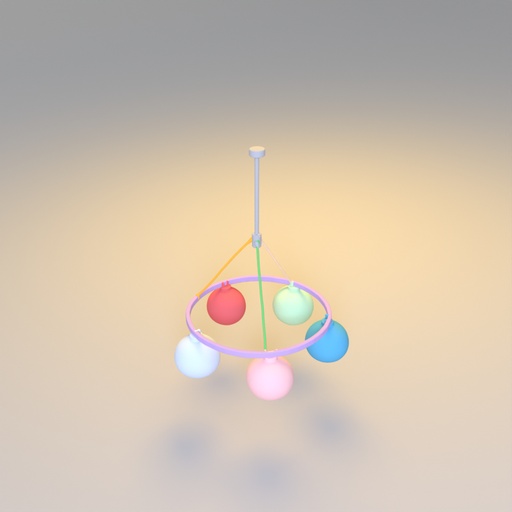} &
\QualGridImage{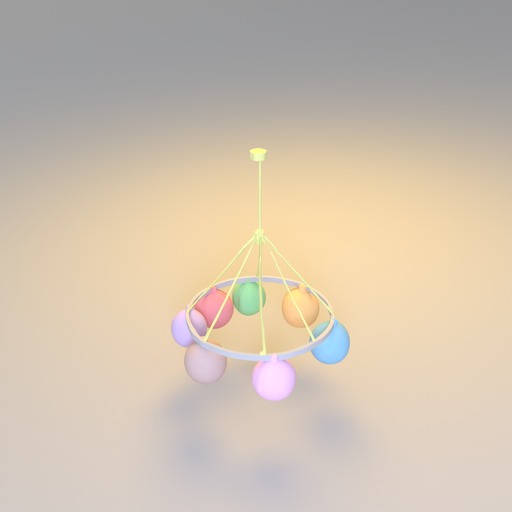} &
\QualGridImage{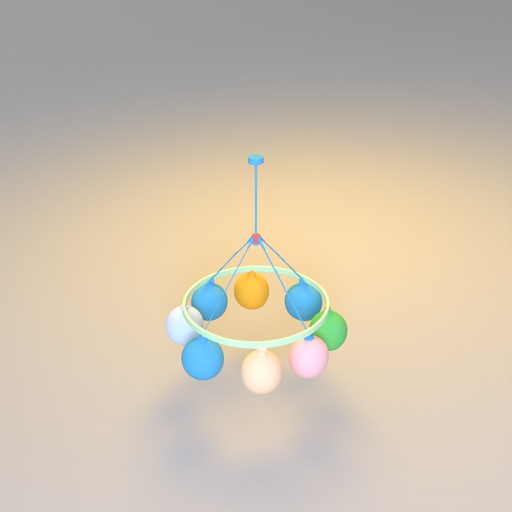} \\
\QualGridImage{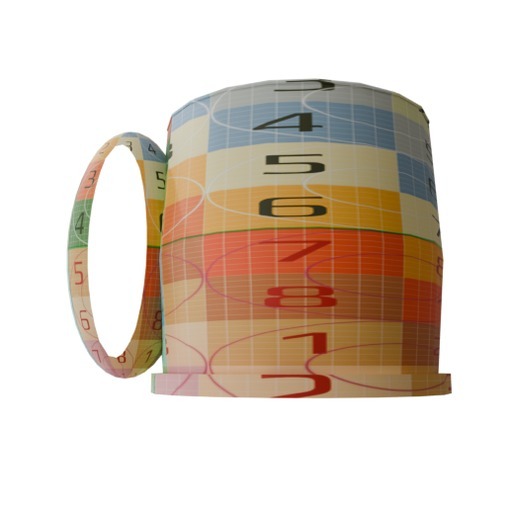} &
\QualGridImage{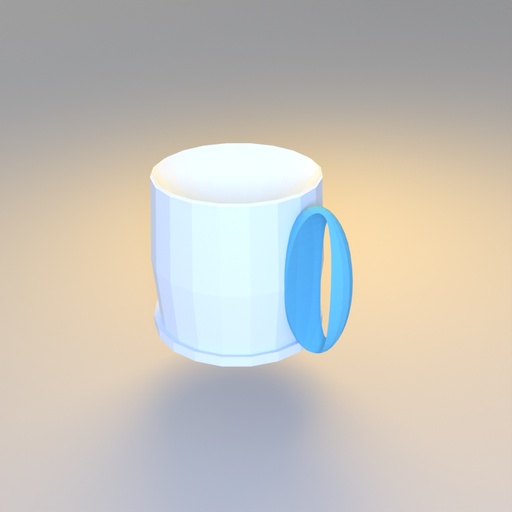} &
\QualGridImage{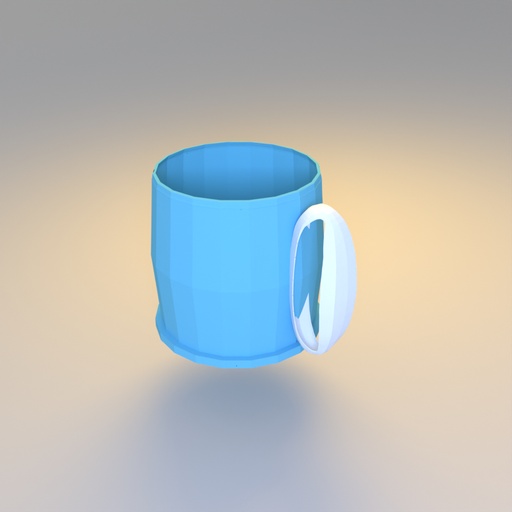} &
\QualGridImage{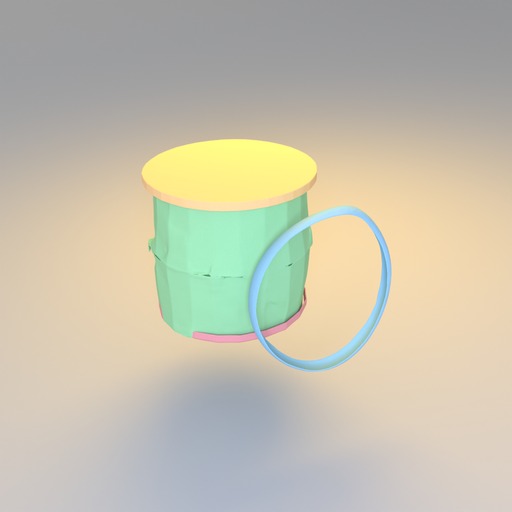} &
\QualGridImage{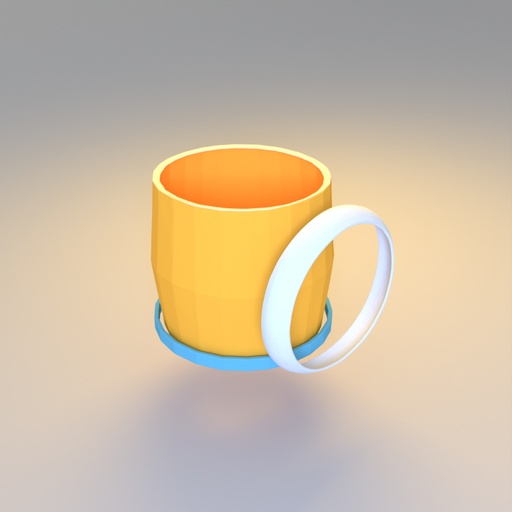} &
\QualGridImage{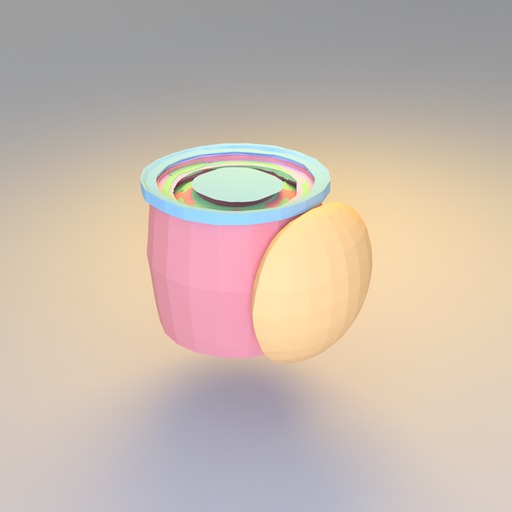} &
\QualGridImage{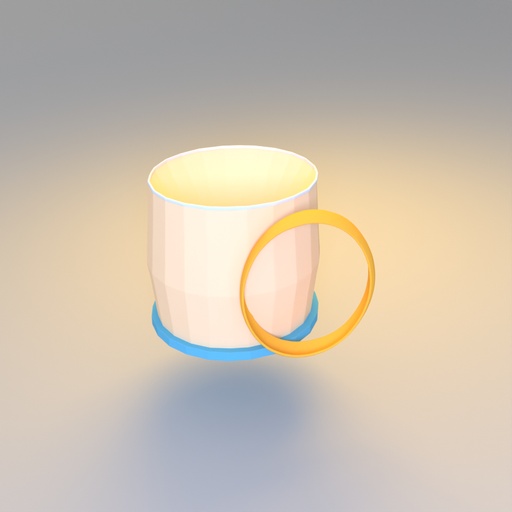} \\
\QualGridImage{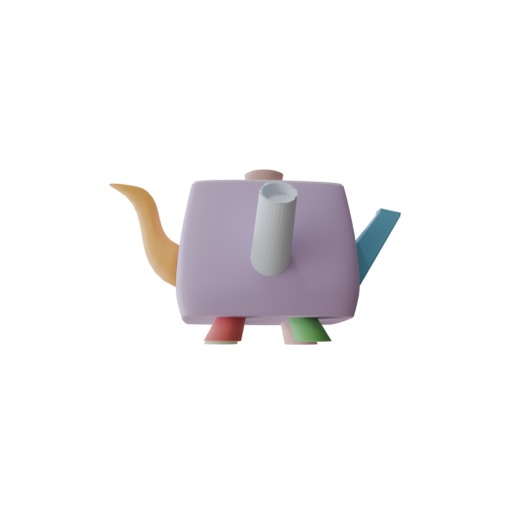} &
\QualGridImage{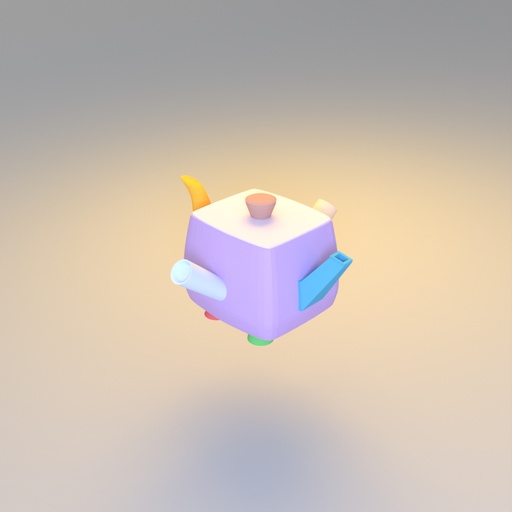} &
\QualGridImage{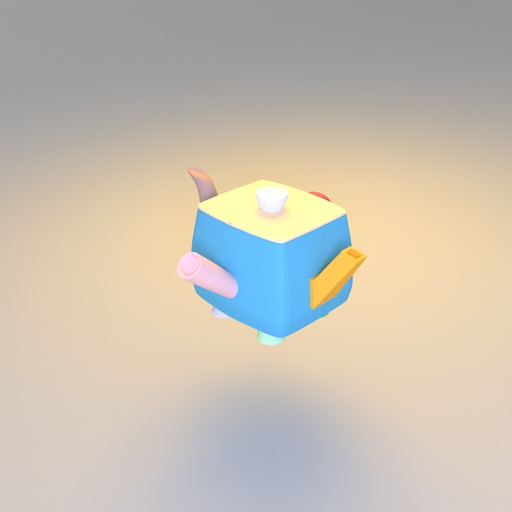} &
\QualGridImage{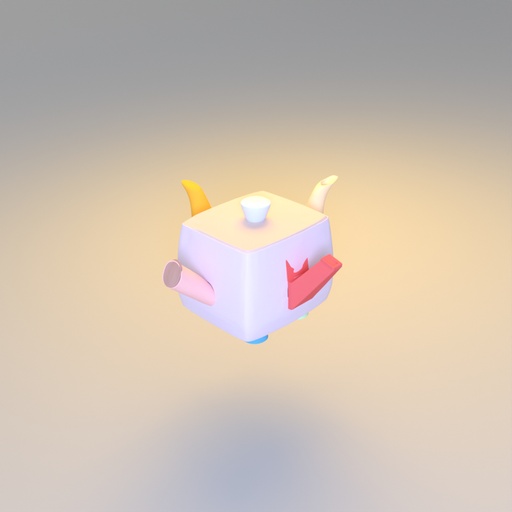} &
\QualGridImage{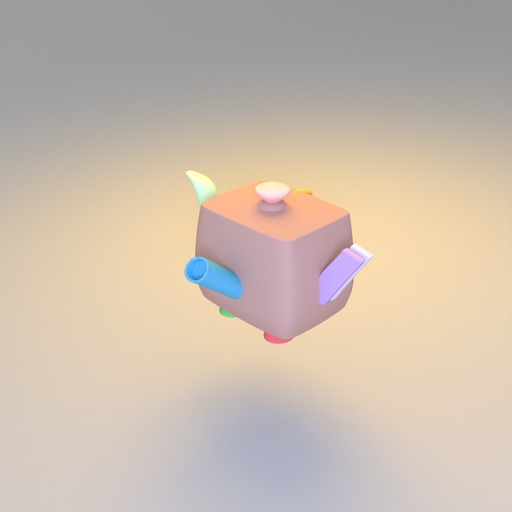} &
\QualGridImage{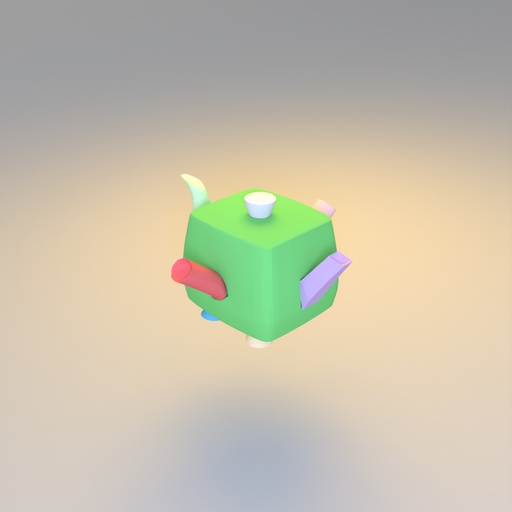} &
\QualGridImage{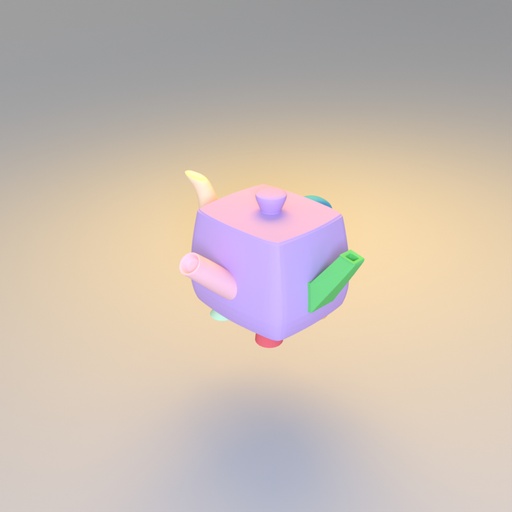} \\
\QualGridImage{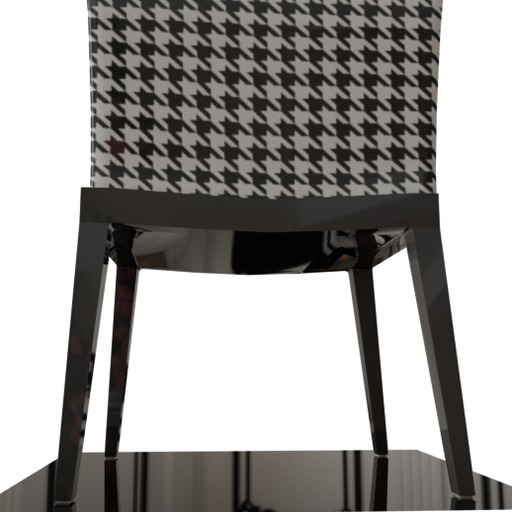} &
\QualGridImage{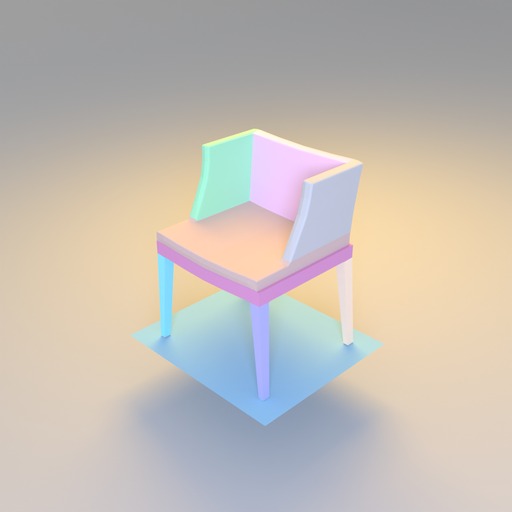} &
\QualGridImage{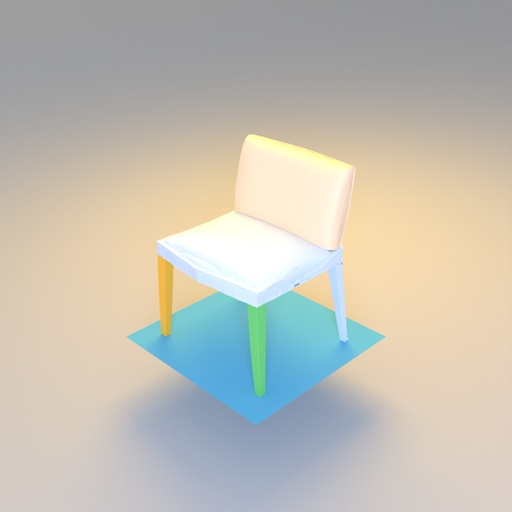} &
\QualGridImage{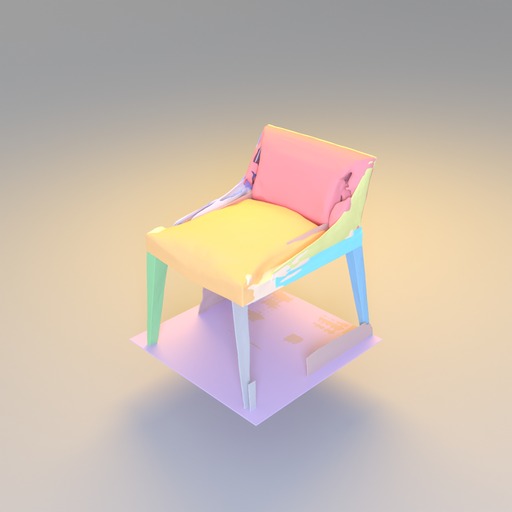} &
\QualGridImage{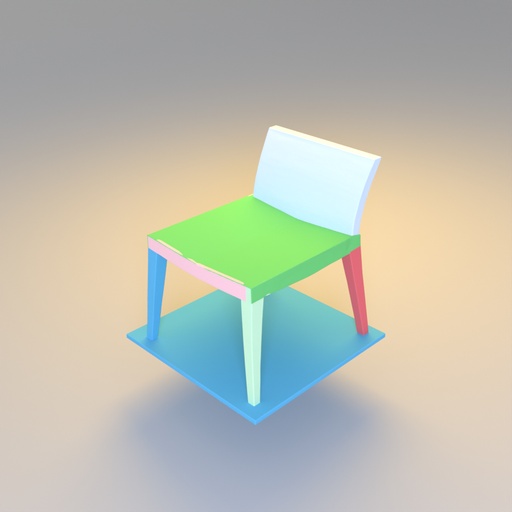} &
\QualGridImage{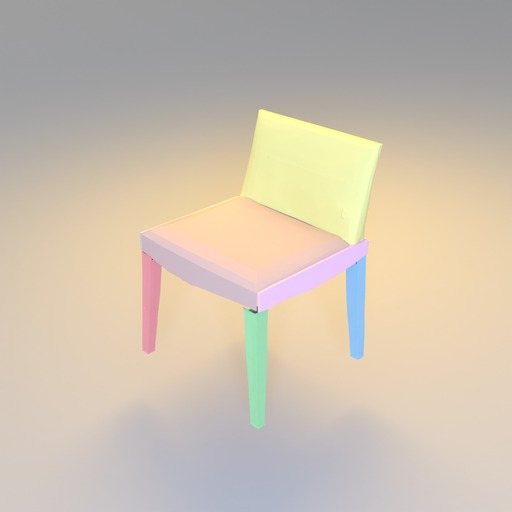} &
\QualGridImage{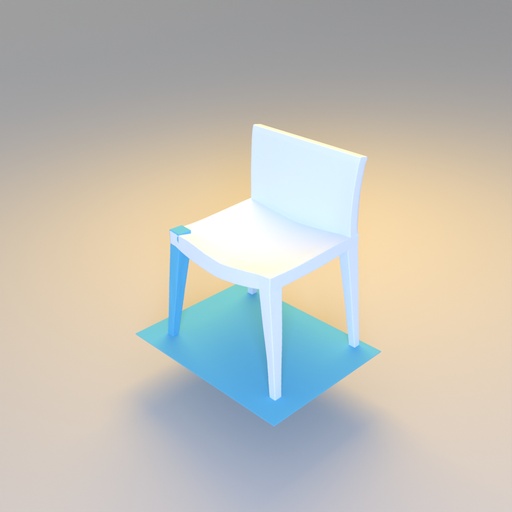} \\
\QualGridImage{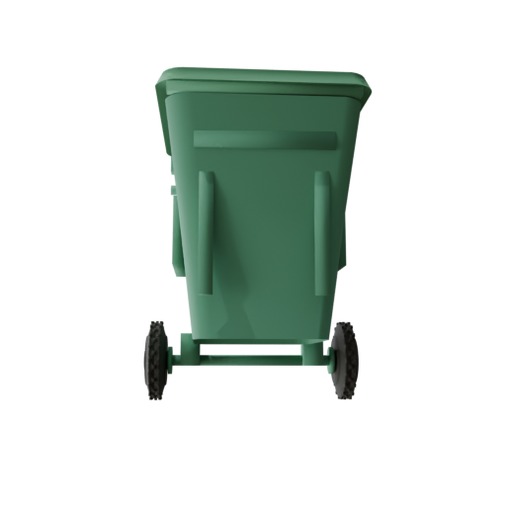} &
\QualGridImage{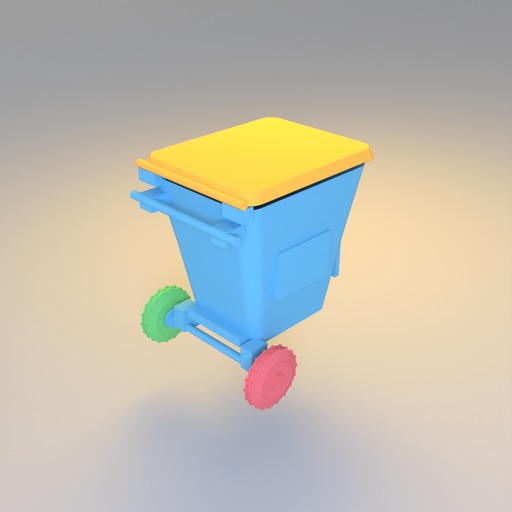} &
\QualGridImage{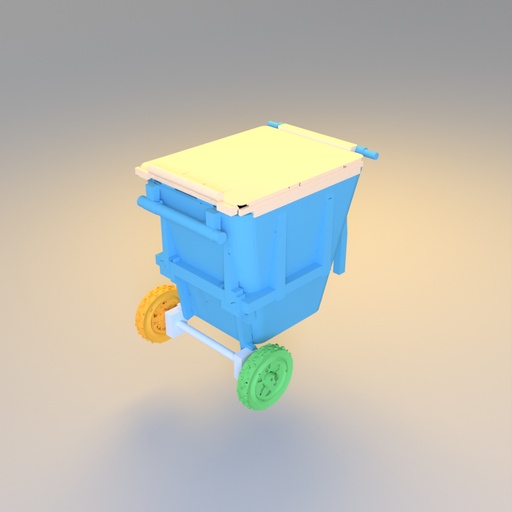} &
\QualGridImage{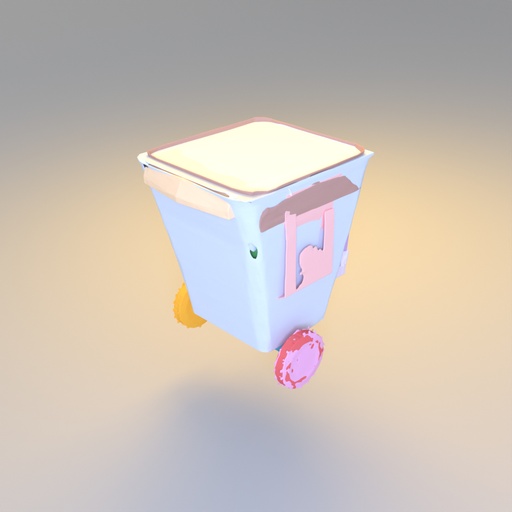} &
\QualGridImage{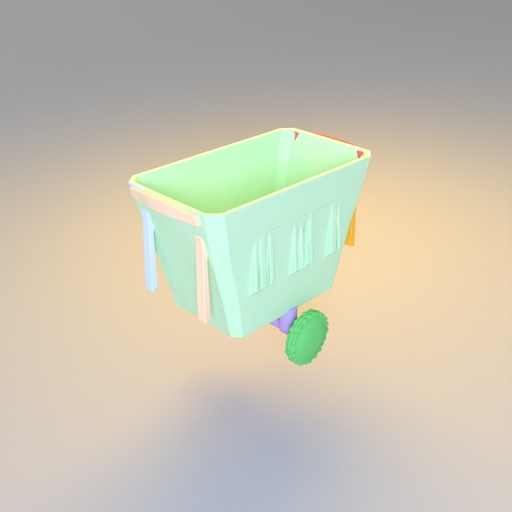} &
\QualGridImage{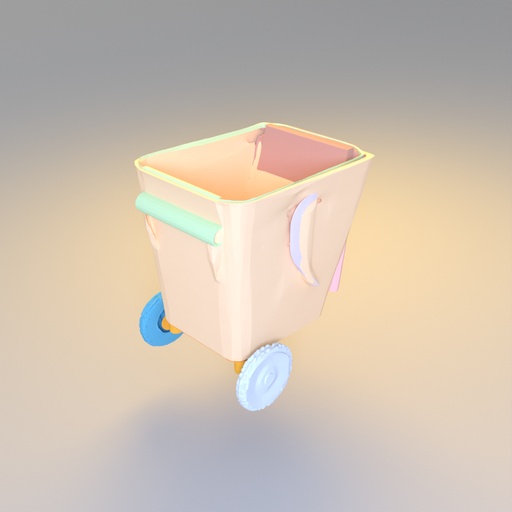} &
\QualGridImage{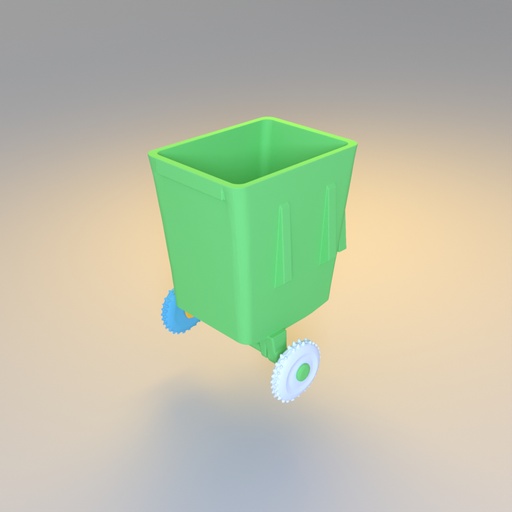} \\
\end{tabular}
\caption{\textbf{Training corpus.} Same columns, coloring and alignment as
Figure~\ref{fig:qual-test}.}
\label{fig:qual-train}
\end{figure*}

\begin{figure*}[!tp]
\centering
\QualGridSetup{7}{\linewidth}
\begin{tabular}{ccccccc}
\QualGridHeader{Input} &
\QualGridHeader{Ground truth} &
\QualGridHeader{KaiNinja\\(ours)} &
\QualGridHeader{TRELLIS.2\\+X-Part} &
\QualGridHeader{Hunyuan3D-2.1\\+X-Part} &
\QualGridHeader{OmniPart} &
\QualGridHeader{PartPacker} \\
\QualGridImage{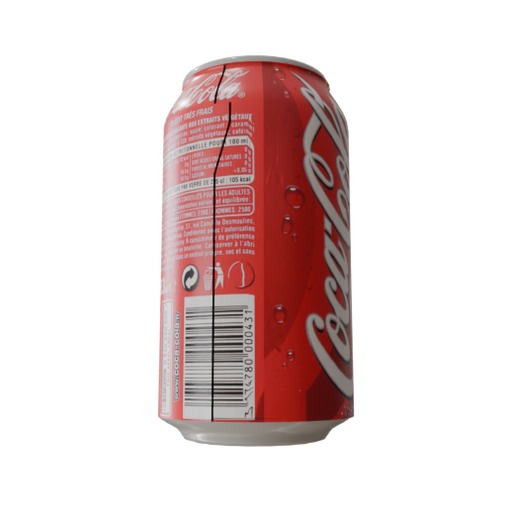} &
\QualGridImage{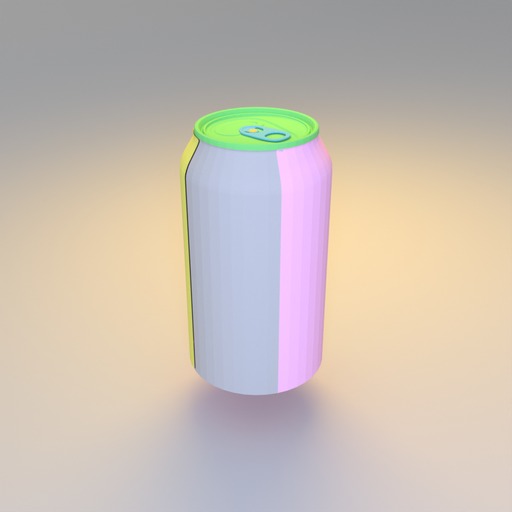} &
\QualGridImage{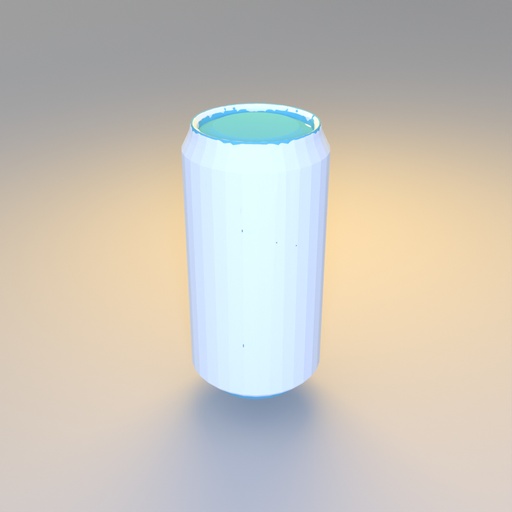} &
\QualGridImage{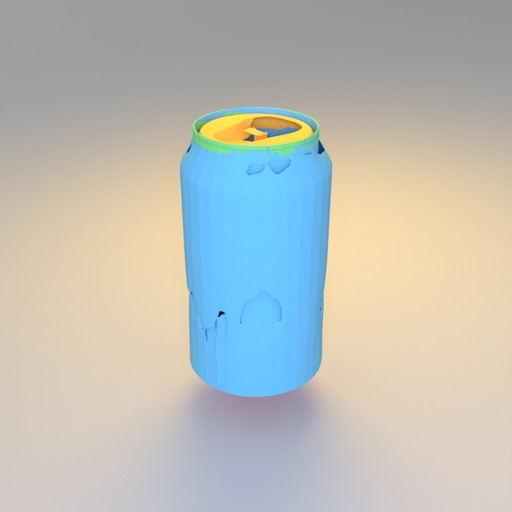} &
\QualGridImage{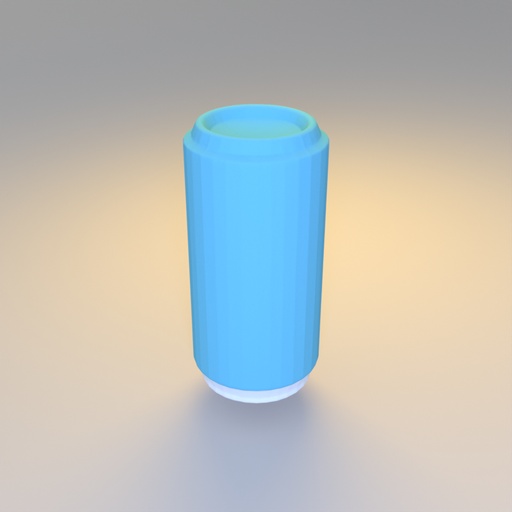} &
\QualGridImage{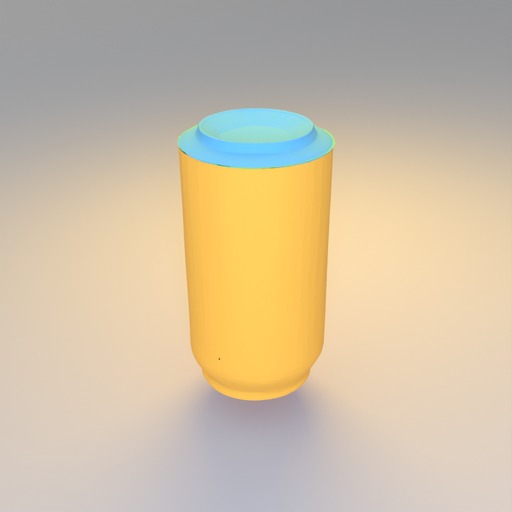} &
\QualGridImage{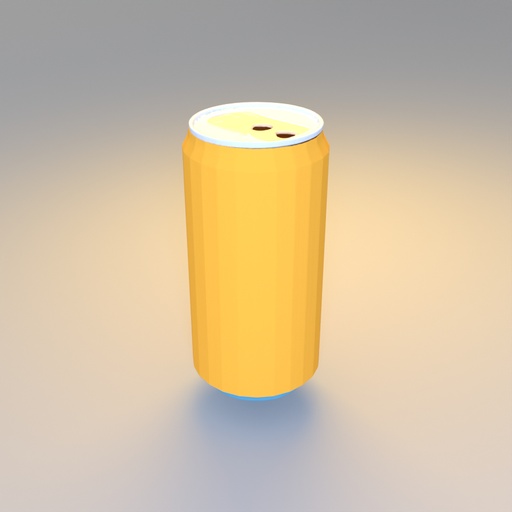} \\
\QualGridImage{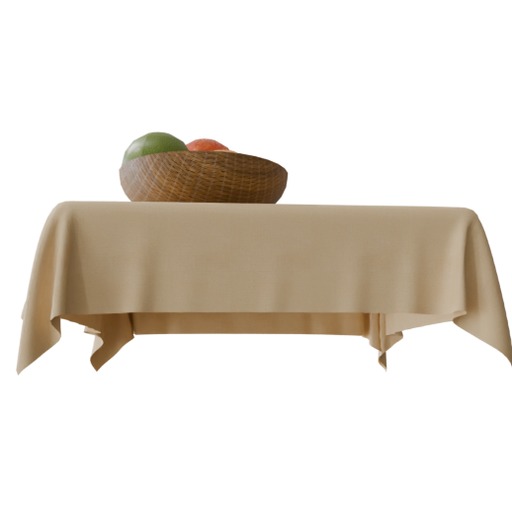} &
\QualGridImage{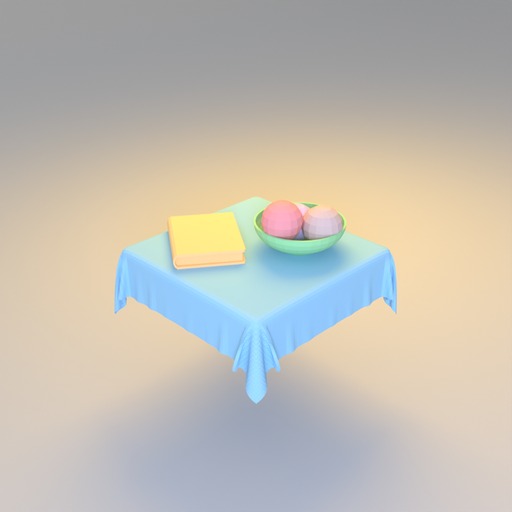} &
\QualGridImage{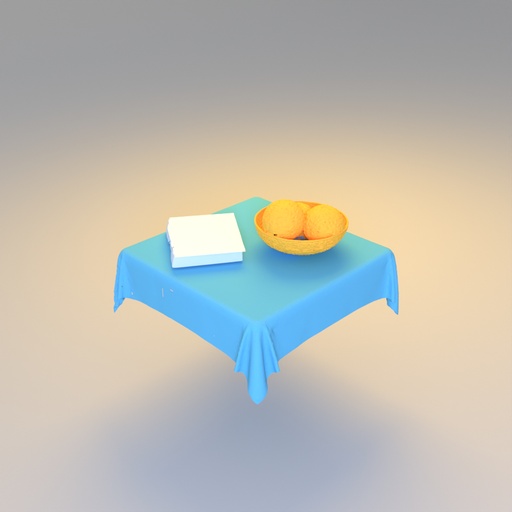} &
\QualGridImage{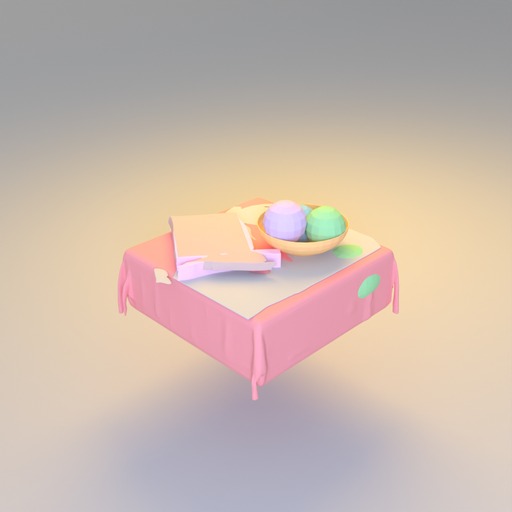} &
\QualGridImage{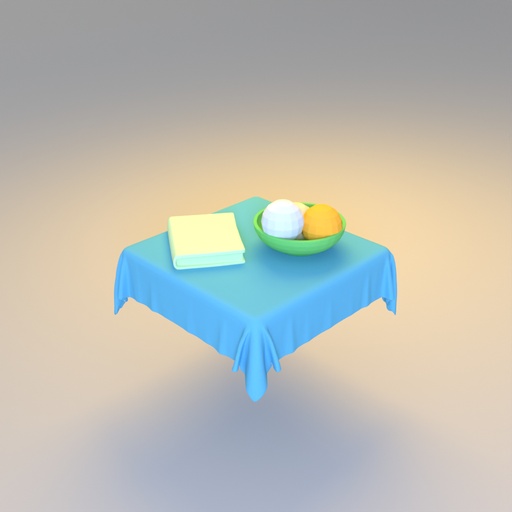} &
\QualGridImage{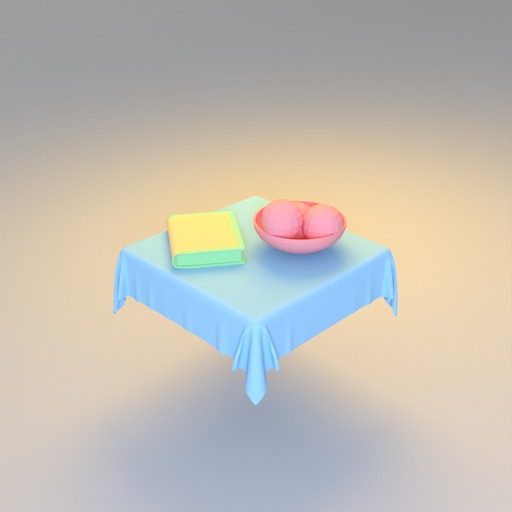} &
\QualGridImage{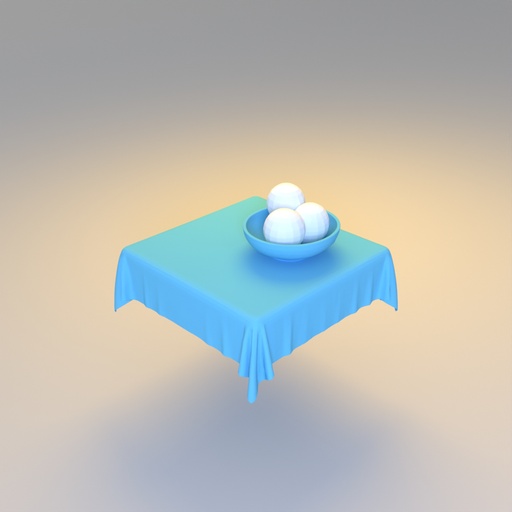} \\
\QualGridImage{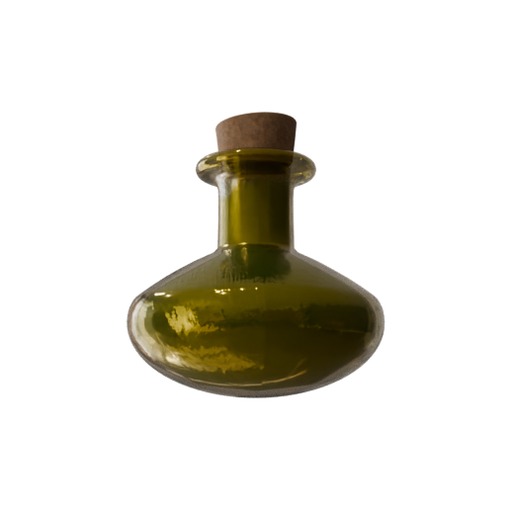} &
\QualGridImage{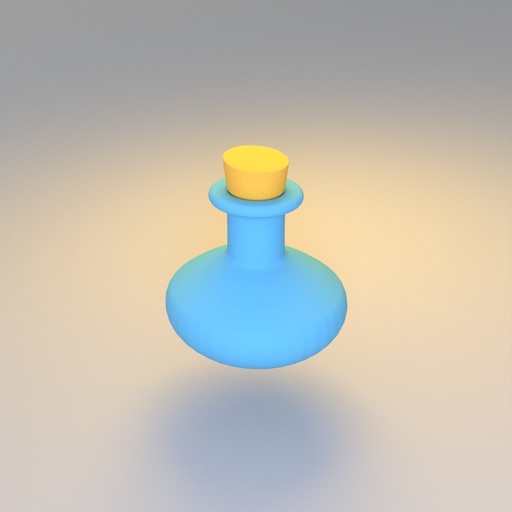} &
\QualGridImage{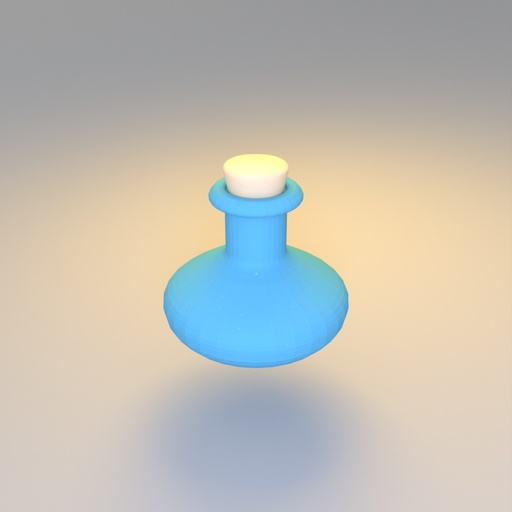} &
\QualGridImage{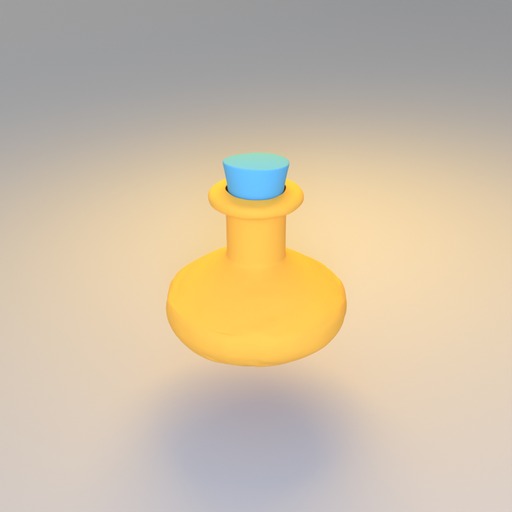} &
\QualGridImage{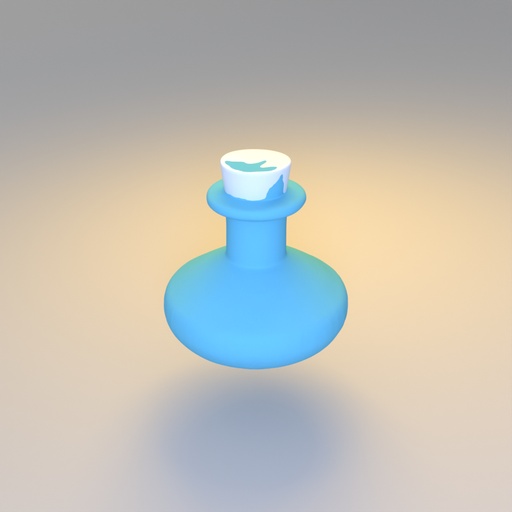} &
\QualGridImage{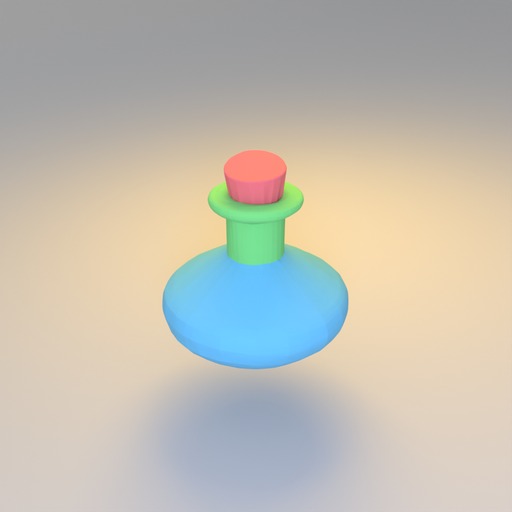} &
\QualGridImage{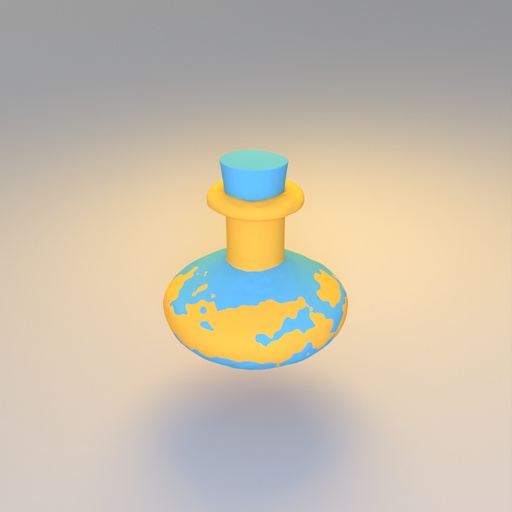} \\
\QualGridImage{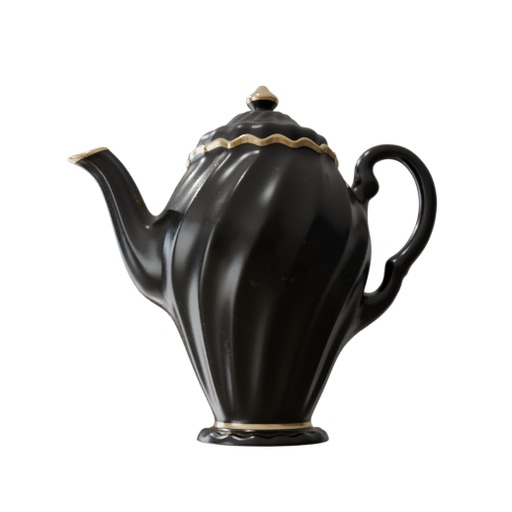} &
\QualGridImage{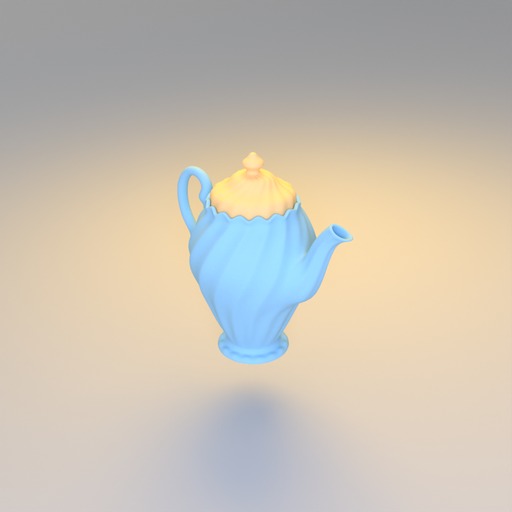} &
\QualGridImage{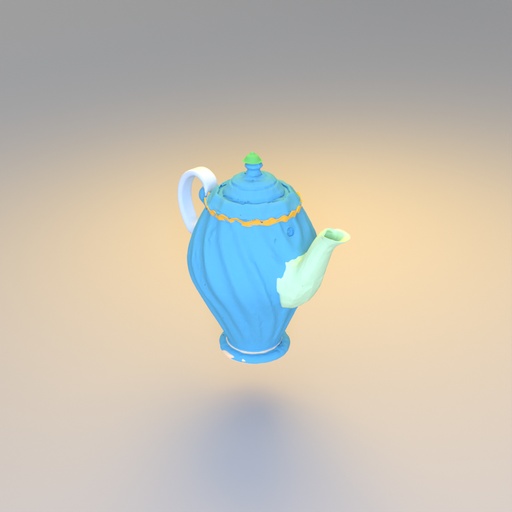} &
\QualGridImage{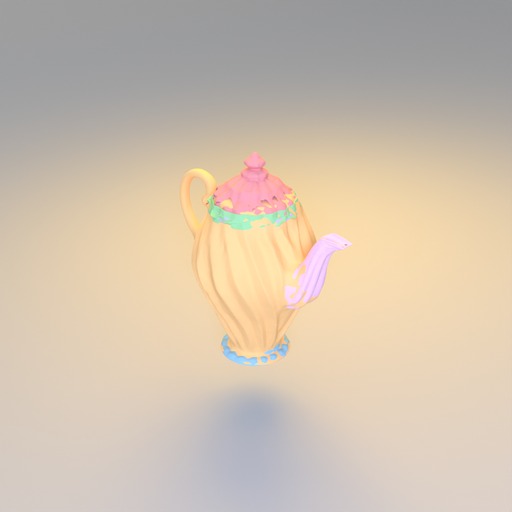} &
\QualGridImage{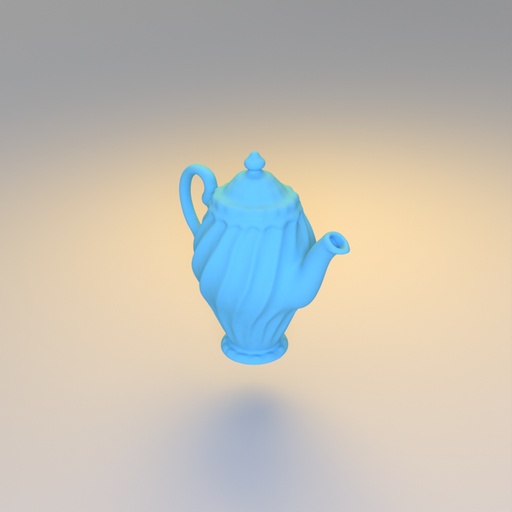} &
\QualGridImage{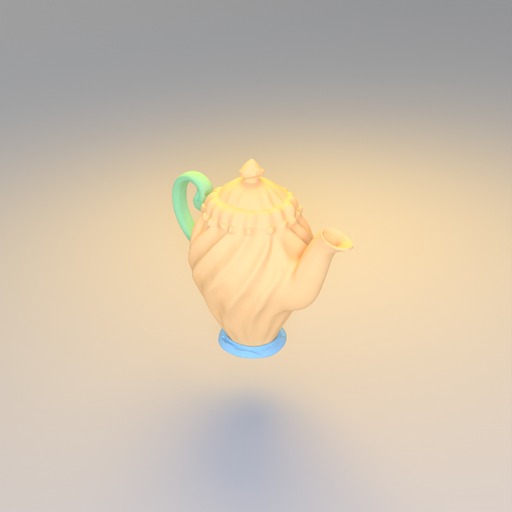} &
\QualGridImage{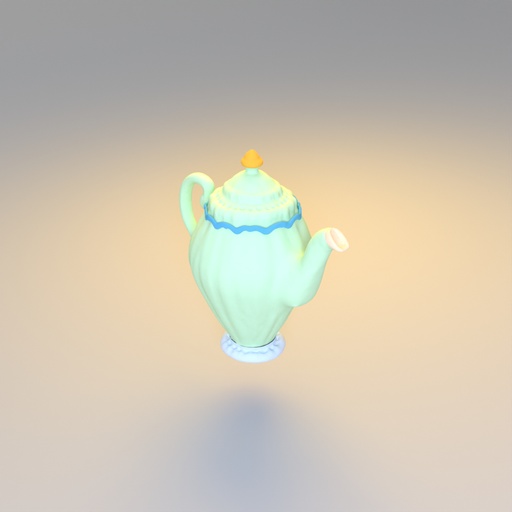} \\
\QualGridImage{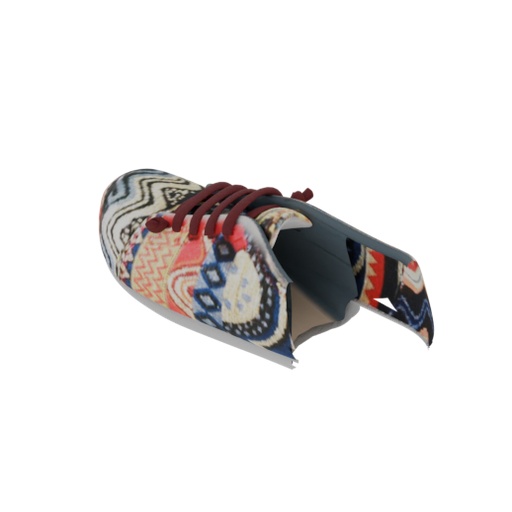} &
\QualGridImage{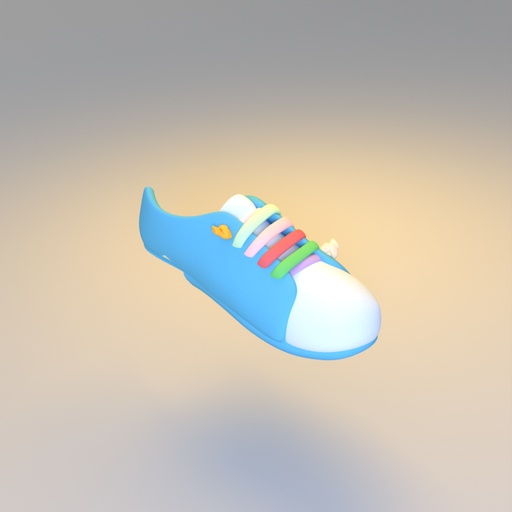} &
\QualGridImage{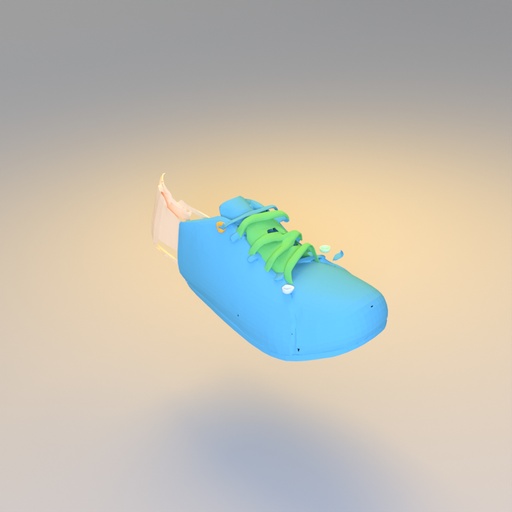} &
\QualGridImage{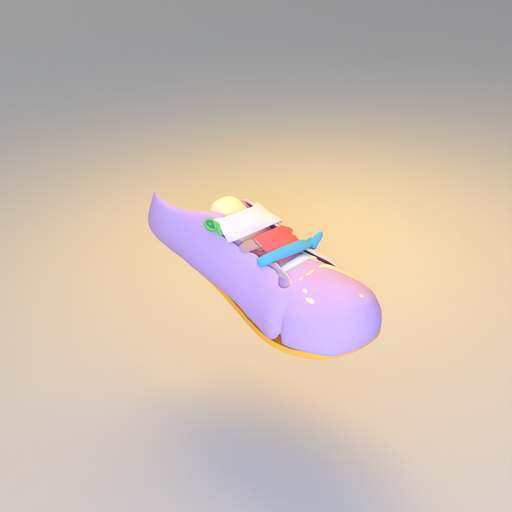} &
\QualGridImage{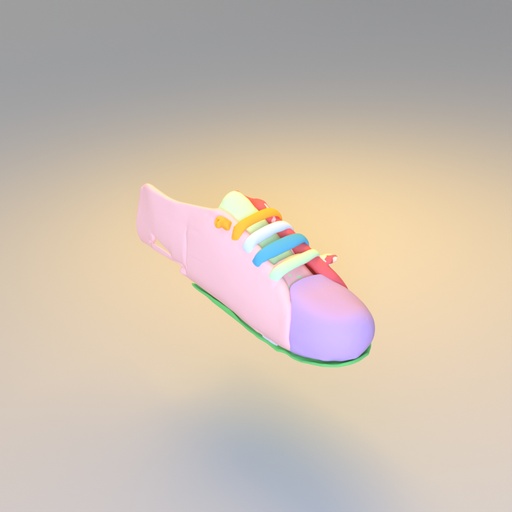} &
\QualGridImage{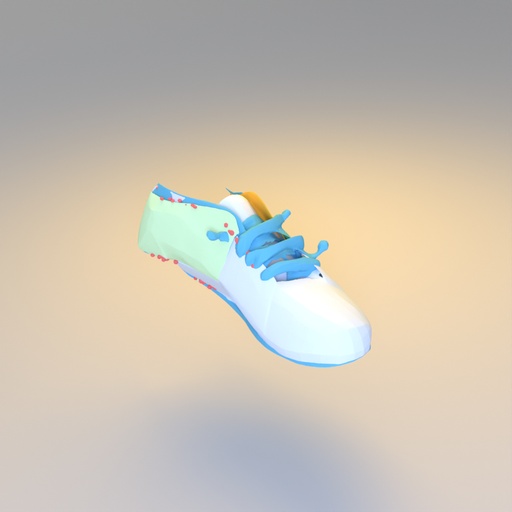} &
\QualGridImage{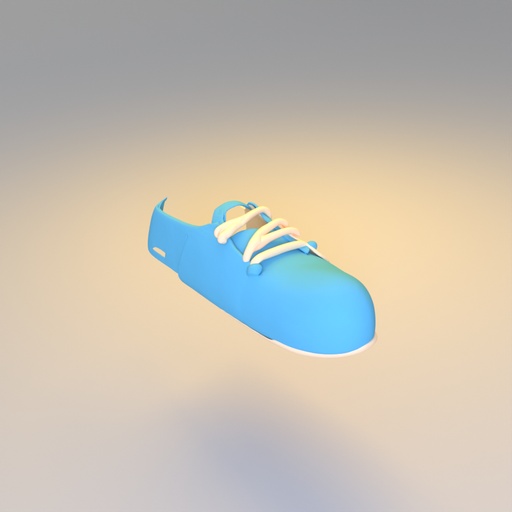} \\
\QualGridImage{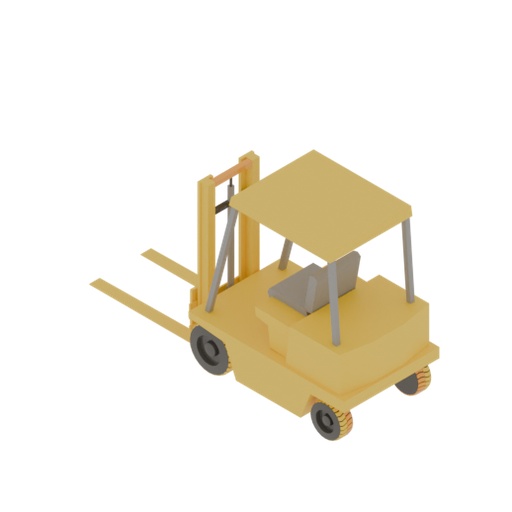} &
\QualGridImage{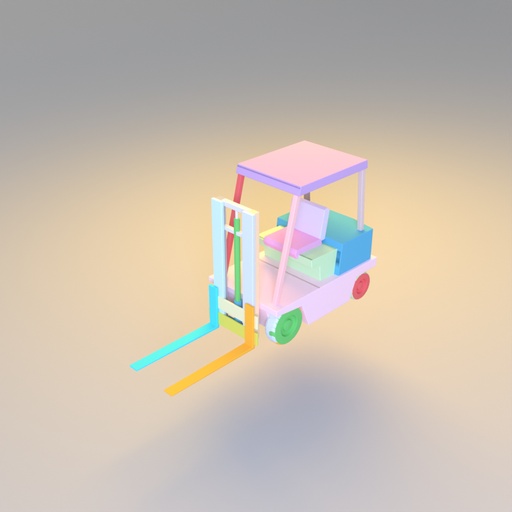} &
\QualGridImage{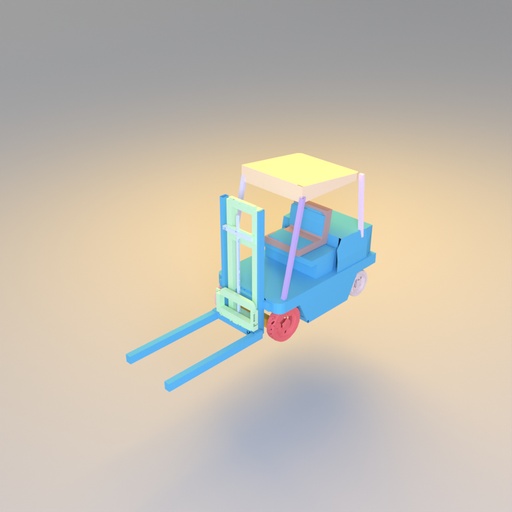} &
\QualGridImage{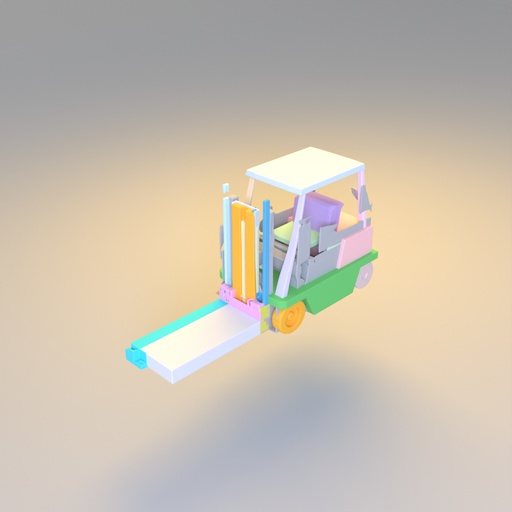} &
\QualGridImage{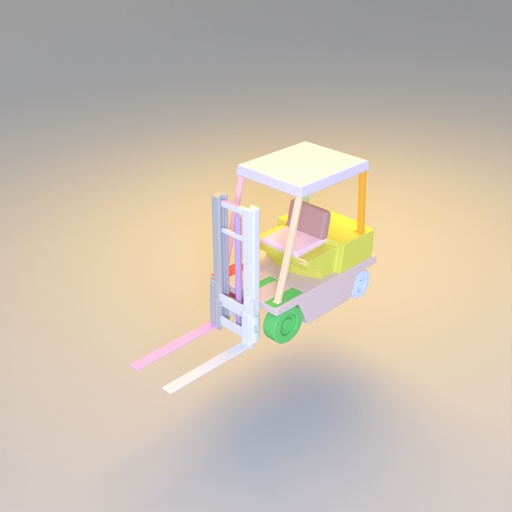} &
\QualGridImage{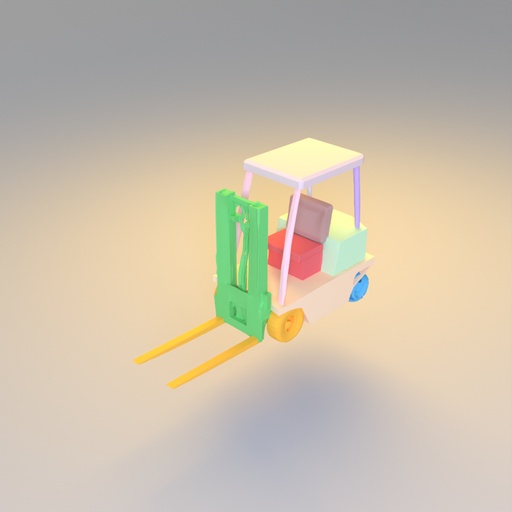} &
\QualGridImage{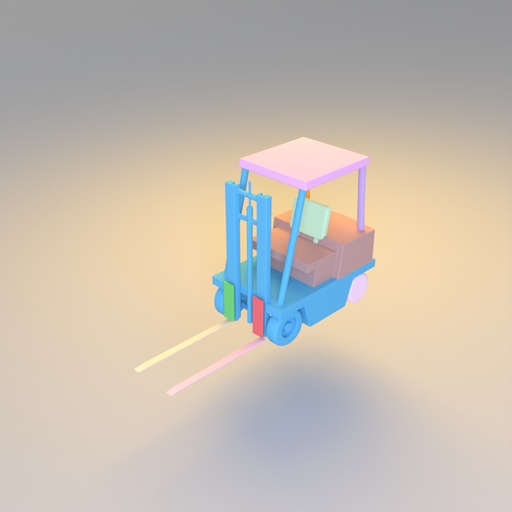} \\
\end{tabular}
\caption{\textbf{Sources not used in training.} Same columns, coloring and
alignment as Figure~\ref{fig:qual-test}.}
\label{fig:qual-nontrain}
\end{figure*}

Figures~\ref{fig:qual-test}, \ref{fig:qual-train} and~\ref{fig:qual-nontrain}
compare separated parts against both X-Part cascades, OmniPart and PartPacker
on three groups of objects: the held-out test set, the training corpus, and
objects whose source is not used in training at all. Every prediction is
rotated into the ground truth frame by its best axis permutation, the same
alignment the metrics use, so orientations are directly comparable, and
colors mark parts rather than materials. KaiNinja gives clean and coherent
parts on rigid CAD shapes, on articulated furniture and on organic objects.
It also keeps the part count close to the ground truth, whereas the baselines
tend to fuse parts, as X-Part does on the drinks can, or to shatter them, as
OmniPart does on the bundle of pipes.

\subsection{Ablation}
\label{sec:exp-abl}

Three choices decide how parts come out of this model: how many volumes the
packing uses, whether the layout flow is penalized for putting both volumes
in the same place, and what the post-processing does with the result. We take
them in that order.

\providecolor{pxLavM}{HTML}{D2C7E6}
\providecolor{pxLavT}{HTML}{55437E}
\providecolor{pxLavD}{HTML}{6C5BA6}
\providecolor{ptA}{HTML}{B95A64}
\providecolor{ptAd}{HTML}{8E3D47}
\providecolor{ptAl}{HTML}{D48A92}
\providecolor{ptB}{HTML}{6FA3C7}
\providecolor{ptBd}{HTML}{3F7195}
\providecolor{ptBl}{HTML}{9FC6DE}
\providecolor{ptC}{HTML}{D9A05B}
\providecolor{ptCd}{HTML}{A06B28}
\providecolor{ptCl}{HTML}{EFC98E}
\providecolor{ptShd}{HTML}{5D5D68}
\providecolor{ptShl}{HTML}{E4E4EA}
\begin{figure*}[t]
\centering
\resizebox{0.82\textwidth}{!}{%
\begin{tikzpicture}[
  font=\sffamily\small,
  >={Stealth[length=2.4mm, width=2.2mm]},
  pill/.style={rounded corners=2.4pt, draw=black!45, fill=white, line width=1pt,
               inner sep=0.5pt},
  vtxt/.style={font=\sffamily\fontsize{7}{7.6}\selectfont\bfseries, rotate=90,
               black!70, align=center},
  gltxt/.style={font=\sffamily\fontsize{8}{8.6}\selectfont\bfseries, rotate=90,
                white, align=center},
  wlbl/.style={font=\sffamily\small\bfseries},
  ttl/.style={font=\sffamily\small\bfseries, black!80, align=center, anchor=north},
  ar/.style={->, line width=1.1pt, black!60},
  lnbox/.style={rounded corners=2pt, line width=1pt, minimum width=17mm,
                minimum height=8.6mm, align=center,
                font=\sffamily\fontsize{6.6}{7.2}\selectfont},
  chipA/.style={rounded corners=1.4pt, fill=ptAl, draw=ptAd, line width=0.6pt,
                minimum size=3.3mm, inner sep=0pt},
  chipB/.style={rounded corners=1.4pt, fill=ptBl, draw=ptBd, line width=0.6pt,
                minimum size=3.3mm, inner sep=0pt},
  chipC/.style={rounded corners=1.4pt, fill=ptCl, draw=ptCd, line width=0.6pt,
                minimum size=3.3mm, inner sep=0pt},
  embA/.style={rounded corners=2pt, fill=ptAl, draw=ptAd, line width=1pt,
               font=\sffamily\footnotesize, inner sep=2pt, minimum width=6.4mm},
  embB/.style={rounded corners=2pt, fill=ptBl, draw=ptBd, line width=1pt,
               font=\sffamily\footnotesize, inner sep=2pt, minimum width=6.4mm},
  circ/.style={circle, draw=black!55, fill=white, line width=0.9pt, inner sep=0.5pt,
               font=\sffamily\footnotesize},
  vemb/.style={rounded corners=2pt, line width=1pt, rotate=90, align=center,
               minimum width=15mm, minimum height=5.2mm, inner sep=1pt,
               font=\sffamily\fontsize{6.6}{7.2}\selectfont},
  dotspic/.pic={\foreach \d in {-1.9,0,1.9}{\fill[black!60] (\d mm,0) circle[radius=0.6mm];}},
]


\coordinate (Pa) at (0,0);
\foreach \lane/\y/\lc/\ld in {A/11mm/ptA/ptAd, B/-11mm/ptB/ptBd}{
  \node[wlbl, \ld] at ($(Pa)+(-4mm,\y)$) {\lane};
  \foreach \i in {0,1,2}{\node[chip\lane] at ($(Pa)+(0,\y)+(0,4mm-\i*4mm)$) {};}
  \node[pill, minimum width=10mm, minimum height=16mm, draw=\lc!80!black, fill=\lc!25]
    (asa\lane) at ($(Pa)+(14mm,\y)$) {};
  \node[vtxt] at (asa\lane) {Volume\\DiT block};
  \pic at ($(Pa)+(23mm,\y)$) {dotspic};
  \coordinate (adot\lane) at ($(Pa)+(25mm,\y)$);
}
\node[pill, minimum width=11mm, minimum height=38mm, draw=pxLavT, fill=pxLavD]
  (agl) at ($(Pa)+(33mm,0)$) {};
\node[gltxt] at (agl) {Global DiT block};
\coordinate (atop) at ($(Pa)+(14mm,25.6mm)$);
\coordinate (abot) at ($(Pa)+(14mm,-25.6mm)$);
\begin{scope}[on background layer]
\node[rounded corners=2.6pt, draw=pxLavT, fill=pxLavM, line width=1.2pt, inner sep=2.4mm,
      fit={(asaA) (asaB) (adotA) (adotB) (agl) (atop) (abot)}] (agrp) {};
\end{scope}
\node[font=\sffamily\small\bfseries, pxLavT, anchor=north west]
  at ($(agrp.north west)+(1.4mm,-0.8mm)$) {$\circlearrowright$ $\times 5$};
\node[ttl] at ($(Pa)+(23.75mm,-30mm)$) {(a) dual volume (ours)};

\coordinate (Pc) at (56mm,0);
\foreach \lane/\y/\lc/\ld in {A/13mm/ptA/ptAd, B/0mm/ptB/ptBd, C/-13mm/ptC/ptCd}{
  \node[wlbl, \ld] at ($(Pc)+(-4mm,\y)$) {\lane};
  \foreach \i in {0,1}{\node[chip\lane] at ($(Pc)+(0,\y)+(0,2mm-\i*4mm)$) {};}
  \node[pill, minimum width=10mm, minimum height=11mm, draw=\lc!80!black, fill=\lc!25]
    (csa\lane) at ($(Pc)+(14mm,\y)$) {};
  \node[vtxt, font=\sffamily\fontsize{6.2}{6.8}\selectfont\bfseries] at (csa\lane) {Volume\\DiT block};
  \pic at ($(Pc)+(23mm,\y)$) {dotspic};
  \coordinate (cdot\lane) at ($(Pc)+(25mm,\y)$);
}
\node[pill, minimum width=11mm, minimum height=38mm, draw=pxLavT, fill=pxLavD]
  (cgl) at ($(Pc)+(33mm,0)$) {};
\node[gltxt] at (cgl) {Global DiT block};
\coordinate (ctop) at ($(Pc)+(14mm,25.6mm)$);
\coordinate (cbot) at ($(Pc)+(14mm,-25.6mm)$);
\begin{scope}[on background layer]
\node[rounded corners=2.6pt, draw=pxLavT, fill=pxLavM, line width=1.2pt, inner sep=2.4mm,
      fit={(csaA) (csaB) (csaC) (cdotA) (cdotC) (cgl) (ctop) (cbot)}] (cgrp) {};
\end{scope}
\node[font=\sffamily\small\bfseries, pxLavT, anchor=north west]
  at ($(cgrp.north west)+(1.4mm,-0.8mm)$) {$\circlearrowright$ $\times 5$};
\node[ttl] at ($(Pc)+(23.75mm,-30mm)$) {(b) tri volume};

\path (108.8mm,0);
\end{tikzpicture}}
\par\vspace{1.6mm}
\caption{\textbf{Ablated Stage-1 variants.} Both variants run the same block
budget: each repeat applies six \emph{Volume DiT blocks} per stream and one
\emph{Global DiT block} that pools the tokens of all streams and attends
across them, for $30$ and $5$ blocks in total. They differ only in how many
volumes the parts are packed into: (a)~two volumes, from the bipartite
packing of Section~\ref{sec:method-repr}; (b)~three volumes, from a greedy
three-coloring of the part-contact graph. Table~\ref{tab:abl} compares
them.}
\label{fig:abl-arch}
\end{figure*}
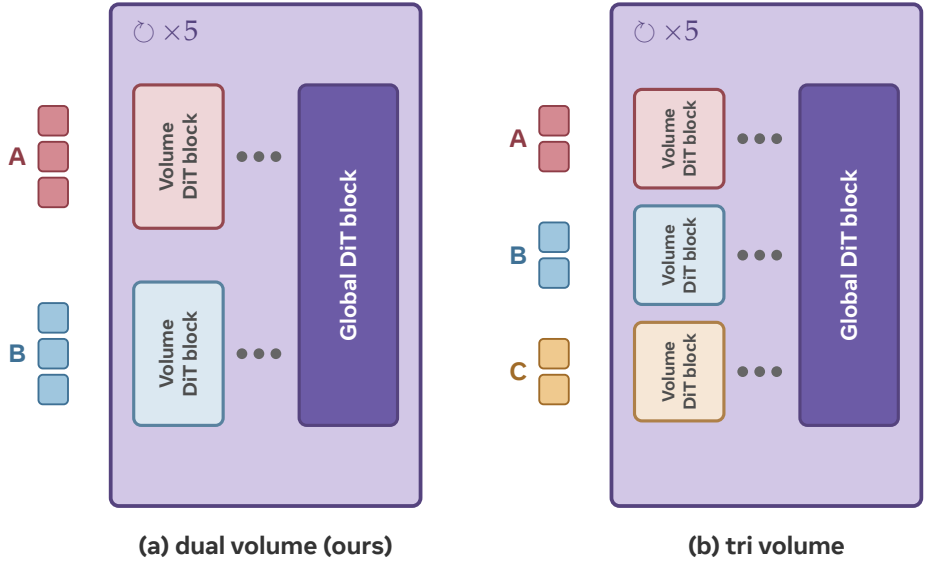

\begin{table}[t]
\centering
\caption{\textbf{Two volumes against three}, scored on Stage-1 occupancy alone
over the $986$ objects both arms complete. Three volumes additionally fail on
$13$, two volumes on none.}
\label{tab:abl}
\tablestyle{4pt}{1.15}
\begin{tabular}{l ccc cccc}
\toprule
Variant & CD$_W\!\downarrow$ & F1$_W^{0.1}$ & F1$_W^{0.05}$ & mIoU$_P$ & CD$_P\!\downarrow$ & F1$_P^{0.1}$ & F1$_P^{0.05}$ \\
\midrule
Tri-volume (3 streams) & 0.0428 & 0.961 & 0.908 & \bnum{0.306} & \bnum{0.0834} & \bnum{0.875} & \bnum{0.795} \\
\rowcolor{rowgray}
Dual-volume (ours) & \bnum{0.0423} & \bnum{0.963} & 0.908 & 0.301 & 0.0901 & 0.861 & 0.779 \\
\bottomrule
\end{tabular}
\end{table}

\paragraph{Number of volumes.}
The obvious generalization is to pack into three volumes. The contact graph
is then three-colored greedily instead of contracted until bipartite, and the
flow runs three streams (Figure~\ref{fig:abl-arch}b). Backbone, data and
recipe are identical, so the only thing that changes is the number of
volumes. Table~\ref{tab:abl} compares them at the end of Stage~1, where the
designs differ. A prediction is the set of occupied voxel centers at $64^3$.
Ground truth is $100$K points from the continuous surface of the whole object
and $20$K from each part, and parts are matched by Hungarian assignment on
voxel-set IoU, with no merge and no relabel. F1$^{0.1}$ saturates in this
setting because one voxel spans $0.016$, so a $0.1$ tolerance is over six
voxels wide, and we read the $0.05$ column instead.

Three volumes are not worse at geometry. The whole object ties, and on parts
three volumes are slightly ahead, by $1.6$ points of F1$_P^{0.05}$ and
$0.0067$ of CD$_P$, both significant under a paired bootstrap. The part counts
suggest why. Three volumes leave $4.73$ connected components per object
against $3.84$ for two, with $5.05$ in the ground truth, so a third volume
gives touching parts one more way to avoid being fused. What three volumes
cost is robustness and supervision. They fail outright on $13$ of $1000$
objects where two volumes never do, and they leave a stream empty on $7.7\%$
of objects against $1.5\%$, so a third of the capacity idles on one object in
thirteen while every step costs half again as much. We choose two volumes for
robustness, not for a fidelity advantage.

\begin{figure*}[t]
\centering
\QualGridSetup{7}{\linewidth}
\begin{tabular}{ccccccc}
{\sffamily\footnotesize\bfseries Input} &
\multicolumn{3}{c}{\sffamily\footnotesize\bfseries With $\mathcal{L}_{\text{ov}}$} &
\multicolumn{3}{c}{\sffamily\footnotesize\bfseries Without $\mathcal{L}_{\text{ov}}$} \\
& {\sffamily\scriptsize Result} & {\sffamily\scriptsize Volume $A$} & {\sffamily\scriptsize Volume $B$} &
{\sffamily\scriptsize Result} & {\sffamily\scriptsize Volume $A$} & {\sffamily\scriptsize Volume $B$} \\
\QualGridImage{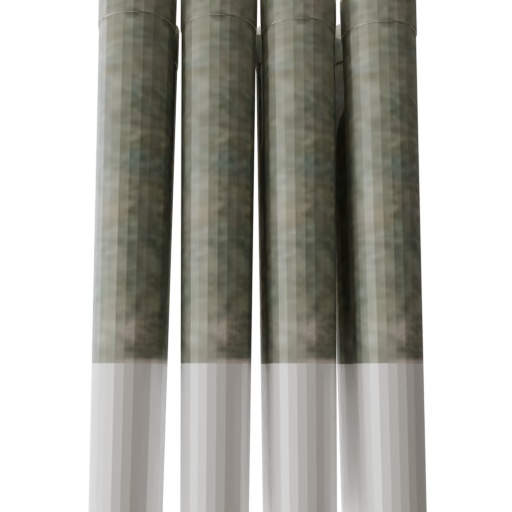} &
\QualGridImage{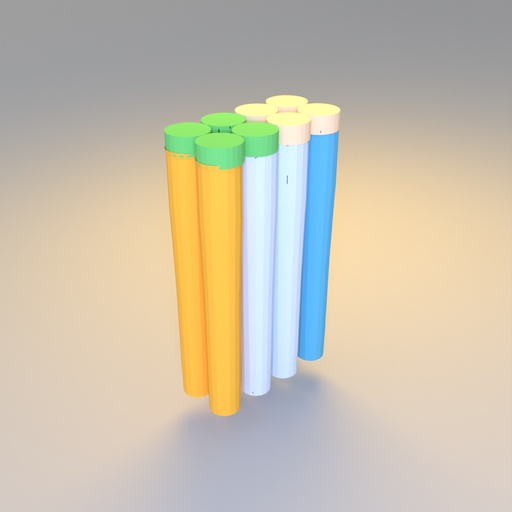} &
\QualGridImage{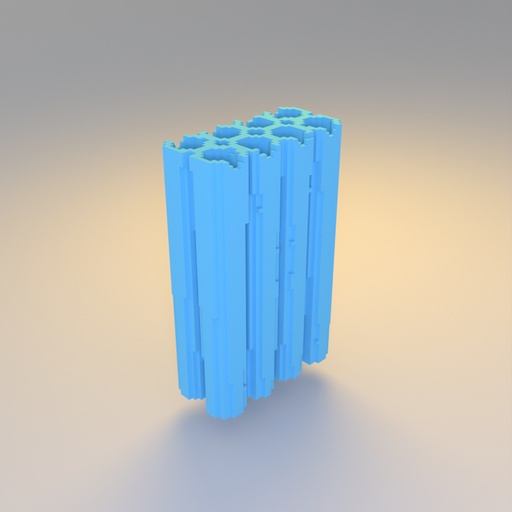} &
\QualGridImage{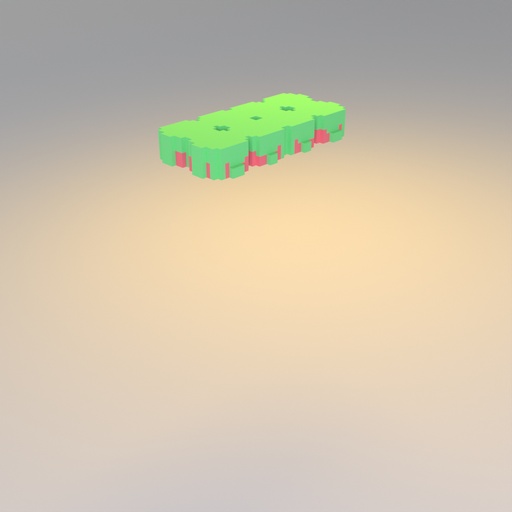} &
\QualGridImage{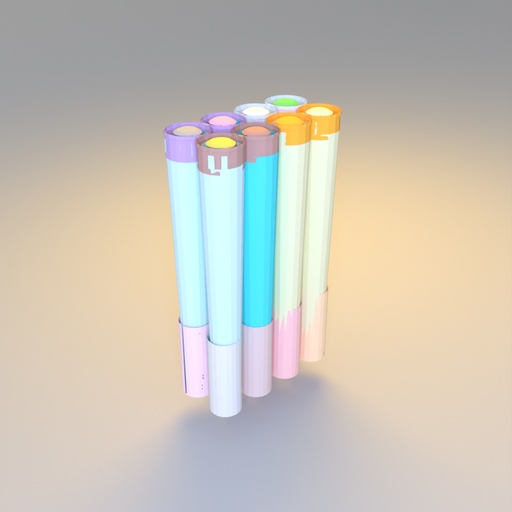} &
\QualGridImage{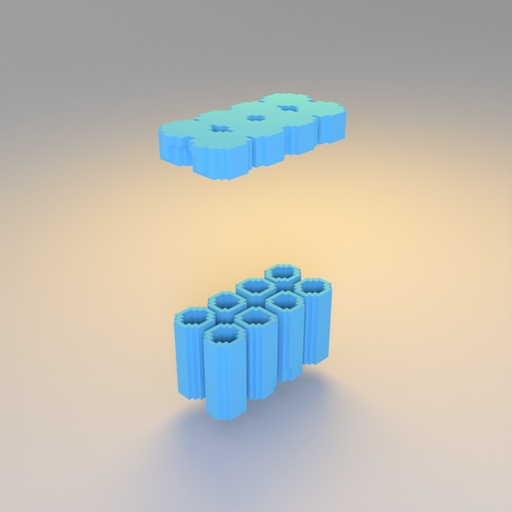} &
\QualGridImage{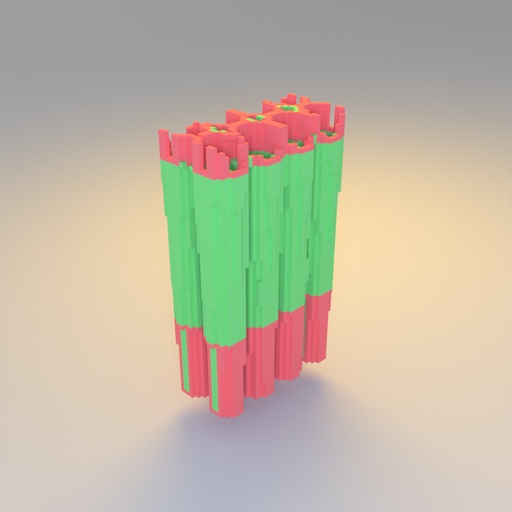} \\
\QualGridImage{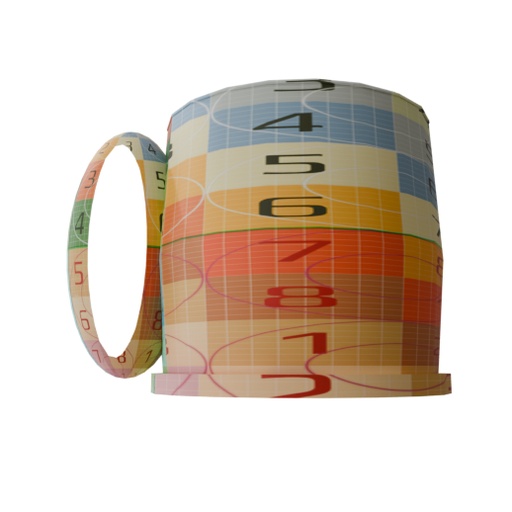} &
\QualGridImage{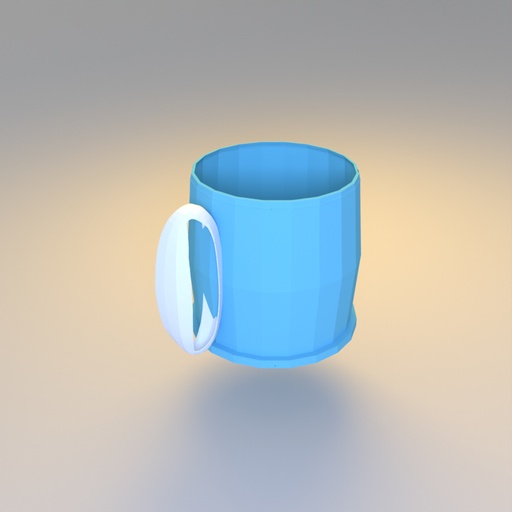} &
\QualGridImage{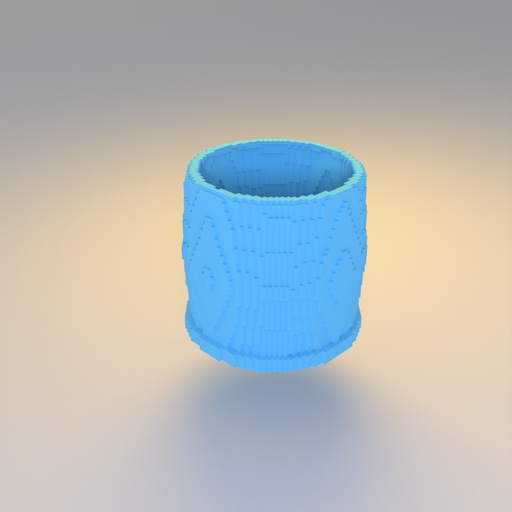} &
\QualGridImage{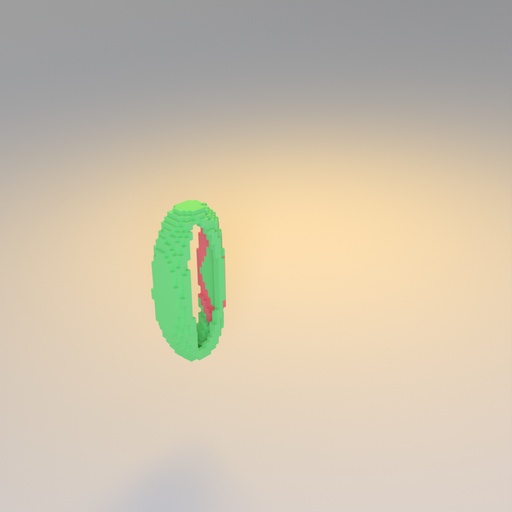} &
\QualGridImage{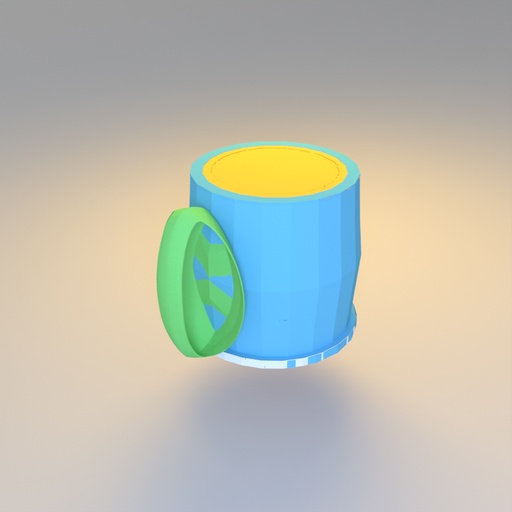} &
\QualGridImage{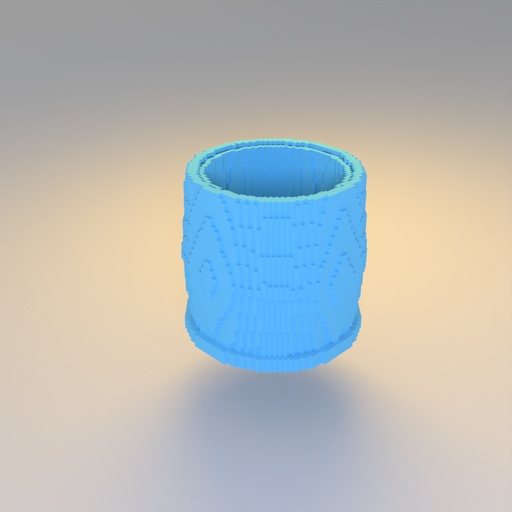} &
\QualGridImage{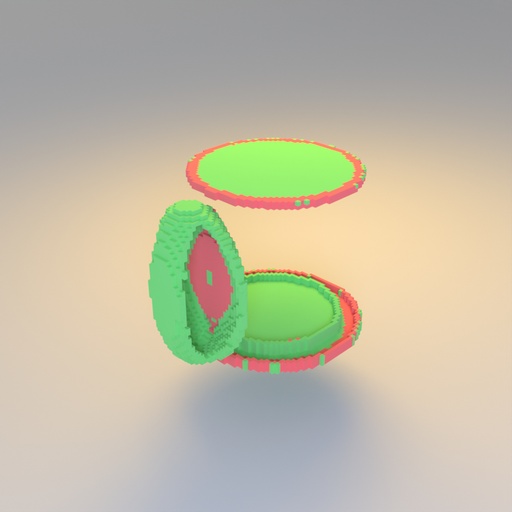} \\
\QualGridImage{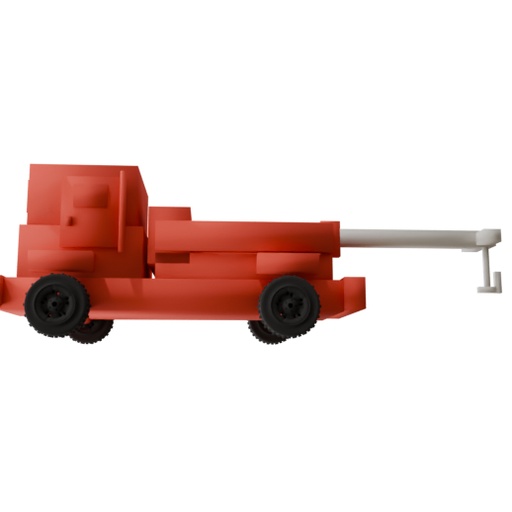} &
\QualGridImage{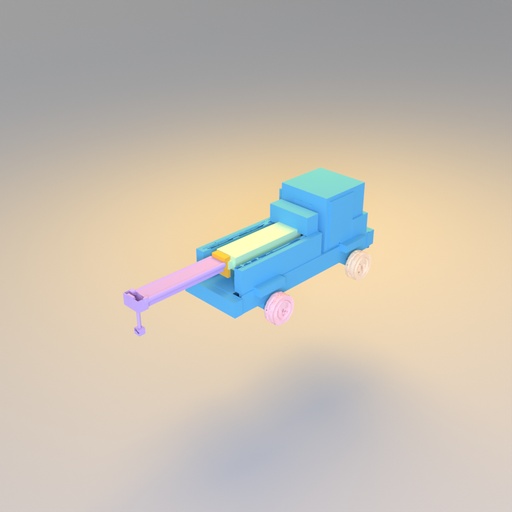} &
\QualGridImage{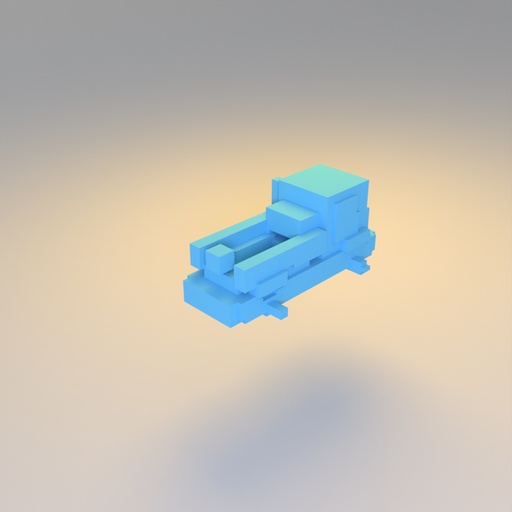} &
\QualGridImage{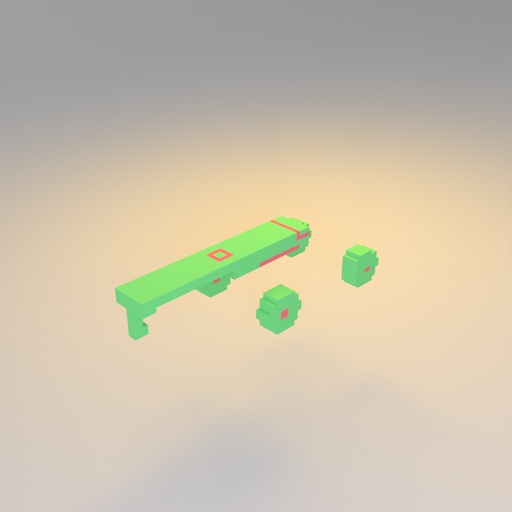} &
\QualGridImage{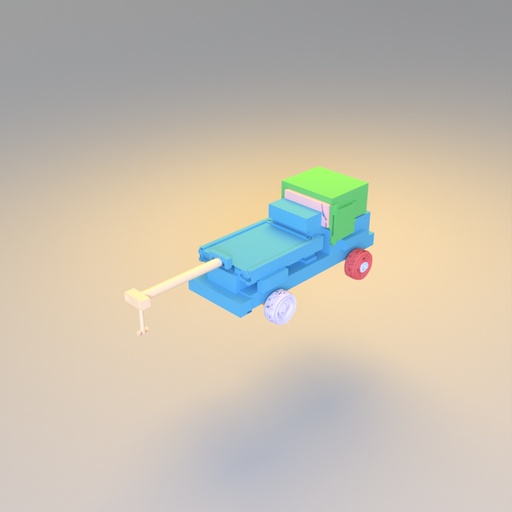} &
\QualGridImage{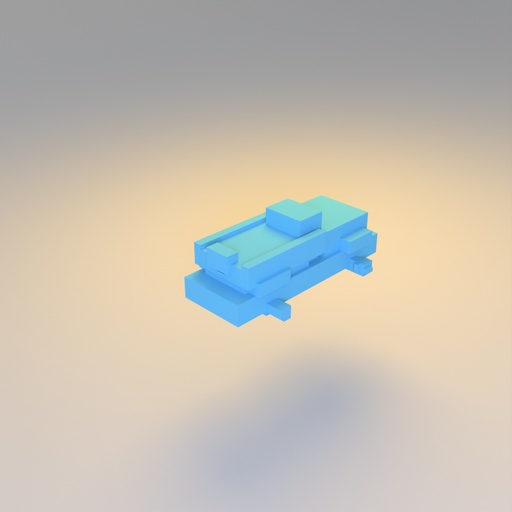} &
\QualGridImage{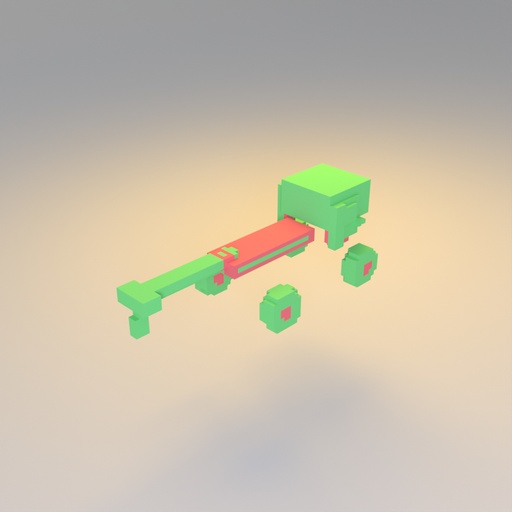} \\
\end{tabular}
\caption{\textbf{Effect of the disjointness penalty.} Three objects from
the qualitative figures, generated with and without
$\mathcal{L}_{\text{ov}}$ from the same image and seed. For each arm we show
the final parts and the Stage-1 occupancy of the two volumes. Volume $A$ is
blue; in the $B$ columns, green marks voxels that only $B$ occupies and red
marks voxels claimed by both volumes. Without the term, volume $B$ repeats
surfaces that already exist in $A$: the whole bundle of pipes, the rims of
the mug, the chassis of the truck. With the term, $B$ carries its own parts.
The overlap $|A\cap B|/\min(|A|,|B|)$ drops from $0.64$ to $0.31$, from
$0.21$ to $0.14$ and from $0.22$ to $0.09$. These objects were chosen for a
visible difference; on many others the two arms are close, and on a few the
term does not help. After post-processing the final parts differ little in
either case, which matches Table~\ref{tab:abl-ov}.}
\label{fig:abl-ov-qual}
\end{figure*}

\begin{table}[t]
\centering
\caption{\textbf{Post-processing and the disjointness penalty}, on the $977$ test
objects all six arms complete. Ground truth averages $5.85$ parts.}
\label{tab:abl-ov}
\tablestyle{4pt}{1.15}
\begin{tabular}{ll ccc cccc c}
\toprule
Post-processing & $\mathcal{L}_{\text{ov}}$ & CD$_W\!\downarrow$ & F1$_W^{0.1}$ & F1$_W^{0.05}$ & mIoU$_P$ & CD$_P\!\downarrow$ & F1$_P^{0.1}$ & F1$_P^{0.05}$ & \#parts \\
\midrule
none & \cmark & 0.0195 & 0.973 & 0.913 & 0.481 & 0.0627 & 0.783 & 0.675 & 14.06 \\
none & \xmark & 0.0191 & 0.975 & 0.917 & 0.482 & 0.0621 & 0.789 & 0.681 & 14.09 \\
\midrule
merge & \cmark & 0.0195 & 0.973 & 0.912 & 0.482 & 0.0627 & 0.783 & 0.675 & 13.98 \\
merge & \xmark & 0.0191 & 0.975 & 0.917 & 0.483 & 0.0621 & 0.788 & 0.679 & 14.06 \\
\midrule
\rowcolor{rowgray}
merge + relabel (ours, full) & \cmark & 0.0186 & 0.974 & 0.918 & 0.520 & 0.0613 & 0.784 & 0.692 & 5.43 \\
merge + relabel & \xmark & 0.0182 & 0.977 & 0.923 & 0.521 & 0.0606 & 0.788 & 0.695 & 5.43 \\
\bottomrule
\end{tabular}
\end{table}

\paragraph{The disjointness penalty.}
Table~\ref{tab:abl-ov} crosses the overlap term $\mathcal{L}_{\text{ov}}$ of
Section~\ref{sec:method-train} with the three levels of post-processing. On
the decoded output the two arms are indistinguishable. At every level they
agree to within $0.006$ on each metric, with the arm trained without the term
consistently and slightly ahead, and the component counts agree to within
$0.1$ of a part. The effect of the term appears one stage earlier, where it
is aimed. Over $250$ held-out objects scored at the Stage-1 occupancy, the
term lowers the median cross-volume overlap from $0.229$ to $0.184$, and it
lowers the fraction of objects in which one volume largely duplicates the
other, overlap above $0.5$, from $17.2\%$ to $14.4\%$.
Figure~\ref{fig:abl-ov-qual} shows three objects under both arms. Without
the term, volume $B$ repeats surfaces that already exist in volume $A$, such
as the whole bundle of pipes or the rims of the mug; with the term, $B$
carries its own parts. The effect is not uniform: across the
objects of the qualitative figures the term lowers the overlap on most,
leaves some unchanged, and raises it on a few. We keep $\mathcal{L}_{\text{ov}}$ in the shipped
model because it acts on the layout, the stage that decides part structure,
and it costs one decoder pass per step.

\paragraph{Post-processing.}
The same table isolates the two post-processing steps of
Section~\ref{sec:method-infer} along its first column. The merge changes the
metrics very little; it is a safeguard against duplicated volumes, which the
trained model sometimes produces. But the relabel matters as the final step, which
attaches floating facets back onto their parts and joins the pieces of one
part. This trick cuts the component count from $14.06$ to $5.43$ against $5.85$
in the ground truth and improves mIoU$_P$ and
F1$_P^{0.05}$.

\section{Conclusion}
\label{sec:conclusion}

We presented KaiNinja, a part-level extension of a native 3D generator.
TRELLIS.2 is extended in place: its O-Voxel representation is packed into two
interleaved volumes, its cascade is adapted stage by stage, and its VAE and
material pipeline stay untouched. The result turns one image into an asset
whose parts are separate meshes, at a small constant increase in generation
cost and with no mask or segmenter anywhere in the pipeline. Two findings
stand out. Extending the generator in place outperforms every pipeline we
compared against that attaches a segmenter outside it, on the whole object
and on its parts alike, and the dual-volume packing improves whole-object
fidelity over the same backbone fine-tuned on the same corpus, which
suggests that packing is a better representation of the same objects rather
than only a container for parts.

\paragraph{Limitations and future work.}
\begin{figure*}[t]
\centering
\QualGridSetup{5}{.785\linewidth}
\begin{tabular}{ccccc}
\QualGridHeader{Input} &
\QualGridHeader{Ground truth} &
\QualGridHeader{KaiNinja (ours)} &
\QualGridHeader{Volume $A$} &
\QualGridHeader{Volume $B$} \\
\QualGridImage{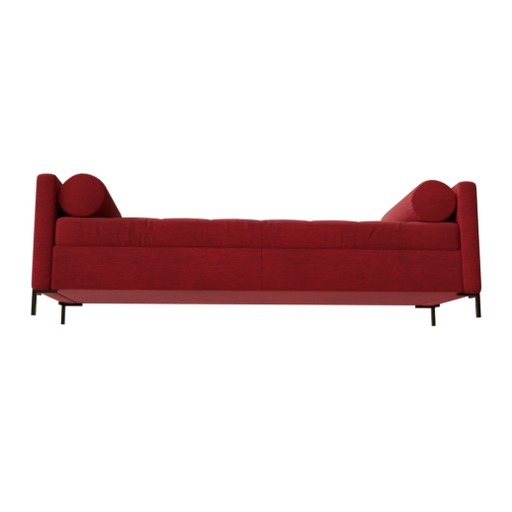} &
\QualGridImage{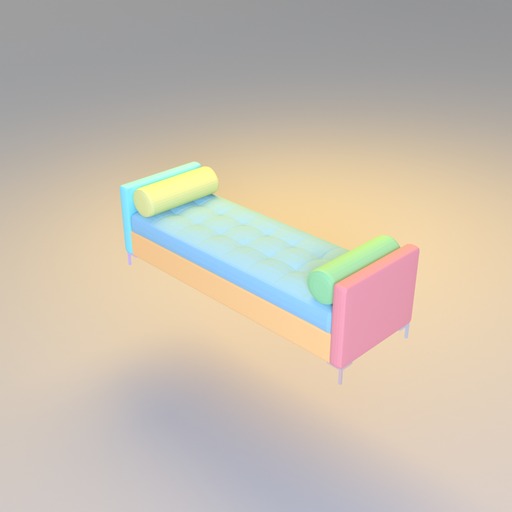} &
\QualGridImage{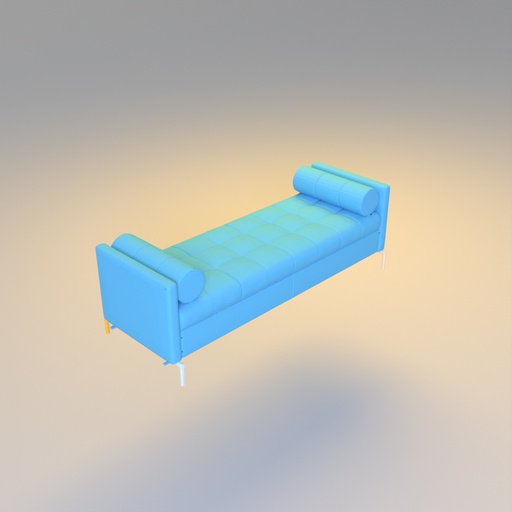} &
\QualGridImage{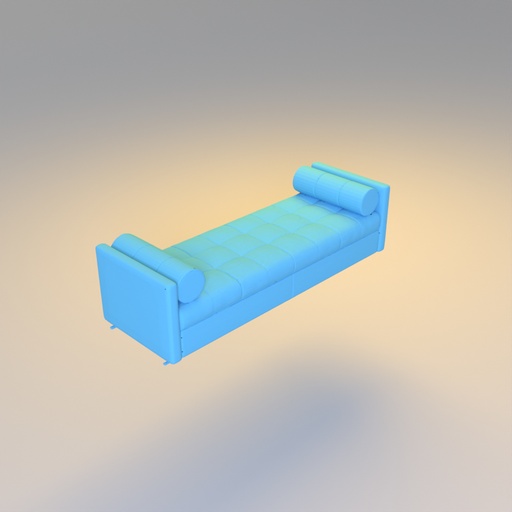} &
\QualGridImage{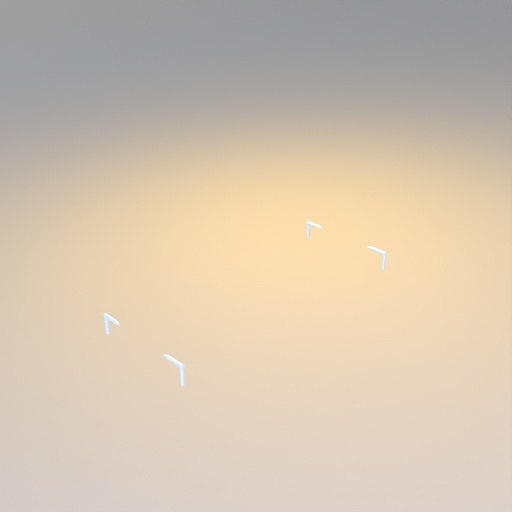} \\
\QualGridImage{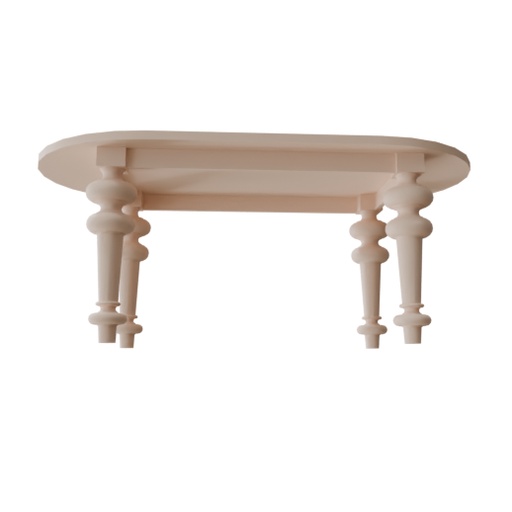} &
\QualGridImage{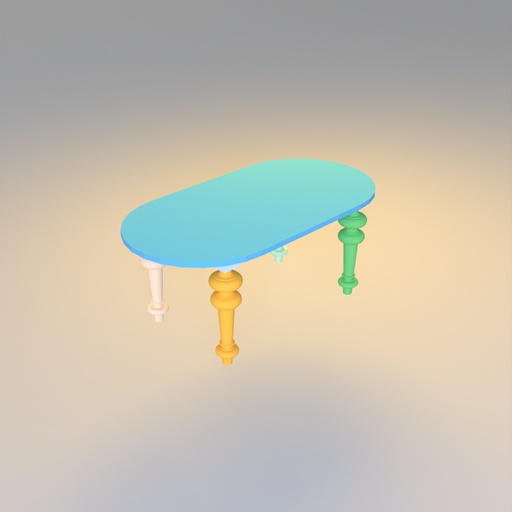} &
\QualGridImage{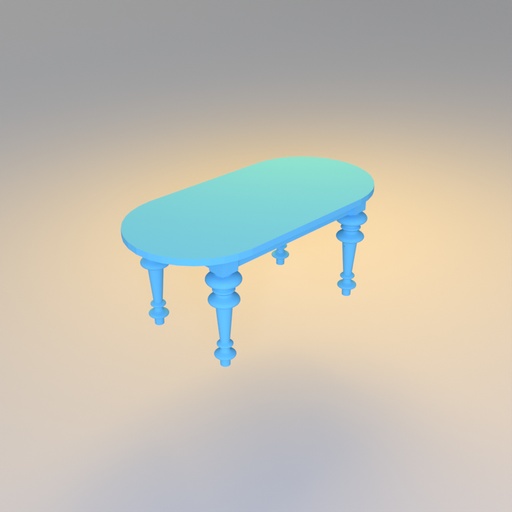} &
\QualGridImage{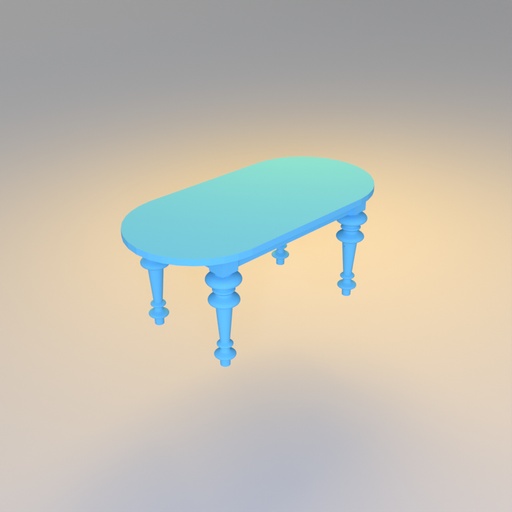} &
\QualGridImage{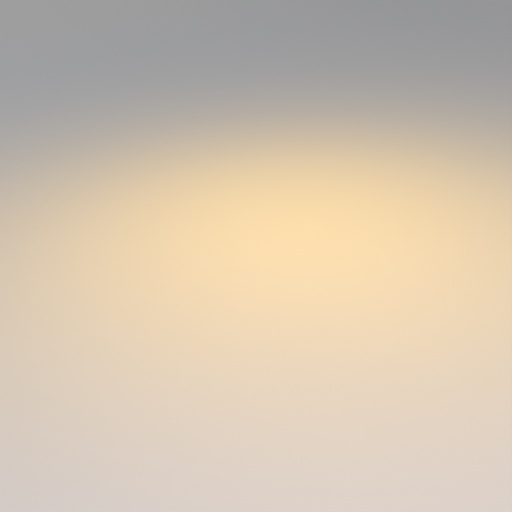} \\
\QualGridImage{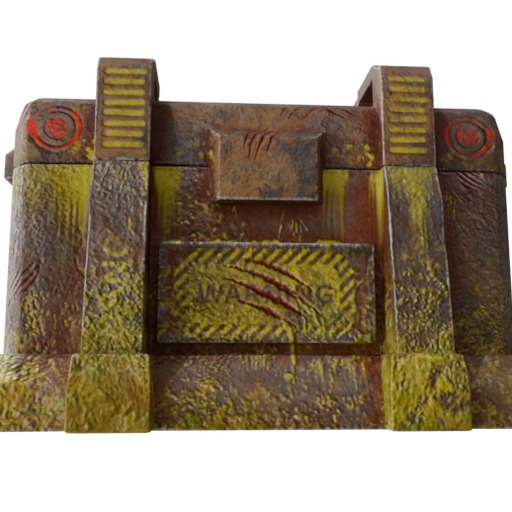} &
\QualGridImage{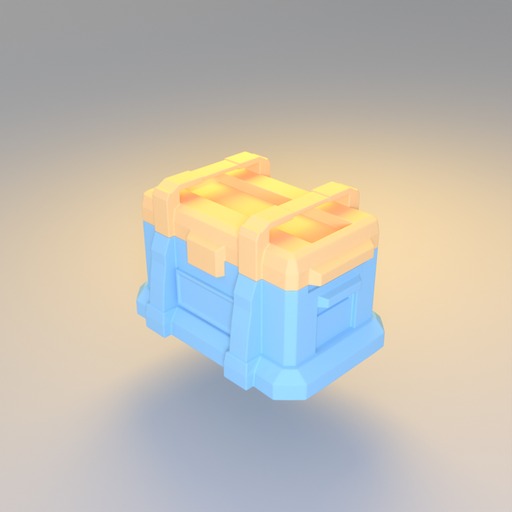} &
\QualGridImage{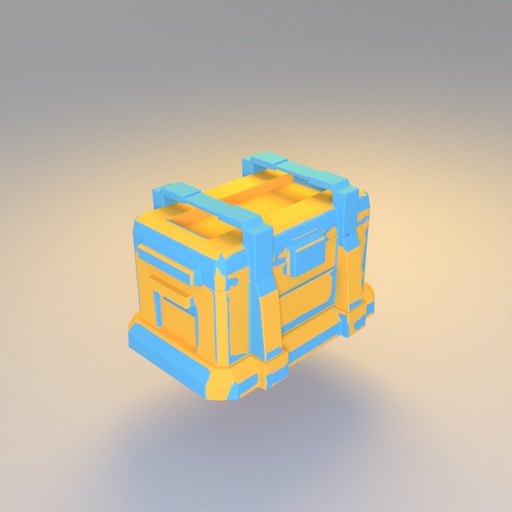} &
\QualGridImage{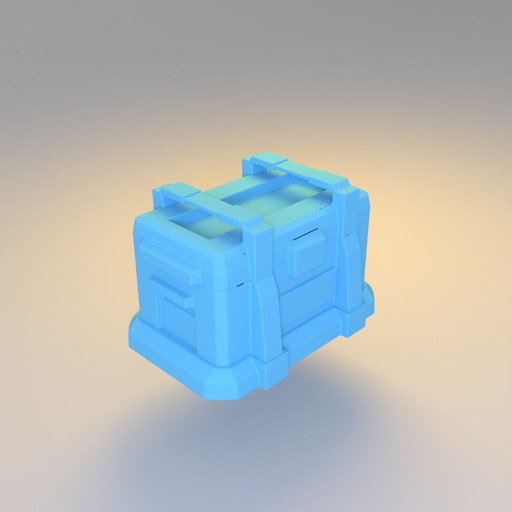} &
\QualGridImage{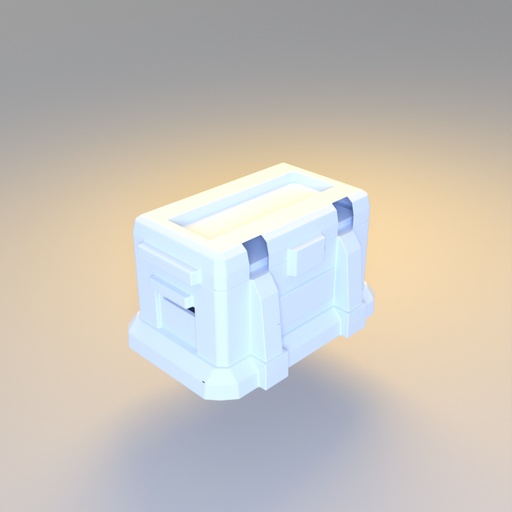} \\
\end{tabular}
\caption{\textbf{Three failure modes, with each volume shown on its own.}
\emph{Top: undersegmentation.} The whole body of the sofa lands in volume $A$
and only the four feet in volume $B$, so base, cushion, arms and bolsters, which
touch and share a volume, come out as one piece. \emph{Middle: single-volume
collapse.} Volume $B$ receives no surface at all; the geometry is right but the
table comes out whole. This happens on $9$ of the $1000$ test objects.
\emph{Bottom: duplicated volumes.} Both streams generate the whole crate, with
$63\%$ of each volume's surface within two fine voxels of the other's, so the
assembled object holds two nearly coincident copies that the part metrics read
as many small pieces. Thin shells like this one sometimes escape the merge
step between the two stages; the disjointness penalty alleviates this mode but
does not resolve it.}
\label{fig:failure}
\end{figure*}

The extension inherits the two-volume assumption from
PartPacker~\cite{partpacker}. When the contact graph is far from bipartite,
parts must be merged until it two-colors, which undersegments objects whose
parts interlock densely, and part quality depends on the per-part annotations
the corpus provides. Like any generator, KaiNinja is stochastic, so the
decomposition varies with the seed. Three failure modes recur
(Figure~\ref{fig:failure}). First, the layout flow sometimes undersegments.
It assigns nearly the whole object to one volume and only a few small parts
to the other, and parts that touch inside the crowded volume are fused,
because nothing within a single volume tells them apart. Second, it sometimes
collapses entirely onto one volume and leaves the other empty, so the object
comes out whole with nothing to separate. Third, it sometimes does the
opposite and generates the whole object in both volumes at once, so the
assembled result carries two nearly coincident copies that the part metrics
read as many small pieces. Although a post-processing step between the two
stages merges overlapping occupancy, thin shells sometimes escape the filter
and lead to this failure mode; the disjointness penalty alleviates it but
does not resolve it. We read all three as the limit of inheritance. The
pretrained weights were shaped by undivided objects, and nothing in them asks
for a balanced split or discourages the two streams from converging on the
same surface. Training on packed volumes from the start would test this
reading, but the compute and data it would take put it beyond this work.
Separately, the Stage-2 decoder drops facets now and then and leaves small
holes, so the released pipeline still runs a light geometric cleanup before
delivery. Natural next steps are packing into more than two volumes, exposing
explicit control over part count and granularity, learning to reassemble the
separated parts, and applying the same inheritance recipe to future
whole-object backbones.

\bibliographystyle{abbrv}
\bibliography{references}

\begin{thebibliography}{10}

\bibitem{besicovitch1947}
A.~S. Besicovitch.
\newblock On crum's problem.
\newblock {\em Journal of the London Mathematical Society}, 22:285--287, 1947.

\bibitem{partgen}
M.~Chen, R.~Shapovalov, I.~Laina, T.~Monnier, J.~Wang, D.~Novotny, and
  A.~Vedaldi.
\newblock {PartGen}: Part-level 3d generation and reconstruction with
  multi-view diffusion models.
\newblock In {\em CVPR}, 2025.

\bibitem{autopartgen}
M.~Chen, J.~Wang, R.~Shapovalov, T.~Monnier, H.~Jung, D.~Wang, R.~Ranjan,
  I.~Laina, and A.~Vedaldi.
\newblock {AutoPartGen}: Autogressive 3d part generation and discovery.
\newblock {\em arXiv preprint arXiv:2507.13346}, 2025.

\bibitem{dora}
R.~Chen, J.~Zhang, Y.~Liang, G.~Luo, W.~Li, J.~Liu, X.~Li, X.~Long, J.~Feng,
  and P.~Tan.
\newblock {Dora}: Sampling and benchmarking for 3d shape variational
  auto-encoders.
\newblock In {\em CVPR}, 2025.

\bibitem{meshanything}
Y.~Chen, T.~He, D.~Huang, W.~Ye, S.~Chen, J.~Tang, X.~Chen, Z.~Cai, L.~Yang,
  G.~Yu, et~al.
\newblock {MeshAnything}: Artist-created mesh generation with autoregressive
  transformers.
\newblock In {\em ICLR}, 2025.

\bibitem{comboverse}
Y.~Chen, T.~Wang, T.~Wu, X.~Pan, K.~Jia, and Z.~Liu.
\newblock {ComboVerse}: Compositional 3d assets creation using spatially-aware
  diffusion guidance.
\newblock In {\em ECCV}, 2024.

\bibitem{objaversexl}
M.~Deitke, R.~Liu, M.~Wallingford, H.~Ngo, O.~Michel, A.~Kusupati, A.~Fan,
  C.~Laforte, V.~Voleti, S.~Y. Gadre, et~al.
\newblock {Objaverse-XL}: A universe of 10m+ 3d objects.
\newblock In {\em NeurIPS}, 2023.

\bibitem{objaverse}
M.~Deitke, D.~Schwenk, J.~Salvador, L.~Weihs, O.~Michel, E.~VanderBilt,
  L.~Schmidt, K.~Ehsani, A.~Kembhavi, and A.~Farhadi.
\newblock {Objaverse}: A universe of annotated 3d objects.
\newblock In {\em CVPR}, 2023.

\bibitem{sd3}
P.~Esser, S.~Kulal, A.~Blattmann, R.~Entezari, J.~M{\"u}ller, H.~Saini,
  Y.~Levi, D.~Lorenz, A.~Sauer, F.~Boesel, D.~Podell, T.~Dockhorn, Z.~English,
  and R.~Rombach.
\newblock Scaling rectified flow transformers for high-resolution image
  synthesis.
\newblock In {\em ICML}, 2024.

\bibitem{sdmnet}
L.~Gao, J.~Yang, T.~Wu, Y.-J. Yuan, H.~Fu, Y.-K. Lai, and H.~Zhang.
\newblock {SDM-NET}: Deep generative network for structured deformable mesh.
\newblock {\em ACM Transactions on Graphics}, 38(6), 2019.

\bibitem{bipartitecontraction}
P.~Heggernes, P.~van~'t Hof, D.~Lokshtanov, and C.~Paul.
\newblock Obtaining a bipartite graph by contracting few edges.
\newblock {\em SIAM Journal on Discrete Mathematics}, 27(4):2143--2156, 2013.

\bibitem{spaghetti}
A.~Hertz, O.~Perel, R.~Giryes, O.~Sorkine-Hornung, and D.~Cohen-Or.
\newblock {SPAGHETTI}: Editing implicit shapes through part aware generation.
\newblock {\em ACM Transactions on Graphics}, 41(4), 2022.

\bibitem{cfg}
J.~Ho and T.~Salimans.
\newblock Classifier-free diffusion guidance.
\newblock {\em arXiv preprint arXiv:2207.12598}, 2022.

\bibitem{img2threejs}
{img2threejs contributors}.
\newblock img2threejs: Rebuilding the object in a reference image as a
  code-only, procedural three.js model.
\newblock GitHub repository, \url{https://github.com/img2threejs/img2threejs},
  2026.
\newblock Apache-2.0 license; accessed 2026-09-01.

\bibitem{gwn}
A.~Jacobson, L.~Kavan, and O.~Sorkine-Hornung.
\newblock Robust inside-outside segmentation using generalized winding numbers.
\newblock {\em ACM Transactions on Graphics}, 32(4), 2013.

\bibitem{dualcontouring}
T.~Ju, F.~Losasso, S.~Schaefer, and J.~Warren.
\newblock Dual contouring of hermite data.
\newblock {\em ACM Transactions on Graphics}, 21(3):339--346, 2002.

\bibitem{sam}
A.~Kirillov, E.~Mintun, N.~Ravi, H.~Mao, C.~Rolland, L.~Gustafson, T.~Xiao,
  S.~Whitehead, A.~C. Berg, W.-Y. Lo, et~al.
\newblock Segment anything.
\newblock In {\em ICCV}, 2023.

\bibitem{salad}
J.~Koo, S.~Yoo, M.~H. Nguyen, and M.~Sung.
\newblock {SALAD}: Part-level latent diffusion for 3d shape generation and
  manipulation.
\newblock In {\em ICCV}, 2023.

\bibitem{articulateanything}
L.~Le, J.~Xie, W.~Liang, H.-J. Wang, Y.~Yang, Y.~J. Ma, K.~Vedder, A.~Krishna,
  D.~Jayaraman, and E.~Eaton.
\newblock Articulate-anything: Automatic modeling of articulated objects via a
  vision-language foundation model.
\newblock In {\em ICLR}, 2025.

\bibitem{craftsman}
W.~Li, J.~Liu, H.~Yan, R.~Chen, Y.~Liang, X.~Chen, P.~Tan, and X.~Long.
\newblock {CraftsMan3D}: High-fidelity mesh generation with 3d native
  generation and interactive geometry refiner.
\newblock {\em arXiv preprint arXiv:2405.14979}, 2024.

\bibitem{triposg}
Y.~Li, Z.-X. Zou, Z.~Liu, D.~Wang, Y.~Liang, Z.~Yu, X.~Liu, Y.-C. Guo,
  D.~Liang, W.~Ouyang, et~al.
\newblock {TripoSG}: High-fidelity 3d shape synthesis using large-scale
  rectified flow models.
\newblock {\em arXiv preprint arXiv:2502.06608}, 2025.

\bibitem{infinitemobility}
X.~Lian, Z.~Yu, R.~Liang, Y.~Wang, L.~R. Luo, K.~Chen, Y.~Zhou, Q.~Tang, X.~Xu,
  Z.~Lyu, B.~Dai, and J.~Pang.
\newblock Infinite mobility: Scalable high-fidelity synthesis of articulated
  objects via procedural generation.
\newblock {\em arXiv preprint arXiv:2503.13424}, 2025.

\bibitem{partcrafter}
Y.~Lin, C.~Lin, P.~Pan, H.~Yan, Y.~Feng, Y.~Mu, and K.~Fragkiadaki.
\newblock {PartCrafter}: Structured 3d mesh generation via compositional latent
  diffusion transformers.
\newblock In {\em NeurIPS}, 2025.

\bibitem{flowmatching}
Y.~Lipman, R.~T. Chen, H.~Ben-Hamu, M.~Nickel, and M.~Le.
\newblock Flow matching for generative modeling.
\newblock In {\em ICLR}, 2023.

\bibitem{part123}
A.~Liu, C.~Lin, Y.~Liu, X.~Long, Z.~Dou, H.-X. Guo, P.~Luo, and W.~Wang.
\newblock Part123: Part-aware 3d reconstruction from a single-view image.
\newblock In {\em ACM SIGGRAPH}, 2024.

\bibitem{cage}
J.~Liu, H.~I.~I. Tam, A.~Mahdavi-Amiri, and M.~Savva.
\newblock Cage: Controllable articulation generation.
\newblock In {\em CVPR}, 2024.

\bibitem{partfield}
M.~Liu, M.~A. Uy, D.~Xiang, H.~Su, S.~Fidler, N.~Sharp, and J.~Gao.
\newblock {PartField}: Learning 3d feature fields for part segmentation and
  beyond.
\newblock In {\em ICCV}, 2025.

\bibitem{one2345}
M.~Liu, C.~Xu, H.~Jin, L.~Chen, {Mukund Varma T}, Z.~Xu, and H.~Su.
\newblock One-2-3-45: Any single image to 3d mesh in 45 seconds without
  per-shape optimization.
\newblock In {\em NeurIPS}, 2023.

\bibitem{partslip}
M.~Liu, Y.~Zhu, H.~Cai, S.~Han, Z.~Ling, F.~Porikli, and H.~Su.
\newblock {PartSLIP}: Low-shot part segmentation for 3d point clouds via
  pretrained image-language models.
\newblock In {\em CVPR}, 2023.

\bibitem{pact}
Q.~Liu, X.~Yao, S.~Zhang, Y.~Deng, G.~Liu, Z.~Liu, and K.~Jia.
\newblock Pact: Part-decomposed single-view articulated object generation.
\newblock {\em arXiv preprint arXiv:2602.14965}, 2026.

\bibitem{rectifiedflow}
X.~Liu, C.~Gong, and Q.~Liu.
\newblock Flow straight and fast: Learning to generate and transfer data with
  rectified flow.
\newblock In {\em ICLR}, 2023.

\bibitem{lato2}
H.~Long, T.~Zhao, J.~Lin, Y.~Zhang, H.~Guo, R.~Liang, J.~Xu, J.~Hladk{\'y},
  M.~Nie{\ss}ner, Y.~Hu, and W.~Yang.
\newblock {LATO.2}: Factorized 3d mesh generation with vertex and topology
  flow.
\newblock {\em arXiv preprint arXiv:2607.10623}, 2026.

\bibitem{wonder3d}
X.~Long, Y.-C. Guo, C.~Lin, Y.~Liu, Z.~Dou, L.~Liu, Y.~Ma, S.-H. Zhang,
  M.~Habermann, C.~Theobalt, et~al.
\newblock {Wonder3D}: Single image to 3d using cross-domain diffusion.
\newblock In {\em CVPR}, 2024.

\bibitem{adamw}
I.~Loshchilov and F.~Hutter.
\newblock Decoupled weight decay regularization.
\newblock In {\em ICLR}, 2019.

\bibitem{p3sam}
C.~Ma, Y.~Li, X.~Yan, J.~Xu, Y.~Yang, C.~Wang, Z.~Zhao, Y.~Guo, Z.~Chen, and
  C.~Guo.
\newblock {P3-SAM}: Native 3d part segmentation.
\newblock {\em arXiv preprint arXiv:2509.06784}, 2025.

\bibitem{structurenet}
K.~Mo, P.~Guerrero, L.~Yi, H.~Su, P.~Wonka, N.~J. Mitra, and L.~J. Guibas.
\newblock {StructureNet}: Hierarchical graph networks for 3d shape generation.
\newblock {\em ACM Transactions on Graphics}, 38(6), 2019.

\bibitem{dreamfusion}
B.~Poole, A.~Jain, J.~T. Barron, and B.~Mildenhall.
\newblock {DreamFusion}: Text-to-3d using 2d diffusion.
\newblock In {\em ICLR}, 2023.

\bibitem{infinigen}
A.~Raistrick, L.~Lipson, Z.~Ma, L.~Mei, M.~Wang, Y.~Zuo, K.~Kayan, H.~Wen,
  B.~Han, Y.~Wang, A.~Newell, H.~Law, A.~Goyal, K.~Yang, and J.~Deng.
\newblock Infinite photorealistic worlds using procedural generation.
\newblock In {\em CVPR}, 2023.

\bibitem{sam2}
N.~Ravi, V.~Gabeur, Y.-T. Hu, R.~Hu, C.~Ryali, T.~Ma, H.~Khedr, R.~R{\"a}dle,
  C.~Rolland, L.~Gustafson, et~al.
\newblock {SAM 2}: Segment anything in images and videos.
\newblock In {\em ICLR}, 2025.

\bibitem{zero123pp}
R.~Shi, H.~Chen, Z.~Zhang, M.~Liu, C.~Xu, X.~Wei, L.~Chen, C.~Zeng, and H.~Su.
\newblock Zero123++: A single image to consistent multi-view diffusion base
  model.
\newblock {\em arXiv preprint arXiv:2310.15110}, 2023.

\bibitem{meshgpt}
Y.~Siddiqui, A.~Alliegro, A.~Artemov, T.~Tommasi, D.~Sirigatti, V.~Rosov,
  A.~Dai, and M.~Nie{\ss}ner.
\newblock {MeshGPT}: Generating triangle meshes with decoder-only transformers.
\newblock In {\em CVPR}, 2024.

\bibitem{dinov3}
O.~Sim{\'e}oni, H.~V. Vo, M.~Seitzer, et~al.
\newblock {DINOv3}.
\newblock {\em arXiv preprint arXiv:2508.10104}, 2025.

\bibitem{samesh}
G.~Tang, W.~Zhao, L.~Ford, D.~Benhaim, and P.~Zhang.
\newblock Segment any mesh.
\newblock {\em arXiv preprint arXiv:2408.13679}, 2024.

\bibitem{partpacker}
J.~Tang, R.~Lu, Z.~Li, Z.~Hao, X.~Li, F.~Wei, S.~Song, G.~Zeng, M.-Y. Liu, and
  T.-Y. Lin.
\newblock Efficient part-level 3d object generation via dual volume packing.
\newblock In {\em NeurIPS}, 2025.

\bibitem{hunyuan3d21}
{Team Hunyuan3D}, S.~Yang, M.~Yang, Y.~Feng, X.~Huang, S.~Zhang, Z.~He, D.~Luo,
  H.~Liu, Y.~Zhao, et~al.
\newblock {Hunyuan3D} 2.1: From images to high-fidelity 3d assets with
  production-ready pbr material.
\newblock {\em arXiv preprint arXiv:2506.15442}, 2025.

\bibitem{tietze1905}
H.~Tietze.
\newblock {\"U}ber das problem der nachbargebiete im raum.
\newblock {\em Monatshefte f{\"u}r Mathematik und Physik}, 16:211--216, 1905.

\bibitem{nexus}
H.~Wang, Y.-T. Liu, Y.-C. Guo, Q.-Y. Feng, Z.-X. Zou, D.~Liang, B.~Zhang, and
  Y.-P. Cao.
\newblock Nexus: Native mesh generation with diffusion.
\newblock {\em ACM Transactions on Graphics}, 45(4), 2026.

\bibitem{partnext}
P.~Wang, Y.~He, X.~Lv, Y.~Zhou, L.~Xu, J.~Yu, and J.~Gu.
\newblock {PartNeXt}: A next-generation dataset for fine-grained and
  hierarchical 3d part understanding.
\newblock In {\em NeurIPS Datasets and Benchmarks Track}, 2025.
\newblock arXiv:2510.20155.

\bibitem{fusion360}
K.~D.~D. Willis, Y.~Pu, J.~Luo, H.~Chu, T.~Du, J.~G. Lambourne,
  A.~Solar-Lezama, and W.~Matusik.
\newblock Fusion 360 gallery: A dataset and environment for programmatic {CAD}
  construction from human design sequences.
\newblock {\em ACM Transactions on Graphics}, 40(4), 2021.

\bibitem{direct3d}
S.~Wu, Y.~Lin, Y.~Zeng, F.~Zhang, J.~Xu, P.~Torr, X.~Cao, and Y.~Yao.
\newblock {Direct3D}: Scalable image-to-3d generation via 3d latent diffusion
  transformer.
\newblock In {\em NeurIPS}, 2024.

\bibitem{trellis2}
J.~Xiang, X.~Chen, S.~Xu, R.~Wang, Z.~Lv, Y.~Deng, H.~Zhu, Y.~Dong, H.~Zhao,
  N.~J. Yuan, and J.~Yang.
\newblock Native and compact structured latents for 3d generation.
\newblock {\em arXiv preprint arXiv:2512.14692}, 2025.

\bibitem{trellis}
J.~Xiang, Z.~Lv, S.~Xu, Y.~Deng, R.~Wang, B.~Zhang, D.~Chen, X.~Tong, and
  J.~Yang.
\newblock Structured 3d latents for scalable and versatile 3d generation.
\newblock In {\em CVPR}, 2025.

\bibitem{instantmesh}
J.~Xu, W.~Cheng, Y.~Gao, X.~Wang, S.~Gao, and Y.~Shan.
\newblock {InstantMesh}: Efficient 3d mesh generation from a single image with
  sparse-view large reconstruction models.
\newblock {\em arXiv preprint arXiv:2404.07191}, 2024.

\bibitem{meshyt2}
J.~Xu, R.~Liang, Y.~Long, S.~Shen, Z.~Xian, X.~Wu, Z.~Xu, and Y.~Hu.
\newblock Meshy t2: Fast native mesh generation with flow matching.
\newblock {\em arXiv preprint arXiv:2607.28675}, 2026.

\bibitem{phycage}
H.~Yan, M.~Zhang, Y.~Li, C.~Ma, and P.~Ji.
\newblock {PhyCAGE}: Physically plausible compositional 3d asset generation
  from a single image.
\newblock {\em arXiv preprint arXiv:2411.18548}, 2024.

\bibitem{xpart}
X.~Yan, J.~Xu, Y.~Li, C.~Ma, Y.~Yang, C.~Wang, Z.~Zhao, Z.~Lai, Y.~Zhao,
  Z.~Chen, and C.~Guo.
\newblock {X-Part}: High fidelity and structure coherent shape decomposition.
\newblock {\em arXiv preprint arXiv:2509.08643}, 2025.

\bibitem{holopart}
Y.~Yang, Y.-C. Guo, Y.~Huang, Z.-X. Zou, Z.~Yu, Y.~Li, Y.-P. Cao, and X.~Liu.
\newblock {HoloPart}: Generative 3d part amodal segmentation.
\newblock {\em arXiv preprint arXiv:2504.07943}, 2025.

\bibitem{sampart3d}
Y.~Yang, Y.~Huang, Y.-C. Guo, L.~Lu, X.~Wu, E.~Y. Lam, Y.-P. Cao, and X.~Liu.
\newblock {SAMPart3D}: Segment any part in 3d objects.
\newblock {\em arXiv preprint arXiv:2411.07184}, 2024.

\bibitem{omnipart}
Y.~Yang, Y.~Zhou, Y.-C. Guo, Z.-X. Zou, Y.~Huang, Y.-T. Liu, H.~Xu, D.~Liang,
  Y.-P. Cao, and X.~Liu.
\newblock {OmniPart}: Part-aware 3d generation with semantic decoupling and
  structural cohesion.
\newblock In {\em SIGGRAPH Asia}, 2025.

\bibitem{hi3dgen}
C.~Ye, Y.~Wu, Z.~Lu, J.~Chang, X.~Guo, J.~Zhou, H.~Zhao, and X.~Han.
\newblock {Hi3DGen}: High-fidelity 3d geometry generation from images via
  normal bridging.
\newblock In {\em ICCV}, 2025.

\bibitem{shape2vecset}
B.~Zhang, J.~Tang, M.~Niessner, and P.~Wonka.
\newblock {3DShape2VecSet}: A 3d shape representation for neural fields and
  generative diffusion models.
\newblock {\em ACM Transactions on Graphics (TOG)}, 42(4), 2023.

\bibitem{clay}
L.~Zhang, Z.~Wang, Q.~Zhang, Q.~Qiu, A.~Pang, H.~Jiang, W.~Yang, L.~Xu, and
  J.~Yu.
\newblock {CLAY}: A controllable large-scale generative model for creating
  high-quality 3d assets.
\newblock {\em ACM Transactions on Graphics (TOG)}, 43(4), 2024.

\bibitem{hunyuan3d}
Z.~Zhao, Z.~Lai, Q.~Lin, Y.~Zhao, H.~Liu, S.~Yang, Y.~Feng, M.~Yang, S.~Zhang,
  X.~Yang, et~al.
\newblock Hunyuan3d 2.0: Scaling diffusion models for high resolution textured
  3d assets generation.
\newblock {\em arXiv preprint arXiv:2501.12202}, 2025.

\bibitem{michelangelo}
Z.~Zhao, W.~Liu, X.~Chen, X.~Zeng, R.~Wang, P.~Cheng, B.~Fu, T.~Chen, G.~Yu,
  and S.~Gao.
\newblock Michelangelo: Conditional 3d shape generation based on
  shape-image-text aligned latent representation.
\newblock In {\em NeurIPS}, 2023.

\bibitem{articraft}
M.~Zhou, R.~Li, X.~Lyu, Z.~Song, Z.~Huang, C.~Zheng, C.~Rupprecht, A.~Vedaldi,
  and S.~Wu.
\newblock Articraft: An agentic system for scalable articulated 3d asset
  generation.
\newblock {\em arXiv preprint arXiv:2605.15187}, 2026.

\end{thebibliography}

\clearpage
\beginappendix

\section{Hyperparameters}
\label{app:hparams}

Both flows use a width of $1536$, $30$ blocks, $12$ heads and rotary position
encoding. They are trained in bfloat16 with AdamW~\cite{adamw}
($\beta=(0.9,0.95)$, weight decay $0.01$), an EMA rate of $0.9999$, adaptive
gradient clipping at norm $1.0$, a logit-normal timestep schedule~\cite{sd3},
and $p_{\text{uncond}}=0.1$ for classifier-free guidance. The global batch
size is $32$ throughout.

Stage~1, the layout flow, works on a $16^3$ sparse structure latent with $8$
channels per stream and places a cross-volume attention block after depths
$\{6,12,18,24,29\}$. Its frozen warmup runs $10$K steps at learning rate
$10^{-4}$ with $500$ warmup steps. Joint fine-tuning runs $100$K steps at
$5\times10^{-5}$ with $2000$ warmup steps and the disjointness penalty at
$\lambda_{\text{ov}}=5$.

Stage~2, the refinement flow, works on the $32$-channel structured latent at
resolution $32$ and merges the attention pools at every third block, depths
$\{2,5,8,\dots,29\}$. Its frozen warmup runs $30$K steps and full fine-tuning
$100$K steps, both at learning rate $3\times10^{-5}$. The structured latent
autoencoder is the released TRELLIS.2 one and is never trained.

\section{Two Volumes and Graph Coloring}
\label{app:coloring}
Two-coloring is a heuristic, not a guarantee, because it works only when the
contact graph is bipartite. PartPacker suggests lifting this restriction by
making the connectivity graph planar and applying the four color theorem, but
that step does not work in general. A 3D contact graph need not be planar.
Tietze~\cite{tietze1905} showed that any number of regions in space can be
pairwise adjacent, and Besicovitch~\cite{besicovitch1947} showed the same for
convex polyhedra, so the complete graph on any number of parts is a 3D
contact graph and no planar-map argument applies. Making such a graph planar
means deleting edges. An edge can be deleted in two ways. Merging its two
parts is what we and PartPacker do, at the cost of a coarser decomposition.
Breaking the contact instead, by moving one part away from the other, keeps
both parts but changes the assembled position of the object, so the model
would generate an exploded object and a dedicated network would likely be
needed to predict the displacements and reassemble it; we leave this route to
future work. Deleting an edge without either of these puts two touching parts
back into the same volume, which defeats the packing. In fact, the chromatic number
of a 3D contact graph is unbounded, so no fixed number of volumes is safe in
the worst case. PartPacker and KaiNinja both merge parts until the graph is
bipartite, which undersegments objects whose parts interlock densely.
Therefore, packing does not give a correct coloring in general. What it does
give is a volume in which no part touches another, so each part can be
generated without overwriting its neighbors.

\end{document}